\documentclass[12pt]{article}
\pdfoutput=1
\usepackage{bbm}
\usepackage{color}
\usepackage{diagbox}
\usepackage{authblk}
\usepackage{latexsym}
\usepackage{epsfig,amssymb,euscript, mathrsfs,cite}
\usepackage{amsmath}
\usepackage{mathrsfs} 
\usepackage{enumerate}
\usepackage{slashed}
\definecolor{MyDarkBlue}{rgb}{0.15,0.15,0.45}
\usepackage[linktocpage=true]{hyperref}
\hypersetup{
colorlinks=true,
citecolor=MyDarkBlue,
linkcolor=MyDarkBlue,
urlcolor=MyDarkBlue,
pdfauthor={},
pdftitle={},
pdfsubject={hep-th}
}

\newcounter{savefootnote}
\newcounter{symfootnote}
\newcommand{\symfootnote}[1]{%
   \setcounter{savefootnote}{\value{footnote}}%
   \setcounter{footnote}{\value{symfootnote}}%
   \ifnum\value{footnote}>8\setcounter{footnote}{0}\fi%
   \let\oldthefootnote=\thefootnote%
   \renewcommand{\thefootnote}{\fnsymbol{footnote}}%
   \footnote{#1}%
   \let\thefootnote=\oldthefootnote%
   \setcounter{symfootnote}{\value{footnote}}%
   \setcounter{footnote}{\value{savefootnote}}%
}

\usepackage{mathtools}

\newsavebox{\ns}
\newsavebox{\dbrane}
\newsavebox{\dbshort}

\def\be{\begin{equation}}
\def\ee{\end{equation}}
\def\bea{\begin{eqnarray}}
\def\eea{\end{eqnarray}}

\newcommand{\nn}{\notag \\}

\def\cO{{\mathcal O}}
\def\eq#1 { \begin{equation} #1 \end{equation} }

\def\Lt{{\tilde L}}

\def\cE{\mathcal{E}}

\newcommand{\vol}{\mathrm{vol}}

\newlength{\sswidth}

\def\eq#1 { \begin{equation} #1 \end{equation} }

\def\lny{\ln[\tfrac{y}{\Lambda}]}
\def\a{\alpha}
\def\b{\beta}
\def\g{\gamma}
\def\cT{\mathcal{T}}
\newcommand{\vev}[1]{\left< #1 \right>} 
\newcommand{\diag}[1]{{\mathrm{diag}\{ #1 \}}}
\def\cA{\mathcal{A}}
\newcommand{\atanh}{{\mathrm{arctanh}}}

\newcommand{\phis}{\varphi_{s}}
\newcommand{\phiv}{\varphi_{v}}
\newcommand{\alphas}{\alpha_{s}}
\newcommand{\alphav}{\alpha_{v}}
\newcommand{\betas}{\beta_{s}}
\newcommand{\betav}{\beta_{v}}

\numberwithin{equation}{section}       

\begin{document}

\begin{titlepage}

\vfill

\begin{flushright}
APCTP Pre2024-005\\
CCTP-2024-9,  ITCP-2024/9
\end{flushright}

\begin{center}
   \baselineskip=16pt
   {\Large\bf Superconformal Monodromy Defects\\ in $\mathcal{N}$=4 SYM and LS theory}
  \vskip 1cm
Igal Arav$^1$,  Jerome P. Gauntlett$^2$, Yusheng Jiao$^2$\\
Matthew M. Roberts$^{3,4}$ and Christopher Rosen$^5$\\
     \vskip .5cm     
                          \begin{small}
                                \textit{$^1$KU Leuven, Instituut voor Theoretische Fysica,\\
                                Celestijnenlaan 200D, B-3001 Leuven, Belgium}
        \end{small}\\
        \begin{small}\vskip .3cm
      \textit{$^2$Blackett Laboratory, 
  Imperial College\\ Prince Consort Rd., London, SW7 2AZ, U.K.}
        \end{small}\\
                \begin{small}\vskip .3cm
      \textit{$^3$Asia Pacific Center for Theoretical Physics,\\
      Pohang, 37673, Korea}
        \end{small}\\
                \begin{small}\vskip .3cm
      \textit{$^4$Department of Physics, Pohang University of Science and Technology,\\
      Pohang 37673, Korea}
        \end{small}\\
             \begin{small}\vskip .3cm
      \textit{$^5$Crete Center for Theoretical Physics, Department of Physics, University of Crete,\\
71003 Heraklion, Greece}
        \end{small}\\
                       \end{center}
\vfill

\begin{center}
\textbf{Abstract}
\end{center}
\begin{quote}
We study type IIB supergravity solutions that are dual to two-dimensional superconformal defects in $d=4$ SCFTs which preserve
$\mathcal{N}=(0,2)$ supersymmetry. 
We consider solutions dual to defects in $\mathcal{N}=4$ SYM theory that have non-trivial monodromy
for $U(1)^3\subset SO(6)$ global symmetry and we also allow for 
the possibility of conical singularities.  In addition, we consider the addition of fermionic and bosonic mass terms 
that have non trivial dependence on the spatial directions transverse to the defect, while preserving the superconformal symmetry of the defect.
We compute various physical quantities including the central charges of the defect expressed as a function of the monodromy,
the on-shell action as well as
associated supersymmetric Renyi entropies. Analogous computations are carried out for superconformal defects
in the $\mathcal{N}=1$, $d=4$ Leigh-Strassler SCFT. 
We also show that the defects of  the two SCFTs are connected by a line of bulk marginal mass deformations and argue that they are also related by bulk RG flow.

\end{quote}

\vfill

\end{titlepage}

\tableofcontents

\newpage

\section{Introduction}\label{sec:intro}

Defects are objects of fundamental interest in quantum field theory.
It is particularly fruitful to consider defects in the context of supersymmetric conformal field theories (SCFTs) with 
the defect
preserving some residual conformal supersymmetry.
In this paper, we study $d=4$ SCFTs with $d=2$ defects that preserve $\mathcal{N}=(0,2)$ superconformal symmetry.

More specifically, we consider co-dimension two monodromy defects in flat spacetime that are characterised by
the presence of flat, background gauge fields for an abelian subgroup of the global symmetry, with non-trivial monodromy as one circles around the linear defect in space. Equivalently, there is a delta function magnetic flux for the global symmetry at the location of the defect. 
We also allow for the possibility of a conical singularity at the origin.
We study such defects both in $\mathcal{N}=4$ SYM theory as well as in the 
$\mathcal{N}=1$ Leigh-Strassler (LS) SCFT \cite{Leigh:1995ep}, and we recall that the latter can be obtained as the IR limit
of an RG flow from a supersymmetric
mass deformation of the former.

By carrying out a Weyl transformation of flat spacetime, we can also consider the defects in the context of SCFTs
on $AdS_3\times S^1$. In this setting, effectively, the defect has been pushed out to infinity and instead
there is a non-trivial monodromy of the global symmetry around the $S^1$. Furthermore, the data associated with the 
conical singularity is replaced with the ratio of the radii of the $AdS_3$ and $S^1$ factors. General aspects of SCFTs on $AdS_3\times S^1$
have been discussed in \cite{Aharony:2015hix}. 

We analyse these monodromy defects in type IIB supergravity using holography and
compute several observables including the
conformal weight of the defect, $h_D$, the $b, d_1$ and $d_2$ central charges (reviewed below), as well as the on-shell action. 
For the case of $\mathcal{N}=4$ SYM theory, with $SU(4)$ R-symmetry, we consider a family of $U(1)^3\subset SU(4)$ monodromy defects, extending the investigations\footnote{Co-dimension two monodromy defects have been studied holographically in various contexts e.g.\cite{Gutperle:2018fea,Chen:2020mtv,Gutperle:2020rty,
Giombi:2021uae,Gutperle:2022pgw,Gutperle:2023yrd,Capuozzo:2023fll}. We also note that
the monodromy defects that we study differ from the surface operators considered in 
\cite{Gukov:2006jk,Gomis:2007fi,Drukker:2008wr} which
are parametrised by a conjugacy class of the gauge group.} of
\cite{Gutperle:2019dqf} who considered defects preserving $\mathcal{N}=(2,2)$ supersymmetry. 
In addition, we also make a comparison with defects in free $\mathcal{N}=4$ SYM theory, finding values
for $b,d_2,h_D$ that are the same at weak and strong coupling.
Recall that the LS SCFT has $SU(2)\times U(1)$ global symmetry, where the $U(1)$ is the R-symmetry, and we consider defects in this theory with monodromy for $U(1)^2\subset SU(2)\times U(1)$.

\begin{figure}[htbp!]
\begin{center}
\includegraphics[scale=.35]{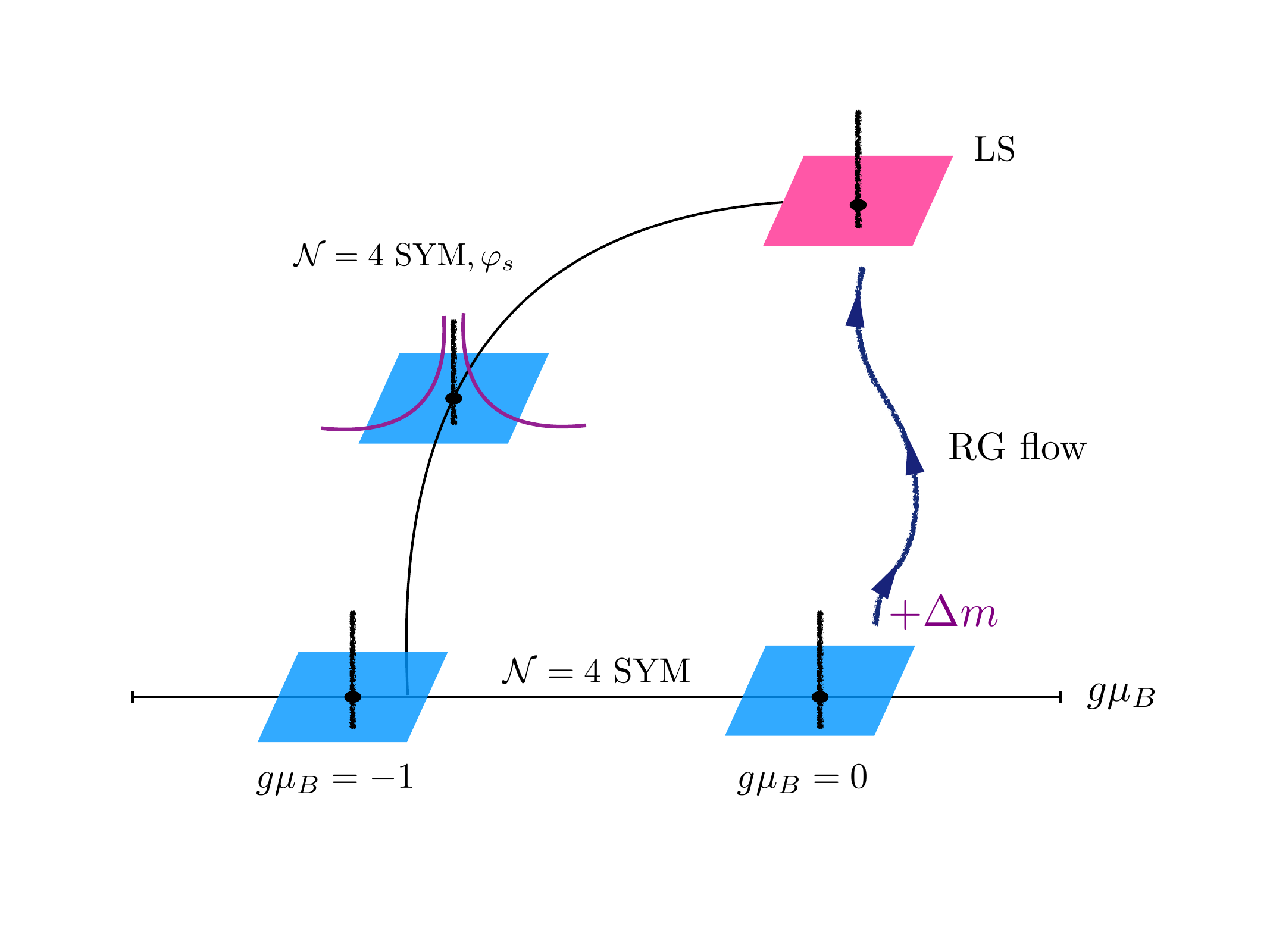}~
\caption{An illustration of part of the solution space of $\mathcal{N}=(0,2)$
defects in flat spacetime (time suppressed)
with no conical singularities ($n=1$) and hence vanishing monodromy for the R-symmetry, $g\mu_R=0$.
The horizontal line corresponds to $\mathcal{N}=4$ SYM defect solutions parametrised by monodromy parameters 
$g\mu_B$ and $g\mu_F$ (suppressed). For $g\mu_B=-1$ a branch of solutions emerges, parametrised by
$\varphi_s$ and associated with spatially dependent mass deformations preserving the superconformal invariance of the defect.
As $\varphi_s\to\infty$ these asymptote to the solutions describing defects in the LS theory, which have $g\mu_B=0$ and 
parametrised by $g\mu_F$. RG flows driven by homogeneous mass deformations ($\Delta m$) away from the 
$\mathcal{N}=4$ SYM defects with $g\mu_B=0$ and $\mathcal{N}=(2,2)$ supersymmetry,
and breaking bulk conformal invariance, to defects in the LS theory are also indicated.
}
\label{fig:sumsolutions}
\end{center}
\end{figure}

For the case of $\mathcal{N}=4$ SYM we also study monodromy defects in flat spacetime which, in addition, 
have supersymmetric mass deformations that depend on the spatial coordinates transverse to the defect. 
The bosonic and fermionic mass operators, have protected scaling dimensions $\Delta=2,3$ and deformations with spatial dependence of the form $\rho^{\Delta-4}$, where $\rho$ is the spatial distance from the defect in flat space, 
will preserve the superconformal symmetry of the defect.
These deformations are a natural generalisation of the spatially dependent mass
deformations transverse to co-dimension one interfaces that were studied in \cite{Arav:2018njv,Arav:2020asu,Arav:2020obl}, and they have also been discussed more generally in \cite{Herzog:2019bom} where they are referred to as `bulk marginal deformations for the defect'. Similar to \cite{Arav:2020obl} one can determine which of these spatially dependent mass deformations
preserve supersymmetry and superconformal invariance by coupling $\mathcal{N}=4$ SYM to $\mathcal{N}=4$ conformal supergravity, as in \cite{Maxfield:2016lok}. In the $AdS_3\times S^1$ Weyl frame these mass deformations are simply constant mass deformations. 
We find a one-parameter family of such defect solutions of $\mathcal{N}=4$ SYM theory. Interestingly, we show that for large values of the deformation the defect solutions approach that of the LS theory, as illustrated in figure \ref{fig:sumsolutions}. 
Remarkably, this feature, combined with the fact that the on-shell action is independent of the deformation,
allows us to deduce properties of the LS defects just from the properties of 
defects of $\mathcal{N}=4$ SYM theory.
The figure also illustrates the existence of RG flows
that are driven by homogeneous (in flat spacetime) mass deformations of $\mathcal{N}=4$ SYM theory, away from the defect, that arrive at
the LS defects in the infrared.

We study the type IIB supergravity equations of motion using a sub-truncation of $D=5$ maximal $SO(6)$ gauged supergravity. By suitably analysing the supergravity BPS equations we compute one-point functions for
the stress tensor and the conserved currents and also relate these to the monodromy sources. As we shall see, somewhat remarkably, we obtain many results without
having access to an analytic solution, but just by carefully examining the boundary conditions and utilising various features of the BPS equations. We also compute the holographic on-shell action and, after a simple analytic continuation, this allows us to compute supersymmetric Renyi entropies (SREs) for the $d=4$ SCFTs associated with
a spherical entangling surface \cite{Nishioka:2013haa,Nishioka:2014mwa}. For the case of $\mathcal{N}=4$ SYM without mass deformations we make contact with previous computations \cite{Huang:2014pda,Crossley:2014oea}, as well as highlighting that there is some ambiguity in how
one defines the supersymmetric Renyi entropy in this context, that was not discussed in \cite{Huang:2014pda,Crossley:2014oea}. 
We also compute the on-shell action for defects of $\mathcal{N}=4$ SYM theory with 
mass deformations as well as for the LS theory and, in addition, compute SREs for defects in the LS theory.
Before summarising the structure of the paper, we first 
review some background material regarding conformal defects that are relevant for
our analysis.

\subsection{Comments on defect central charges}
Consider a SCFT on a four dimensional ambient spacetime, $M_4$,
and a $d=2$ conformal defect localised on $\Sigma\subset M_4$. For flat $M_4$ and flat $\Sigma$, the conformal defect breaks the $SO(4,2)$ conformal symmetry to $SO(2,2)\times SO(2)$ and we are interested in the case that the preserved supersymmetry is a chiral $\mathcal{N}=(0,2)$ supersymmetry in $d=2$. To begin we assume that
there are no spatially dependent deformations transverse to the defect, but we do allow for
a conical singularity at the location of the defect in flat spacetime parametrised by $n$, with
metric given by
\begin{align}\label{flatmet}
ds^2=-dt^2+dx^2+d\rho^2+n^2\rho^2 dz^2\,,
\end{align}
with $\Delta z=2\pi$. 
With $n< 1$, there is a conical deficit angle, while if $n> 1$, there is a conical excess angle.

The presence of the defect breaks translations normal to $\Sigma$. This
gives rise to a displacement operator $\mathcal{D}^\mu$, a $d=4$ vector normal to the defect,
and we have
\begin{align}\label{defopdef}
\nabla_\mu T^{\mu i}=\delta(\Sigma)\mathcal D^i\,,
\end{align}
where $x^i$ are labelling directions transverse to the defect. 

The trace anomaly of the stress tensor includes the usual contributions associated with the ambient CFT plus a delta function that is localised on $\Sigma$
leading to three\footnote{This ignores possible parity odd terms as discussed in
\cite{Cvitan:2015ana,Chalabi:2021jud}.}
 further ``central charges".
In the conventions of e.g. \cite{Bianchi:2021snj} the terms depending on the background metric and embedding are given by 
\begin{align}\label{confanomalydefect}
\langle T^\mu{}_\mu\rangle |_\Sigma=-\frac{1}{24\pi}\left(b R+d_1 \tilde K^\mu_{ab}\tilde K_\mu^{ab}-d_2 W_{ab}{}^{ab}\right)\,,
\end{align}
where 
$\tilde K^\mu_{ab}$ is the traceless part of the second fundamental form of $\Sigma$ and 
$W_{abcd}$ is the pull back of the Weyl tensor to the defect. In the language of \cite{Deser:1993yx},
$b$ is a type A anomaly, while $d_1,d_2$ are type B; as such $b$ cannot depend
on defect marginal couplings while $d_1,d_2$ can. In general $b,d_1$ and $d_2$ can depend on bulk marginal couplings
\cite{Herzog:2019rke,Bianchi:2019umv}, but
for two-dimensional defects preserving $\mathcal{N}=(0,2)$ supersymmetry,
it has been argued that $b$ does not depend on some specific bulk marginal couplings \cite{Bianchi:2019umv}.
In appendix \ref{app:freefield} we will argue that this could also be true for $d_2$ (and hence $h_D$, discussed below).
In addition, it is known that $b$ satisfies a ``$b$-theorem" for defect RG flows i.e. flows driven by deformations localised on the defect \cite{Jensen:2015swa,Casini:2016fgb}, with $b$ decreasing from the UV to the IR
(see also \cite{Shachar:2022fqk}). We also note that for superconformal defects there is a notion of $b$-extremisation \cite{Wang:2020xkc}.
In this paper
we will discuss RG flows that are driven by spatially homogeneous deformations that are \emph{not} localised on the defect; 
interestingly we find that in some of the cases we consider $b$ \emph{increases}\footnote{Such increasing behaviour has also seen in non-holographic examples \cite{Shachar:2024ubf}.} from the UV to the IR 
(but it can also decrease as well as stay the same value). 
However, the on-shell action in our examples is proportional to $12an-b$, where $a$ is the central charge of the SCFT and $n$ parametrises a possible conical singularity (see \eqref{flatmet}), and remarkably we find that this \emph{decreases} in the RG flows we study
and we conjecture that this result may hold more generally for such flows.

For flat defects embedded in flat spacetime the one-point function of the stress tensor can be written, in mostly plus signature, as
\begin{align}\label{teeaich}
\langle T^{ab}\rangle =-\frac{h_D}{2\pi}\frac{\eta^{ab}}{\rho^4}\,,\qquad
\langle T^{ij}\rangle =\frac{h_D}{2\pi}\frac{3\delta^{ij}-4\frac{x^ix^j}{\rho^2}}{\rho^4}\,,
\end{align}
where $h_D$ is the conformal weight of the defect \cite{Kapustin:2005py}.
Here $x^a$ are coordinates tangent to $\Sigma$, $x^i$ are coordinates transverse to $\Sigma$ and $\rho^2=x^i x^i$.
By extension of the results in 
\cite{Lewkowycz:2014jia,Bianchi:2015liz,Jensen:2018rxu,Bianchi:2021snj} we have that $h_D$ is related to $d_2$ in \eqref{confanomalydefect} via
\begin{align}\label{d2hd}
d_2=18\pi n h_D\,.
\end{align}
When $n=1$ it has been argued that $h_D\ge 0$ \cite{Jensen:2018rxu} when the ANEC holds. However, if $n<1$ then it is possible for
$h_D<0$ as seen for free fields \cite{Dowker:1987mn,Dowker:2015qta} and in holographic models \cite{Bianchi:2016xvf,Baiguera:2022sao}; we will
see additional (supersymmetric) 
examples in this paper too. It would be interesting to understand why the ANEC argument of \cite{Jensen:2018rxu} fails for
these $n<1$ and $h_D<0$ cases.
It is also known that $d_1$ in \eqref{confanomalydefect} is determined by the coefficient of the two-point function of the displacement operator in flat space \cite{Bianchi:2015liz,Bianchi:2021snj}, $C_D$, at least when $n=1$; when this is true one expects $d_1\ge 0$.
However, it is not clear to what extent this holds for $n\ne 1$; in particular it seems that $d_1$ can be negative for
$n<1$ (see the holographic computations in \cite{Bianchi:2016xvf,Baiguera:2022sao}) and it would be interesting to understand why this is.

When the $d=2$ defect in the $d=4$ SCFT preserves $\mathcal{N}=(0,2)$  supersymmetry,  
the case of interest in this paper, the supersymmetry algebra implies $d_1=d_2$ \cite{Bianchi:2019sxz},
when $n=1$. It seems likely that $d_1=d_2$ is also true for general $n$, but this has not yet been proven.
We also note that some evidence has been provided that $h_D$ is related to $C_D$ in supersymmetric and holographic contexts for general $n$ and dimensions, including supersymmetric defect solutions of $D=4$ minimal gauged supergravity \cite{Baiguera:2022sao}.
Therefore, in these cases one expects supersymmetry to also imply the positivity of $h_D$. As noted above, however, we will see examples of D = 4 supersymmetric defects with $h_D < 0$ for $n < 1$, again indicating some tension with the above displacement operator picture for $n<1$.
The $\mathcal{N}=(0,2)$ supersymmetry also implies that the one point function of the R-symmetry current 
appearing in the supersymmetry algebra is also fixed by $h_D$ \cite{Bianchi:2019sxz}. 
Using spatial polar coordinates $(\rho,z)$ normal to the defect as in \eqref{flatmet}, then\footnote{{For $n=1$, see
(2.25) of \cite{Bianchi:2019sxz} (and one should take into account their (2.18)); the appearance of $n$ in \eqref{jhrelation}
arises from the fact we are using the coordinate $z$ with period $2\pi$ rather than $nz$ with period $2\pi n$.}  }
\begin{align}\label{jhrelation}
\langle J_R^{a}\rangle =\langle J_R^{\rho}\rangle=0\,,\qquad
\langle (J_R)_{z}\rangle =\frac{n h_D}{2\pi \rho^2}
\,.
\end{align}
There is a defect anomaly for the R-symmetry current parametrised by a coefficient $k_R$, 
as well as anomalies mixed with other flavour symmetries and Lorentz symmetry (see (1.4) of \cite{Wang:2020xkc}).
As discussed in \cite{Wang:2020xkc}, $b=3k_R -k_g/2$, with $k_g$ parametrising the gravitational
defect anomaly 
$\langle \nabla_\mu T^{\mu a}\rangle=\delta(\Sigma)\frac{k_g}{24\pi} \epsilon^{ab}\nabla_b R$, where
$x^a$ are coordinates on $\Sigma$ and $R$ is the Ricci scalar of the induced two-dimensional metric on $\Sigma$ (see (1.5) of \cite{Wang:2020xkc}).

If $d_2$ is known then $b$ can also be determined by computing the defect contribution to entanglement entropy
\cite{Jensen:2018rxu}. For a spherical entanglement region of radius $\ell$ that the defect pierces along the axis
we have (at least for $n=1$)
\begin{align}\label{entent}
\Delta S_A=\frac{1}{3}\left(b-\frac{d_2}{3}\right)\log\frac{ \ell}{\epsilon}+\dots \,,
\end{align}
where $\epsilon$ is a short-distance cutoff. In general, the entanglement entropy can change sign under RG flow \cite{Rodgers:2018mvq}.

We now discuss some aspects of monodromy defects that are associated with CFTs with abelian global symmetries,
which are of primary interest in this paper (several results for free theories are presented in \cite{Bianchi:2021snj}). 
Specifically, considering such CFTs in flat spacetime and using 
polar coordinates $(\rho,z)$ normal to the defect 
we turn on a constant background gauge field for one or more $U(1)$ global symmetries of the form
$A_I=\alpha_I dz$, where $\alpha_I$ are constants. 
For $\alpha_I\ne 0$, these are closed but not exact forms. By integrating the gauge field around the defect and using
Stokes theorem, these defects can be viewed as having a delta function magnetic field source localised on the defect.
Under $U(1)$ gauge transformations of the form $g_I=e^{in_I z}$, with $n_I\in \mathbb{Z}$,
we have $\alpha_I\to\alpha_I+n_I$ and hence $\alpha_I$ are periodic variables\footnote{In fact, the precise nature of the periodicity structure is somewhat subtle due to possible choices of boundary conditions placed on free fields as discussed in e.g.\cite{Alford:1989ie,Bianchi:2021snj}. Further discussion in the context of defects in $\mathcal{N}=4$ SYM theory appears
in appendix \ref{appperiod}.}. If one were to carry out a singular gauge transformation to set $\alpha_I=0$ then all fields that are charged under the global symmetry would pick up a
non-trivial monodromy when circling the defect and this would be an equivalent way of defining the defect.

In the presence of a monodromy defect, the stress tensor is not conserved at the location of the defect; from the Ward identity
$\nabla_\mu T^{\mu\nu}=J^\mu F_{\mu}{}^{\nu}$ and comparing with
\eqref{defopdef} we can deduce (e.g. \cite{Bianchi:2021snj}) that the displacement operator $\mathcal{D}^i$
is proportional to
the currents restricted to the defect: $\mathcal{D}^1=-2\pi \sum_I\alpha_I (J^{2}_I)|_\Sigma$ and
$\mathcal{D}^2=2\pi\sum_I\alpha_I  (J^{1}_I)|_\Sigma$. In particular, notice that
if one knows the two point function of the currents $J^i_I$, restricted to $\Sigma$, one
can obtain the two point function of $\mathcal{D}^i$ and hence, as noted above, $d_1$.

For monodromy defects, if the one point functions of the global symmetry currents are written as
\begin{align}\label{bintone}
\langle (J_I)_z\rangle =\frac{C_I(\alpha)}{\rho^2}\,,
\end{align}
then
\begin{align}\label{bcalpha}
\frac{d}{d\alpha_I}b(\alpha)=\frac{1}{n}12\pi^2 C_I(\alpha)\,.
\end{align}
{For $n=1$ this was shown in \cite{Bianchi:2021snj}; the generalisation to other integer values of $n$ is straightforward
and we assume that the associated analytic continuation for real $n$ also holds.}
Thus, given $\langle (J_I)_z\rangle$, we can compute $b$ as a function of monodromy parameters
by integration, with the integration constant fixed by the condition $b=0$ for $\alpha_I=0$. 
Notice that $C_Id\alpha^I$ is an exact 
one-form (for fixed $n$) and we have the integrability condition $\partial_IC_J=\partial_J C_I$.
In order to obtain $b$ as a function of the monodromy parameters and $n$ we will use \eqref{bcalpha}
as well as (see (1.7) of \cite{Lewkowycz:2014jia} and (3.60) of \cite{Bianchi:2021snj}):
\begin{align}\label{derivbwrtn}
\partial_n b=-\frac{1}{n}d_2+12 a=-18\pi h_D+12 a\,,
\end{align}
where $a$ is the $a$-central charge, and we used \eqref{d2hd} to get the second expression.

In the context of the holographic computations for monodromy defects that we present later, we can summarise as follows.
We will compute $\langle T^{\mu\nu}\rangle $ for flat defects in flat spacetime, to obtain $h_D$ and
hence $d_2$ from \eqref{d2hd}. For the $\mathcal{N}=(0,2)$ 
superconformal defects in  $d=4$ SCFTs of interest, 
we then have $d_1=d_2$ (at least for $n=1$). For $n=1$ we find $h_D\ge 0$, as expected.  We also find
$h_D\ge 0$ for $n\ge 1$, but there are examples with $h_D<0$ when $n<1$. 
In principle one could obtain $d_1$ directly by calculating $\langle J^I J^J\rangle|_\Sigma $, but we won't explicitly pursue that here.
We can also check the expression for $h_D$ by calculating the one point function of the R-symmetry current using \eqref{jhrelation}.
For monodromy defects, we will obtain $b$ by calculating the one-point functions $\langle J^i_I \rangle_\Sigma $ 
and $h_D$ as a function of the monodromy parameters $\alpha_I$ and the parameter $n$, and then integrating 
\eqref{bcalpha}, \eqref{derivbwrtn}. 
We find that for $n\ge 1$ we always have $b\ge 0$, but for $n<1$ there are examples with $b<0$.
Although we will not do so here, by calculating the entanglement entropy we could also obtain $b$ from \eqref{entent}.

Within holography we will consider monodromy defects for both $\mathcal{N}=4$ SYM theory and the 
$\mathcal{N}=1$ LS SCFT in flat space time.
In order to preserve supersymmetry, as we shall see, there is a non-trivial monodromy for the flavour $U(1)$'s but not the $U(1)$ R-symmetry appearing in the
$\mathcal{N}=(0,2)$ supersymmetry algebra.
However, we will also consider monodromy defects with non-trivial monodromy for the $U(1)$ R-symmetry, but in order to preserve
supersymmetry there is then necessarily a conical singularity in the two-dimensional metric transverse to the defect. 
At this point it is very helpful to recall that $\mathbb{R}^{1,3}$ is conformal to $AdS_3\times S^1$. Using Poincar\'e coordinates we can write
\begin{align}\label{ads3poincint}
ds^2(AdS_3)+n^2 dz^2
=\frac{1}{\rho^2}[-dt^2+dx^2+d\rho^2+n^2 \rho^2dz^2]\,,
\end{align}
where $n$ is a positive constant.
With $\Delta z=2\pi$, we see that $AdS_3\times S^1$ is always regular,
with $n$ specifying the ratio of the radius of the $S^1$ to that of $AdS_3$,
but there will be a conical singularity in $\mathbb{R}^{1,3}$ except in the special case that $n=1$.
We find it convenient to carry out most of our holographic analysis
using the $AdS_3\times S^1$ frame.

For $\mathcal{N}=4$ SYM theory we will also consider the possibility of switching on spatially dependent bosonic and fermion mass deformations that preserve superconformal invariance. 
These
`bulk marginal deformations' for the co-dimension two defect have been little studied as yet\footnote{A holographic study for
such deformations in the context of co-dimension one defects preserving conformal invariance i.e. Janus-type interfaces, has been made for $d=3,4$ SCFTs in
\cite{Arav:2018njv,Arav:2020asu,Arav:2020obl}. }, with much of the above discussion not immediately applicable, making them particularly interesting.
In a flat boundary Weyl frame, 
these deformations are parametrised by spatially dependent sources of the form $\phi_s  \rho^{\Delta-4}$, for $\Delta=2,3$, which become
singular at the location of the defect. Notice that in the $AdS_3\times S^1$ frame, however, the source terms $\phi_s$ are constants, making this a convenient frame.
Such source terms give rise to additional terms in the Ward identity for the conservation of the
stress tensor as well as for the trace. 
As in e.g. \cite{Arav:2020obl} there are additional terms appearing in the bulk conformal anomaly\footnote{The mass sources might also give rise to additional defect anomaly terms, but as the mass sources are diverging at the defect (in flat slicing), this analysis is delicate and is left for future work.} depending
on the boson and fermion mass sources; see
\eqref{trwid}, \eqref{trwidanom}. Interestingly, we will see that the bulk conformal anomaly vanishes for the 
BPS configurations which we shall study.  

\subsection{Outline of the paper}

We begin in section \ref{sec:sugramodel} by introducing the $U(1)^3$
$D=5$ gauged supergravity theory that we consider in the remainder of the paper; it can be simply described as the STU model
with an additional complex scalar field. In section \ref{sec:ads3} we introduce the ansatz associated with the defects and analyse the BPS equations assuming that the complex scalar is non-vanishing; this covers both defects of $\mathcal{N}=4$ SYM with non-vanishing mass sources and also defects in the LS theory. For such cases, several of our main results are presented in section \ref{sec:Iscore}. In section \ref{sectstu} we study 
solutions of the STU model (i.e. with vanishing complex scalar) associated with monodromy defects in
$\mathcal{N}=4$ SYM theory with no additional mass sources. In section \ref{sec:bcalc}
we present expressions for the
central charge $b$ as a function of the monodromy parameters and the conical deficit parameter for both
$\mathcal{N}=4$ SYM and the LS cases. 
We also show (in appendix \ref{app:freefield}) that for defects of $\mathcal{N}=4$ SYM theory our computations for $b$ and $h_D$,
when there is no conical singularity,
 are in agreement with
a computation in the free limit using the results of \cite{Bianchi:2021snj}.
In section \ref{susyrenyi} we compute the on-shell action and discuss supersymmetric Renyi entropies for the various cases 
and in section \ref{RGflowsec} we discuss how the $b$ central charge and the on-shell action behave under
the RG flows mentioned above. In section \ref{Fincomments} we conclude with some discussion, 
including explaining how we can obtain the $b$ central charge of
the monodromy defects in LS theory just from the properties of defects of $\mathcal{N}=4$ SYM theory.

We have several appendices. Appendix \ref{susyvarstext} contains some analysis of the BPS equations. In appendices
\ref{FGN=4} and \ref{FGLS} we discuss holographic renormalisation with finite counterterms, taking supersymmetry into account, 
for both $\mathcal{N}=4$ SYM and the LS cases. Some details on the computation of the on-shell action appear in appendix \ref{secosact}.
As noted, appendix \ref{app:freefield} discusses free field computations for defects in 
$\mathcal{N}=4$ SYM theory while appendix \ref{appperiod} discusses subtleties concerning periodicity structures with respect to the monodromy parameters.
Finally, in appendix \ref{mingsugrasols} we present the explicit supergravity solutions for minimal
gauged supergravity, associated with the STU model, in the conventions of this paper. This helps to clarify
that for the general supergravity solutions, both for the STU model and also for solutions with non-vanishing complex scalar
field, there are in general two branches of solutions, only one of which is continuously connected with solutions
with no conical singularity.

\section{The supergravity model}\label{sec:sugramodel}
We will use a $U(1)^3\subset SO(6)$ consistent truncation of maximal gauged supergravity in $D=5$ that 
keeps a metric, three gauge fields $A^{1},A^{2},A^{3}$, two real and neutral scalars $\alpha,\beta$ 
and a single complex scalar field $\zeta\equiv \varphi e^{i\theta}$ which is charged with respect to a specific linear combination of the three $U(1)$'s. This model has been used in \cite{Bobev:2014jva,Arav:2022lzo} and can be obtained as a truncation of a more general class of models with four charged scalar fields that was 
presented in \cite{Khavaev:2000gb,Bobev:2010de}. As in \cite{Arav:2022lzo} the truncation is called the 
\emph{extended LS truncation}. Upon setting $\zeta=0$ we obtain the STU model.

The bosonic part of the Lagrangian, in a \emph{mostly minus} signature, is given by 
\begin{align}\label{model1text}
\mathcal{L} =& -\tfrac{1}{4} R
 + \tfrac{1}{2}(\partial \varphi)^2+  \tfrac{1}{8}\sinh^2 2\varphi \left(D\theta \right)^2
 + 3 (\partial \alpha)^2 + (\partial \beta)^2   - \mathcal{P} \nn
&
 - \tfrac{1}{4}\left[ e^{4\alpha-4\beta} F^{1}_{\mu\nu}F^{1\mu\nu}
+ e^{4\alpha+4\beta} F^{2}_{\mu\nu}F^{2\mu\nu}
+ e^{-8\alpha} F^{3}_{\mu\nu}F^{3\mu\nu} \right]\nn
&+\tfrac{1}{2}\epsilon^{\mu\nu\rho\sigma\delta}F^{1}_{\mu\nu}F^{2}_{\rho\sigma}A^{3}_\delta\,,
\end{align}
where
\begin{align}\label{dthetastext}
 D\theta&\equiv d \theta +g \left(A^{1} + A^{2} - A^{3} \right)
 \equiv d \theta +g A_B \,.
\end{align}
The scalar potential $\mathcal{P}$ is given by
\begin{equation}\label{model2text}
\mathcal{P} = \frac{g^2}{8} \left[ \left( \frac{\partial W}{\partial \varphi} \right)^2 + \frac{1}{6}  \left( \frac{\partial W}{\partial \alpha} \right)^2 + \frac{1}{2} \left( \frac{\partial W}{\partial \beta} \right)^2 \right] - \frac{g^2}{3} W^2 \, ,
\end{equation}
where $W$ is defined as
\begin{align}\label{superpottext}
W = -\frac{1}{4} \Big[ 
 &( e^{-2\alpha-2\beta}+e^{-2\alpha+2\beta } - e^{ 4\alpha } ) \cosh2\varphi 
 + ( e^{-2\alpha-2\beta}+e^{-2\alpha+2\beta } +3 e^{ 4\alpha } )   \Big] \, .
\end{align}

In order that a solution preserves some supersymmetry of the maximal gauged supergravity theory
we require
\begin{align}\label{firstgravpass2text}
&\Big(\nabla_\mu -iQ_\mu-\frac{ig}{6}W\gamma_\mu
-\frac{1}{12}H_{\nu\rho}(\gamma^{\nu\rho}\gamma_\mu+2\gamma^\nu\delta^\rho_\mu)\Big)\epsilon=0\,,
\end{align}
where $\epsilon$ is a complex $D=5$ Dirac spinor, $\nabla_\mu=\partial_\mu+\tfrac{1}{4}\omega_{\mu ab}\gamma^{ab}$ and
\begin{align}\label{HQdefstext}
H_{\mu\nu} &\equiv  e^{2\alpha-2\beta} F^{1}_{\mu\nu} + e^{2\alpha + 2\beta} F^{2}_{\mu\nu} + e^{-4\alpha} F^{3}_{\mu\nu}  \, ,\nn
Q_\mu&\equiv -\frac{g}{2}(A^{1}_\mu+A^{2}_\mu+A^{3}_\mu)
-\frac{1}{4}(\cosh2\varphi-1)D_\mu\theta\,.
\end{align}
In addition, we also require 
\begin{align}\label{gauginovarstext}
[\gamma^\mu\partial_\mu \alpha +\frac{ig}{12}\partial_\alpha W-\frac{1}{12}(e^{2\alpha-2\beta} F^{1}_{\mu\nu} + e^{2\alpha+2\beta} F^{2}_{\mu\nu} - 2 e^{-4\alpha} F^{3}_{\mu\nu} )\gamma^{\mu\nu}]\epsilon&=0\,,\nn
{}[\gamma^\mu\partial_\mu \beta +\frac{ig}{4}\partial_\beta W-\frac{1}{4}(-e^{2\alpha-2\beta} F^{1}_{\mu\nu} +e^{2\alpha+2\beta} F^{2}_{\mu\nu})\gamma^{\mu\nu}]\epsilon&=0\,,\nn
{}[\gamma^\mu\partial_\mu \varphi +\frac{ig}{2}\partial_{\varphi}W+i\partial_{\varphi} Q_\mu\gamma^\mu]\epsilon&=0\,.
\end{align}
Instead of $g$ we sometimes find it convenient to use $L$ defined by $L\equiv\frac{2}{g}$.

\subsection{The $\mathcal{N}$=4 SYM $AdS_5$ vacuum}
This model admits the maximally supersymmetric $AdS_5$ vacuum solution with vanishing matter fields and the $AdS_5$ metric having radius equal to 
\begin{align}
R_{\mathcal{N}=4}=L=\frac{2}{g}\,.
\end{align}
Within the associated dual $\mathcal{N}=4$ SYM theory we can identify the scalar fields $\alpha,\beta$ with specific bosonic mass operators, of conformal dimension $\Delta=2$ and transforming in the ${\bf 20}'$ of $SO(6)$, while $\zeta$ is dual to a fermionic mass 
operator, with scaling dimension $\Delta=3$, transforming in the ${\bf 10}$ of $SO(6)$.

The three gauge fields $A^{1},A^{2},A^{3}$ are dual to $U(1)^3\subset SU(4)$ R-symmetry currents $J^i$. 
If we decompose $SU(4)\to U(1)_R\times SU(3) $
it is natural to consider the $SU(4)$ generators $diag(\tfrac{1}{3},\tfrac{1}{3},\tfrac{1}{3},-1)$, $diag(0,1,-1,0)$, $diag(1,-\tfrac{1}{2},-\tfrac{1}{2},0)$, associated, respectively, with gauge fields 
$A_R, A_F ,A_{F'}$ given by
\begin{align}\label{lincsn4gfnatbas}
A_R&\equiv A^1+A^2+A^3\,,\qquad
A_F\equiv A^1-A^2\,,\qquad
A_{F'}\equiv \frac{2}{3}(A^1+A^2-2A^3)\,,\nn
J^{\mathcal{N}=4}_R&\equiv \frac{1}{3}(J^1+J^2+J^3)\,,\quad
J_F\equiv \frac{1}{2}(J^1-J^2)\,,\quad
J_{F'}\equiv \frac{1}{4}(J^1+J^2-2J^3)\,,
\end{align}
with $A^iJ_i=A_RJ^{\mathcal{N}=4}_R+A_FJ_F+A_{F'}J_{F'}$ and the $J$'s are the associated currents in this basis. 
Notice that the R-symmetry generator is invariant under $SU(3)$ and furthermore,
that $A_R$ appears in $Q$ in \eqref{HQdefstext}. This is a natural basis to study
$\mathcal{N}=4$ solutions with $\varphi=0$ (as too is the basis associated with $A^i$, $J^i$).

For $\mathcal{N}=4$ solutions with $\varphi\ne 0$, associated with mass deformations
that break a $U(1)\in SU(3)$, recalling \eqref{dthetastext}, a natural alternative basis is the $SU(4)$ generators
$diag(0,\tfrac{1}{2},\tfrac{1}{2},-1)$, $diag(0,1,-1,0)$, $diag(1,-\tfrac{1}{2},-\tfrac{1}{2},0)$ associated with
\begin{align}\label{lincsn4gf0}
A_R&\equiv A^1+A^2+A^3\,,\qquad
A_F\equiv A^1-A^2\,,\qquad
A_B\equiv A^1+A^2-A^3\,,\nn
 {J}_R^{\varphi} &\equiv \frac{1}{4}({J}_1 + {J}_2 + 2 {J}_3)\,,
\quad {J}_F \equiv \frac{1}{2}({J}_1 - {J}_2) \,,\quad
{J}_B \equiv \frac{1}{4}({J}_1 + {J}_2 - 2 {J}_3) \,,
\end{align}
and $A^iJ_i=A_RJ^{\varphi}_R+A_FJ_F+A_{B}J_{B}$. 
When $\varphi=0$, all currents in \eqref{lincsn4gfnatbas} and \eqref{lincsn4gf0} are conserved, but when $\varphi\ne 0$
the conserved currents are $ {J}_R^{\varphi}$ and ${J}_F $, with ${J}_B$ not conserved.
It will be important later to note that $J^{\mathcal{N}=4}_R= {J}_R^{\varphi} +\frac{1}{3}{J}_B $.
We also note that $J_B=J_{F'}$.

\subsection{The LS $AdS_5$ vacuum}\label{subseclsvac}

The supergravity model also admits a supersymmetric LS $AdS_5$ solution \cite{Khavaev:1998fb} with 
\begin{align}\label{lsfpsc}
R_{LS}\equiv \frac{3}{2^{2/3}g}=\frac{3L}{2^{5/3}},\qquad e^{6\alpha}=2,\qquad e^{2\varphi}={3}\,,\qquad \beta=0\,,
\end{align}
where $R_{LS}$ is the radius of the $AdS_5$,
and vanishing gauge fields. 
This solution preserves $SU(2)\times U(1)_R$ global symmetry and is dual
to the $d=4$, $\mathcal{N}=1$ LS SCFT \cite{Leigh:1995ep}. In this background $\beta$ is dual to an operator with
$\Delta=2$. The scalar fields $\alpha, \varphi$ mix and are dual
to two scalar operators with scaling dimension $\Delta =1+\sqrt{7}$ and $\Delta=3+\sqrt{7}$.

In this case it is convenient to define 
\begin{align}\label{lincsLSgf}
A^{LS}_R&\equiv \tfrac{4}{3}(A^1+A^2+\tfrac{1}{2}A^3)\,,\qquad
A_F\equiv A^1-A^2\,,\qquad
A_B\equiv A^1+A^2-A^3\,,
\end{align}
corresponding to $SU(4)$ generators given by $diag(0,\tfrac{1}{2},\tfrac{1}{2},-1)$, $diag(0,1,-1,0)$ and $diag(1,-\tfrac{2}{3},-\tfrac{2}{3},\tfrac{1}{3})$. 
Notice in particular that the $R$-symmetry generator for the LS fixed point
is the same as the choice of R-symmetry used above in the case of $\mathcal{N}=4$ with a mass deformation.
The complex scalar $\zeta$ is charged with respect to the gauge field 
$A_B$ and together they give rise to a massive vector with scaling dimension $\Delta=2+\sqrt{7}$.
There are two massless gauge fields, $A^{LS}_R$ dual to the $U(1)_R$ $R$-symmetry current $J^\varphi_R$, 
and $A_F$ dual to a $U(1)_F\subset SU(2)$ flavour symmetry current $J_F$. 

There is an RG flow between $\mathcal{N}=4$ SYM, deformed by a BPS, Poincar\'e invariant
mass deformation, associated with sources for the fields $\alpha,\varphi$, and the LS fixed point;
the corresponding holographic solution was found in \cite{Freedman:1999gp}.

\subsection{Further subtruncations}\label{section:moretruncs}

 There are additional consistent truncations which one can consider. As noted, we can set $\zeta=0$ to recover the STU model; and if we further set 
 $\alpha=\beta=0$ and $A^{1}=A^{2}=A^{3}$ we get minimal gauged supergravity, and each of these
 theories just admits the $AdS_5$ vacuum that uplifts to the $\mathcal{N}=4$ SYM $AdS_5\times S^5$ solution.
 We can also set $A^{1}=A^{2}$ as well as $\beta=0$ and $\varphi\ne 0$, to get a theory containing both the $\mathcal{N}=4$ SYM and the LS $AdS_5$ solutions, which we call
 the \emph{LS truncation}. Solutions of this theory will preserve the $SU(2)$ flavour symmetry of the LS fixed point.
 As noted above, we refer to the general model in \eqref{model1text} as the \emph{extended LS truncation}.
It is also possible to set $A^{1}=A^{2}=\frac{1}{2}A^{3}$ and the set the scalars to the constant values
 in \eqref{lsfpsc}; we again get minimal gauged supergravity\footnote{Solutions of minimal gauged supergravity can also be uplifted to $D=11$ \cite{Gauntlett:2006ai}. Also, solutions with $A^1=A^2$ and $\beta=0$ in the STU
 model, mentioned in the next paragraph, can be uplifted to $D=11$ via a truncation of Romans theory \cite{Gauntlett:2007sm}.} but now with an $AdS_5$ vacuum that uplifts to the LS
 $AdS_5\times S^5$ solution.
 
Within the STU model, if we truncate $A^1=A^2$ and $\beta=0$ we obtain the $N=2$ $SU(2)\times U(1)$ Romans theory, with $A^1=A^2$ identified with the Cartan of $SU(2)$ and $A^3$ with the $U(1)$. 
We also point out that if we set $A^3=0$
in the STU model we get a theory with two gauge fields which arises as a truncation of Romans theory coupled to a single vector multiplet, a model that was considered\footnote{As we shall see, pure Romans theory admits 
$\mathcal{N}=(0,2)$ defect solutions with $A^3\ne0$, but these were not considered in
\cite{Gutperle:2019dqf} as they were just considering $\mathcal{N}=(2,2)$ defects. Within Romans theory
having $\mathcal{N}=(2,2)$ defects requires setting $A^3=0$ and this leads to defect solutions necessarily with conical singularities as seen in \cite{Gutperle:2019dqf} and consistent with the results we find. The $\mathcal{N}=(2,2)$ defects without conical singularities ($n=1$) that were found in Romans theory coupled to a vector multiplet
arise here in the STU model with $A^3=0$; in fact these have $g\mu_1=-g\mu_2$ and are 
the $g\mu_B=0$ defects in figure \ref{fig:sumsolutions}. {We also note that if we set $A^1=A^2=0$ and $\beta=0$, we find solutions of
Romans theory, not considered in \cite{Gutperle:2019dqf}, which are dual to $\mathcal{N}=(4,4)$ defects with conical singularities.}}
 in \cite{Gutperle:2019dqf}.

\section{General setup} \label{sec:ads3}

\subsection{$AdS_3$ ansatz and a first look at boundary conditions}\label{secads3ansbcs}
We are interested in constructing supersymmetric 
solutions using the ansatz
\begin{align}\label{ads3ans}
ds^2&=e^{2V}ds^2(AdS_3)-(f^2dy^2+h^2 dz^2)\,,\nn
A^{i}&=a^{i}dz\,.
\end{align}
Here $ds^2(AdS_3)$ is a unit radius metric on $AdS_3$ and $V,f,h,a^{i}$ are functions of $y$ only. 
The scalar fields $\alpha,\beta,\varphi$ are also functions of $y$ only. 
With $\varphi\ne 0$, the phase of the complex scalar field 
is linear in $z$, $\theta=\bar\theta z$, with $\bar\theta$ a constant, and hence 
we can write $Q_\mu dx^\mu\equiv Q_z(y)dz$. Locally, we can always shift $\bar\theta$ 
by a constant with a gauge transformation.

\begin{figure}[htbp]
\begin{center}
\includegraphics[scale=.55]{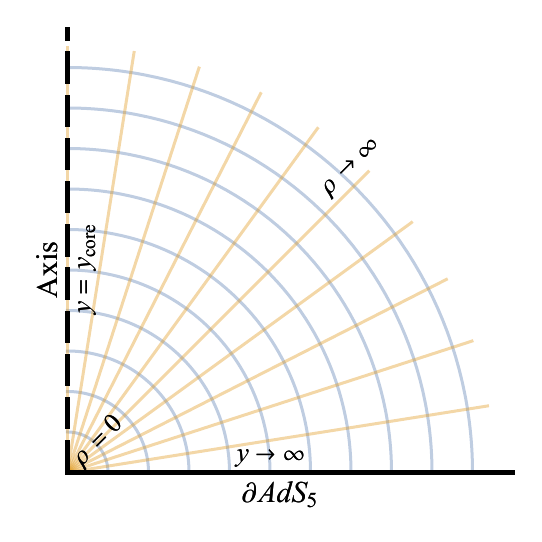}\qquad
\includegraphics[scale=.55]{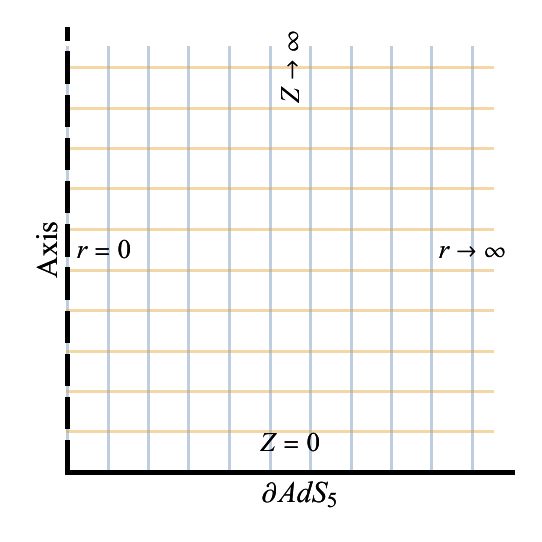}
\caption{Coordinate charts for the defect solutions that asymptote to $AdS_{5}$ and are foliated by
 $AdS_{3}\times S^1$.
The coordinates $(t,x)$ parametrising the defect and the periodic coordinate $z$ transverse to the defect are suppressed. We use coordinates as in
the left hand plot, with boundary $AdS_{3}\times S^1$, where $\rho$ is a coordinate on the $AdS_3$ slices, with $\rho=0$ the boundary of $AdS_3$.  The right hand plot represents a different foliation, with 
$Z$ a Fefferman-Graham type coordinate and flat boundary, with the $r$ coordinate approaching the $\rho$ coordinate 
on the boundary (see \eqref{ads3poinc}). 
}
\label{fig:diff_charts}
\end{center}
\end{figure}

We will take $z$ to be a periodic coordinate with 
\begin{align}
\Delta z=2\pi\,.
\end{align} 
We also take $y\in [y_{core},\infty)$, with $y=y_{core}$ the core of the solution in the bulk where we demand that
the $z$ circle smoothly pinches off so that $(y,z)$ parametrise a cigar geometry. With regular gauge fields, 
regularity of the solutions with non-vanishing $\varphi\ne 0$ requires that we work in a gauge with $\bar\theta=0$.
At the other end of the cigar, $y\to \infty$ is associated with an $AdS_5$ vacuum\footnote{Note that an exact $AdS_5$ metric of radius $R$ is obtained
by setting $e^{2V}=f^2=\frac{R^2}{\cos^2 y}$ and $h^2=R^2\tan^2 y$, with the boundary in these coordinates at $y\to \pi/2$.}, dual to either $\mathcal{N}=4$ SYM or the LS
fixed point. We will be more precise about the boundary conditions at the $AdS_5$ boundary later, but here we note that we will have
$e^{2V}\to e^{2V_0}(y/R)^2$,  $h^2\to e^{2V_0}n^2(y/R)^2$ and
$f^2\to (R/y)^2$, so that the metric approaches one of the two $AdS_5$ vacua in the form 
\begin{align}\label{ads3ansbdy}
ds^2\to -\frac{R^2}{y^2}{dy^2}+\frac{y^2}{R^2}e^{2V_0}[ds^2(AdS_3)-n^2 dz^2] \,,
\end{align}
where $R$ is the radius of the fixed point $AdS_5$ solution and $n>0$ is a positive constant.
The associated four-dimensional boundary is $AdS_3\times S^1$ (for any $n$) which, using Poincar\'e coordinates, can be written in the form
\begin{align}\label{ads3poinc}
e^{2V_0}\left(ds^2(AdS_3)-n^2 dz^2\right)
=e^{2V_0}\left(\frac{1}{\rho^2}[dt^2-dx^2-d\rho^2-n^2 \rho^2dz^2]\right)\,.
\end{align}
Thus, since we take $\Delta z=2\pi$, if $n=1$ the four-dimensional boundary metric is related to the flat space metric on $\mathbb{R}^{3,1}$
by a Weyl transformation. If $n\ne 1$ then the $\mathbb{R}^{3,1}$ boundary has a
co-dimension two conical singularity; with $n< 1$, there is a conical deficit angle, while if $n> 1$, there is a conical excess angle.
Observe that the boundary of the $AdS_3$ space, located at $\rho=0$, gets mapped to the axis of azimuthal symmetry of $\mathbb{R}^{3,1}$; see figure \ref{fig:diff_charts} (\emph{c.f.} \cite{Horowitz:2014gva}).

At the $AdS_5$ boundary the $U(1)^3$ gauge fields will have an expansion of the schematic form 
\begin{align}\label{schematicexp}
a^{i} = & \mu_i +\dots   +j_i\frac{R^2}{y^2} + \ldots\,.
\end{align}
The boundary value $g\mu_i$ is interpreted as the sources for the $U(1)^3$
global symmetry currents in the dual SCFT, and $g\mu_i\ne 0$ is dual to the insertion of a
co-dimension two, monodromy defect in the dual SCFT, which is located at $\rho=0$ in the flat space boundary
coordinates. In particular, in the boundary theory we should view $g\mu_i dz$ as background
gauge fields in the dual field theory which cannot be gauged away.
In the $AdS_3\times S^1$ Weyl frame the gauge fields have non-trivial monodromy about the $S^1$.
Note that we work in a gauge that is regular at the core of the solution in the bulk, as we highlight below.
In addition, the $j_i$ are proportional to the expectation values of the $z$ component of the $U(1)^3$ currents $J_i$,
up to a shift by a finite counterterm which we discuss later, associated, for a flat boundary, with \eqref{bintone}.
A key result of our holographic analysis will be to relate the currents to the sources. We will also be able to express the expectation value of the stress tensor and the 
expectation values for the scalar operators in terms of the currents and hence the sources. We comment on the $AdS_5$ boundary
scalar sources below.

At the core of the solution (in the bulk), located at $y\to y_{core}$, 
we demand that $e^V$ goes to a non-vanishing constant, $h\to 0$ and $f$ is such
that the $y,z$ part of the metric is smooth. We also demand that we have regular scalar fields; for the real scalars
this is achieved by taking $\alpha,\beta\to const.$. For the complex scalar, working
in the gauge $\bar\theta=0$, with $\zeta=\varphi(y)$, we can also demand that $\varphi\to const.$. 
Furthermore, regularity of the gauge fields at the core is imposed by demanding $a_i\to 0$.
As we will see (in section \ref{sec:Iscore}) there is a kind of attractor mechanism at play and there is a universal
behaviour at the core of the solution in the bulk for both the $\mathcal{N}=4$ and LS boundary asymptotics.

For the case of $\mathcal{N}=4$ SYM boundary, we will see that the BPS equations fix the source for the $U(1)$ R-symmetry, $g\mu_R\equiv g\mu_1+g\mu_2+g\mu_3$, with 
$g\mu_R=-2s-n\kappa$ where $s$ is a constant appearing in the phase of the Killing spinor and $\kappa=\pm1$ is a sign 
specifying the chirality of the preserved supersymmetry.
{We will see that there are two branches of solutions, one with $s=-\kappa/2$ which exists for all $0<n$ and
a second branch with $s=+\kappa/2$, which can exist, at most, only for $0<n<\frac{1}{2}$. The branch with $s=-\kappa/2$ is
of more interest and we will refer to it as the ``main branch", since it includes the possibility that there is no conical singularity, $n=1$, in which case the supersymmetry condition
is the vanishing of the R-symmetry monodromy, $g\mu_R=0$.
}

For $\mathcal{N}=4$ SYM boundary, we will consider solutions of the STU model with $\varphi=0$ in section \ref{sectstu}. These solutions, which are known explicitly\footnote{We have presented the minimal gauged supergravity solutions in appendix \ref{mingsugrasols} where, in particular, we discuss the two branches of solutions in this setting.}, preserve 
$U(1)^3$ symmetry and, with $g\mu_R=-2s-n\kappa$, are
specified by two independent sources.
Before considering the STU solutions, though, we will focus on solutions with $\varphi\ne0$ which necessarily break one of the $U(1)^3$ 
symmetries in the bulk. For the case of the $\mathcal{N}=4$ SYM boundary, we will find that the source associated with the broken $U(1)$, 
$g\mu_B \equiv g\mu_1 + g\mu_2 - g\mu_3$,  is non-zero and for the BPS solutions we must have
$g\mu_B = -n\kappa  $. It will also be interesting to consider the limit of these solutions 
as $\varphi\to 0$. This gives rise to a restricted subclass of the
analytic STU defect solutions (which have $\varphi=0$ identically), where we impose a restriction on the source 
$g\mu_B =-n\kappa  $ for 
the associated $U(1)$ current, which is no longer broken, by hand.
We thus call these limiting STU solutions, arising from $\varphi\to 0$, \emph{restricted STU solutions}.
We will also be interested in STU solutions with, instead, $g\mu_B=0$ set by hand, as these are associated with 
RG flows from defects in $\mathcal{N}=4$ SYM theory to defects in the LS theory.
 
 With $\mathcal{N}=4$ SYM boundary, we can allow sources $\alpha_s,\beta_s$ for the
 $\Delta=2$ boson mass operators, as well as a source $\varphi_s$ for the $\Delta=3$ fermion mass operator, but
 the BPS equations imply that $\beta_s=0$ and $\alpha_s=\tfrac{2}{3}\varphi_s^2$. More precisely, 
these mass sources are constants in the Weyl frame with $AdS_3\times S^1$ boundary, the frame we will mostly work with,
but
in the Weyl frame with flat boundary, these sources have the spatially dependent form $\alpha_s/\rho^2$ and $\varphi_s/\rho$ (in the $\bar\theta=0$ gauge), consistent with still preserving the superconformal symmetry of the defect. Notice, in particular, that they are singular at the location
of the defect.

We also study solutions with $\varphi\ne0$ that approach the LS $AdS_5$ boundary. In this case, 
we must have 
\begin{align}
g\mu_B=0\,,
\end{align} 
which essentially arises from diagonalising the mass matrix for the vector fields.
We also find that the BPS equations imply
\begin{align}
g\mu_R^{LS}=g\mu_R=-2s-n\kappa
\end{align}
where the first equality follows from 
\eqref{lincsn4gf0}, \eqref{lincsLSgf}
and imposing $g\mu_B=0$. Notice, in particular, that $g\mu_R$ has exactly the same value for the LS theory 
as it does
for the ${\mathcal{N}=4}$ SYM theory; this is a useful observation
in discussing RG flow between defects of the two SCFTs, as we do later. 
{As for the $\mathcal{N}=4$ SYM case, there are two allowed branches of solutions, the ``main branch" with
$s=-\kappa/2$ which exists for all $n>0$ and a second with $s=+\kappa/2$, which can exist only for $0<n<\frac{1}{2}$.
In particular, when $n=1$ and there is no conical singularity on the flat LS boundary, the BPS equations again imply that $g\mu_R=0$.}
We also note that $g\mu_F$ labels the same source for both the LS
and $\mathcal{N}=4$ SYM cases.
For the LS solutions, one might consider a source for the relevant operator with $\Delta=1+\sqrt{7}$, but we shall not do
so here (it is possible that supersymmetry forbids such a source).

Having made these initial comments,
we continue in the remainder of this section and also in section
\ref{sec:Iscore} focussing on solutions with
\begin{align}\
\varphi\ne0\,,
\end{align}
covering defects in $\mathcal{N}=4$ SYM theory with mass deformations and also defects in the LS SCFT.
In section \ref{sectstu} we return to the simpler STU case, which has $\varphi\equiv 0$,
associated with defects in $\mathcal{N}=4$ SYM theory with no mass deformations.

\subsection{BPS equations}\label{sec3bps}
The BPS equations for the above $AdS_3$ ansatz were already derived in \cite{Arav:2022lzo}, in the context of solutions
where, instead, $(y,z)$ parametrise a compact spindle.
In the obvious orthonormal frame and a convenient set of gamma matrices, the Killing spinor has the form $\epsilon=\psi\otimes\chi$,
with $\psi$ a complex two component spinor on $AdS_3$ satisfying 
\begin{align}
D_m\psi=\frac{i}{2}\kappa\Gamma_m \psi\,,\qquad 
\end{align}
with $\kappa=\pm 1$ determining the chirality of the preserved $d=2$ superconformal symmetry,
i.e. $\mathcal{N}=(0,2)$ or $(2,0)$, and 
\begin{align}\label{chiralityspintext}
\chi  =e^{V/2}e^{is z}\begin{pmatrix}\sin\frac{\xi}{2} \\ \cos\frac{\xi}{2} \end{pmatrix}\,,
\end{align}
where the constant $s$ is the (gauge-dependent) charge of the spinor under the action of $\partial_z$ and recall
we are working in a gauge where the complex scalar is real, $\bar\theta=0$ (see section \ref{sec:bcscore}). 
The BPS equations derived in \cite{Arav:2022lzo} are written in appendix \ref{susyvarstext}
and we assume that $\sin\xi$ is not identically zero.
An integral of the BPS equations implies
\begin{align}\label{hemvszeqtext}
he^{-V}=-n \sin\xi\,,
\end{align}
where $n$ is the constant parametrising the conical singularity in \eqref{ads3poinc}. In particular this shows that at the core of the solution, where $h$ vanishes and $V$ is constant, $\sin\xi$ also vanishes.

With $\varphi\ne 0$, two linear combinations of the gauge field equations of motion can be integrated to give constants of motion (see \eqref{gaugeintmot}). Furthermore, using the BPS equations
we can recast these in the form
\begin{align}\label{conschgestext}
\mathcal{E}_R&=e^{2V}[2ge^V \cos\xi-2\kappa(e^{-4\alpha}+e^{2\alpha}\cosh 2\beta)]\,,\nn
\mathcal{E}_F&=2\kappa e^{2V}e^{2\alpha}\sinh 2\beta\,,
\end{align}
where $\mathcal{E}_R,\mathcal{E}_F$ are constants.
Taking the $\varphi\to 0$ limit, we obtain a restricted STU solution for which the remaining gauge field equation of motion can
also be integrated (see \eqref{extraone}) and for BPS solutions we have
\begin{align}\label{conschgestext2}
\mathcal{E}_B
&=2\kappa e^{2V}(e^{-4\alpha}-e^{2\alpha}\cosh 2\beta)\,,\qquad
(\varphi=0)\,,
\end{align}
with $\mathcal{E}_B$ constant when $\varphi=0$.

Importantly, the BPS equations can also be used\footnote{In \cite{Arav:2022lzo} this was derived in the context of the ``conformal gauge" choice
for the radial coordinate given in \eqref{confgaugef}, but it is true in any gauge.}  to re-express the field strengths of the gauge fields
in the form
\begin{equation}\label{effaiprimetext1}
F^{i}_{yz} = (a^{i})' = (\mathcal{I}^{i})' \, ,
\end{equation}
where
\begin{align}\label{eq:IntegratedFluxesExpr1text}
& \mathcal{I}^{1} \equiv -\frac{n}{2}  e^V \cos\xi \, e^{-2\alpha+2\beta} \, , \quad
\mathcal{I}^{2} \equiv -\frac{n}{2} e^V \cos\xi \, e^{-2\alpha-2\beta} \, , \quad
 \mathcal{I}^{3} \equiv -\frac{n}{2}  e^V \cos\xi \, e^{4\alpha} \, .
\end{align}
This will be crucial in relating the currents of the boundary theory to the sources, as we shall see later.

The BPS equations have  a discrete symmetry $(h,z)\to -(h,z)$ of the BPS equations and so 
without loss of generality we can assume, away from the core of the solution where $h\to0$, that
\begin{align}
h>0\,.
\end{align}

\subsection{Boundary conditions at the $AdS_5$ boundary and $h_D$}
The full expansions of solutions to the BPS equations near the
$AdS_5$ boundary, both for $\mathcal{N}=4$ SYM and the LS fixed point, are given in appendices \ref{FGN=4} and \ref{FGLS}, respectively. Building on the comments made above, at the $AdS_5$ boundary
we take $f\to R/y$ and $e^{2V}\to e^{2V_0}(y/R)^2$, so that $V'\to 1/y$.
Furthermore, for the $\mathcal{N}=4$ SYM case we have $W\to -3/2$ while for the LS case $W\to -2^{2/3}$. In both cases,
we see from the BPS equation for $V$ in \eqref{summbbpsapptext} that we must have $\sin\xi<0$ as we approach the boundary.
Then from \eqref{hemvszeqtext} and $h>0$ we deduce that the constant $n$, defined in \eqref{hemvszeqtext}, is positive:
\begin{align}\label{klesszero}
n>0\,. 
\end{align}

In appendices \ref{FGN=4} and \ref{FGLS}, respectively, we also consider holographic renormalisation, obtaining expressions for the stress tensor and scalar one-point functions in terms of the boundary currents. We now summarise each case in turn, noting that in section \ref{sec:Iscore} we will relate these currents to the monodromy sources.

\subsubsection{$\mathcal{N}=4$ SYM theory asymptotics, $\varphi\ne 0$}
In appendix \ref{FGN=4} we carry out holographic renormalisation for solutions that approach the $\mathcal{N}=4$ SYM $AdS_5$ boundary.
The boundary expansion involves constant source terms $\alpha_s,\beta_s$ and $\varphi_s$ that are dual to sources for the associated boson and fermion mass operators
in $\mathcal{N}=4$ SYM, respectively. The BPS equations
imply that these sources are constrained via
\begin{align}\label{textconsusuyssces}
\betas&=0,\qquad\qquad \alphas = \frac{2}{3} \varphi_s^2\,.
\end{align}
The BPS equations also imply that the sources for the gauge fields are constrained via
\begin{align}\label{n4scebpstext}
g\mu_R&\equiv g\mu_1+g\mu_2+g\mu_3=-n{\kappa }-2s\,, \nn
g\mu_B&\equiv g\mu_1+g\mu_2-g\mu_3 =-n{\kappa }\,,
\end{align}
as noted earlier. The source for the flavour symmetry current
\begin{align}
g\mu_F\equiv g\mu_1- g\mu_2\,,
\end{align}
is not constrained by the BPS equations.
An alternative way to obtain these BPS constraints is to follow the approach of \cite{Arav:2020obl}, extending
 \cite{Maxfield:2016lok}, and couple $\mathcal{N}=4$ SYM to $\mathcal{N}=4$ conformal supergravity. We have carried out this exercise and find precise agreement.

In addition to the counterterms that are required to remove divergences, we also consider a set of finite counterterms, given in \eqref{finitetermsgen}, that depend on five free constants. 
A supersymmetric scheme should fix these constants: in appendix \ref{FGN=4} we discuss
some specific constraints imposed by supersymmetry as well as some other constraints, that we expect are imposed by supersymmetry\footnote{For example, demanding that the $\varphi_s^2$ terms vanish in the stress tensor \eqref{eq:bps_stress_current_relation} so that $h_D$ is fixed by the currents, as in \eqref{stresstextn42}, imposes the
constraint \eqref{eliminphisqterms}, but it can also be obtained by arguing that the on-shell action is independent of $\varphi_s$ as discussed in appendix \ref{appd1osact}.}, which fixes four of the constants, as given in \eqref{bpsscheme}.
 While more refined arguments should also fix the one free constant, $\delta_\beta$, we will not do so here. 
Using this renormalisation scheme, for the BPS configurations of interest in the dual $\mathcal{N}=4$ $SU(N)$ SYM theory, the $z$ component of the currents, $\vev{J_i}\equiv \vev{(J_i)_z}$, associated with $g\mu_i$,
are given by
\begin{align}\label{textbcurrents}
\vev{J_1}&=-\frac{N^2}{2\pi^2}\left[{ L^{-3}j_1}-2\delta_\beta n\kappa   { (L^{-1}\varphi_s)^2}\right]\,,\nn
\vev{J_2}&=-\frac{N^2}{2\pi^2}\left[{ L^{-3}j_2}-2\delta_\beta n\kappa  { (L^{-1}\varphi_s)^2}\right]\,,\nn
\vev{J_3}&=-\frac{N^2}{2\pi^2}\left[{ L^{-3}j_3}+2\delta_\beta n\kappa  { (L^{-1}\varphi_s)^2}\right]\,,
\end{align}
where $j_i$ is the expansion coefficient defined in \eqref{schematicexp} and 
we used \eqref{acentchgen4}. Notice that there is a dependence on the unfixed scheme dependent
parameter, $\delta_\beta$, when $\varphi_s\ne 0$, and also notice the sign difference in the last expression.
Recalling the currents defined in 
\eqref{lincsn4gf0},
\begin{align}\label{neq4currents}
 {J}_R^{\varphi} &\equiv \frac{1}{4}({J}_1 + {J}_2 + 2 {J}_3)\,,
\quad {J}_F \equiv \frac{1}{2}({J}_1 - {J}_2) \,,\quad
{J}_B \equiv \frac{1}{4}({J}_1 + {J}_2 - 2 {J}_3)\,,
\end{align}
we notice that both $\langle{{J}_R^{\varphi}}\rangle$ and $\vev{{J}_F}$, which are the two conserved currents when $\varphi\ne 0$, are independent of $\delta_\beta$.

The components of the stress tensor can be calculated and we find
\begin{align}\label{stresstextn4}
\vev{\cT_{ab}}dx^a dx^b&=\frac{h_D}{2\pi}\left[ds^2(AdS_3)+3n^2 dz^2\right]\,,
\end{align}
where, 
\begin{align}\label{stresstextn42}
h_D=  -\frac{2\pi }{3n \kappa } \left( \vev{J_1} + \vev{J_2} + \vev{J_3}  \right)
&=- \frac{2\pi }{n\kappa }\langle{J}_R^{\mathcal{N}=4}\rangle\nn
&= -\frac{2\pi }{n \kappa } \left( \langle{{J}_R^{\varphi}}\rangle + \frac{1}{3}\vev{J_B}  \right)\,,
\end{align}
and we used \eqref{lincsn4gfnatbas} and \eqref{lincsn4gf0}.
Notice that when $\varphi\ne 0$, supersymmetry is fixing $h_D$ in terms of the
$\mathcal{N}=4$ SYM R-symmetry current ${J}_R^{\mathcal{N}=4}$. In particular it is not being fixed in terms
of the conserved currents ${J}_R^{\varphi}$ and ${J}_F$. As a consequence the value of $h_D$ will depend non-trivially
on $\varphi_s$ and, moreover, to determine its explicit value one needs to solve the BPS equations.

We also find that the expectation values of the scalar operators are given by
\begin{align}\label{vevstextn4}
\vev{\cO_\a} &= -\frac{1}{n\kappa } \left( \vev{J_1}+  \vev{J_2}  - 2 \vev{J_3}  \right)= 
-\frac{4}{n\kappa }  \vev{J_B}\,,\nn
\vev{\cO_\b} &= \frac{1}{n\kappa } \left( \vev{J_1}  -\vev{J_2}   \right)
= \frac{2}{n\kappa }  \vev{J_F}\,,\nn
\vev{\cO_\varphi} &= \frac{4}{3n\kappa }\left( \vev{J_1}  +\vev{J_2}  - 2 \vev{J_3}  \right)(L^{-1}\phis)
=\frac{16}{3n\kappa }\vev{J_B} (L^{-1}\phis)
\,.
\end{align}

As discussed in the appendix \ref{FGN=4}, the conformal anomaly has the form
\begin{align}\label{trwidtext}
\cA=\vev{\cT^a{}_a}  + \vev{\cO_\varphi} (L^{-1}\phis)  + 2\vev{\cO_\a}(L^{-2}\alphas) + 2\vev{\cO_\b} (L^{-2}\betas) \,.
\end{align}
From \eqref{stresstextn4} we deduce $\vev{\cT_{a}^a}=0$. Remarkably, 
from \eqref{textconsusuyssces} and \eqref{vevstextn4}  we then see that the
anomaly vanishes, $\cA=0$, for the BPS configurations with $\varphi_s\ne 0$.

The expectation values given in \eqref{vevstextn4} are associated with $\mathcal{N}=4$ SYM
theory on $AdS_3\times S^1$ with
monodromy for the global symmetry around the $S^1$ and also with mass deformation parametrised by $\varphi_s$.
The expectation values for the scalar operators associated with monodromy defects of $\mathcal{N}=4$ SYM theory on flat spacetime
can be obtained by carrying out a bulk coordinate transformation near the conformal boundary; an analogous computation was carried out in detail for Janus solutions in \cite{Arav:2020obl}. While we have not
carried out this computation, 
the expectation values of the scalar operators will have a power-law dependence on the spatial coordinate transverse to the defect, as required by dimensional analysis, and for $\vev{\mathcal{O}_\alpha}$, $\vev{\mathcal{O}_\varphi}$, associated with
non-vanishing sources $\varphi_s,\alpha_s$, this will be combined with an extra logarithmic dependence
arising from the structure of the conformal anomaly. Similar comments also apply to the currents and the stress tensor, 
with the non-conserved current, and possibly some stress tensor components, also picking up a logarithmic dependence,
but we leave the details to future work.

\subsubsection{LS theory asymptotics}

In appendix \ref{FGLS} we carry out holographic renormalisation for solutions that approach the LS $AdS_5$ boundary.
The complex scalar and gauge field $A_B$ give rise to a massive vector field which is dual to a
vector operator with scaling dimension $\Delta=2+\sqrt{7}$. 
The scalars $\varphi,\a$ mix into scalar operators with scaling dimensions $\Delta=1+\sqrt{7}$ and 
$\Delta=3+\sqrt{7}$. Naturally, we do not consider any source terms for the irrelevant operators, namely the
massive vector and the $\Delta=3+\sqrt{7}$ scalar. To simplify the analysis we also do not 
consider
sources for the 
relevant scalar operator with $\Delta =1+\sqrt{7}$, nor for the scalar operator dual to $\beta$ with $\Delta=2$. 

As previously noted the LS asymptotics imply that
\begin{align}
g\mu_B&\equiv g\mu_1+g\mu_2-g\mu_3 =0\,.
\end{align}
The BPS equations imply that the sources for the gauge fields are also constrained via
\begin{align}\label{LSscebpstext}
g\mu_R^{LS}&\equiv \frac{4}{3}(g\mu_1+g\mu_2+\frac{1}{2}g\mu_3)=-{n\kappa }-2s\,, 
\end{align}
and with $g\mu_B=0$ we also have
\begin{align}
g\mu_R^{LS}=g\mu_R\equiv g\mu_1+g\mu_2+g\mu_3\,,
\end{align}
as also noted earlier. The source for the flavour symmetry current
\begin{align}
g\mu_F\equiv g\mu_1- g\mu_2\,,
\end{align}
is not constrained by the BPS equations.

Associated with the LS solutions, there are
conserved currents $ {J}^{\varphi}_R$, $ {J}_F$ which are dual to 
$A^{LS}_R$, $A_F$ in \eqref{lincsLSgf},
respectively.
The components of the stress tensor can be calculated and we now find
\begin{align}\label{stresstextLS}
\vev{\cT_{ab}}dx^a dx^b&=\frac{h_D}{2\pi}\left[ds^2(AdS_3)+3n^2 dz^2\right]\,,
\end{align}
with
\begin{align}\label{stresstextLS2}
{\frac{h_D}{2\pi}
=-\frac{1}{n\kappa }  \langle{{J}_R^{\varphi}}\rangle}\,.
\end{align}
Thus, supersymmetry is determining $h_D$ in terms of the conserved R-symmetry current of the LS vacuum.
Clearly we have $\vev{\cT_{a}^a}=0$. Since we have no sources for the scalar operators, the anomaly 
is proportional to $\vev{\cT_{a}^a}$ and hence again vanishes for these BPS configurations.
An extension of the holographic renormalisation analysis that we have carried out is needed
to obtain the expectation values of the scalar operators and of the massive vector operator in the LS theory, and we leave this to future work.

\subsection{Boundary conditions for a regular core}\label{sec:bcscore}

We are interested in solutions to the BPS equations that are regular at the core of the 
solution in the bulk, $y\to y_{core}$. We work in a gauge with regular gauge fields and so we require $a^{i}=0$ at the core. 
Recalling that the complex scalar has the form $\zeta=\varphi(y)e^{i\bar\theta z}$,
for solutions with $\varphi\ne0$ at the core of the solution, we should therefore
set
\begin{align}
\bar\theta=0\,.
\end{align}
At the core we then also have
\begin{align}\label{dthdzcoe}
\text{Core:}\qquad D_z\theta=0\,,\qquad Q_z=0\,.
\end{align}

We next examine regularity of the metric at the core. To do this it is convenient to use\footnote{We only use this gauge to analyse the core of the solution and use a different gauge when analysing the $AdS_5$ boundary.} a conformal gauge for the radial coordinate:
\begin{align}\label{confgaugef}
f=e^V\,.
\end{align}
By examining regularity of the metric at the core we deduce that
\begin{align}
\text{Core:}\qquad (n\sin\xi)'=-1,\qquad \cos\xi=(-1)^t\,,
\end{align}
where $t=0,1$.
From the BPS equation \eqref{xiderivnice} we deduce that $(s-Q_z)=\frac{1}{2}(-1)^{t+1}$ at the core and using \eqref{dthdzcoe}
we then deduce that the constant $s$ appearing in the phase of the Killing spinor \eqref{chiralityspintext} is related to $t$ via
\begin{align}\label{esstrelation}
s=-\frac{1}{2}(-1)^t\,.
\end{align}
This condition can also be deduced by ensuring\footnote{To check this one should recast the spinor in an orthonormal frame that
is regular at the core.} that the Killing spinor is smooth at the core of the solution. 
We next note that $\partial_{\varphi}Q_z= -\frac{1}{2}\sinh 2\varphi D_z\theta$ and so
$\partial_{\varphi}Q_z=0$ at the core. From the BPS constraint \eqref{summconstext2}, at the core we thus have 
\begin{align}\label{betbc}
\text{Core:}\qquad\partial_{\varphi}Q_z =\partial_{\varphi}W=0.
\end{align}
From the expression for $W$ in \eqref{superpottext} we then also have
(when $\varphi\ne0$),
\begin{align}\label{betbc2}
\text{Core:}\qquad \cosh 2\beta =\frac{1}{2}e^{6\alpha}\,, \qquad
W=-e^{4\alpha}\,.
\end{align}
Notice that for solutions with $\varphi\ne0$ of the \emph{LS truncation}, which has $\beta=0$ and $A^1=A^2$ (see section \ref{section:moretruncs}), we see that at the core
the scalar field $\alpha$ has the same value\footnote{\label{alphafpcore}{If $\varphi$ takes the LS fixed point value at the core then
the defect solutions have $\alpha,\varphi$ taking the fixed point values everywhere, associated with solutions of minimal gauged supergravity (recall section \ref{section:moretruncs}). For other values of $\varphi$ at the core, then we get solutions with varying $\alpha,\varphi$ which approach the
$\mathcal{N}=4$ $AdS_5$ vacuum at the boundary.}} as in the LS fixed point, $e^{6\alpha}=2$.

\subsection{Evaluating the conserved charges at the core}\label{evalccatcore}

We now want to examine the value of the two conserved charges $\mathcal{E}_R, \mathcal{E}_F$, given in \eqref{conschgestext}, at the core.
As in \cite{Arav:2022lzo}, it is convenient to first define two quantities 
\begin{align}\label{emmbc}
M_{(1)}\equiv g e^{4\alpha} e^V,\qquad M_{(2)} \equiv -2\kappa + 2 M_{(1)} \cos\xi\,,
\end{align}
and for future reference we note that $M_{(1)}>0$.
Using \eqref{betbc2} and
the first BPS constraint equation in \eqref{summconstext2} we deduce that at the core
\begin{align}\label{emmbcpoles}
\text{Core:}\qquad M_{(1)}&= 2 (-1)^{t} \kappa + \frac{1}{n} \,,\qquad
M_{(2)}= 2\kappa + \frac{2}{n} (-1)^{t} \,.
\end{align}
From \eqref{conschgestext} we can thus write
\begin{align}\label{eesemmms}
\text{Core:}\qquad\mathcal{E}_R &= \frac{M_{(1)}^2}{g^2} \left[ -\kappa + M_{(2)} e^{-12\alpha} \right]
\, ,\qquad
\mathcal{E}_F= \pm\frac{M_{(1)}^2}{g^2} \left[ 1 - 4e^{-12\alpha} \right]^{1/2}
\,,
\end{align}
where the $\pm$ is the sign of $\kappa\beta$. Hence, at the core we must have
\begin{align}\label{alphconst}
\text{Core:}\qquad0 < e^{-12\alpha}\leq 1/4\,.
\end{align} 
In particular, for solutions with $\varphi\ne0$ of the LS model (i.e. in the further truncation with $\beta=0$) we have $e^{-12\alpha}=1/4$ at the core,
and necessarily $\mathcal{E}_F=0$.

Taking the $\varphi\to 0$ limit, leading to the restricted STU solutions (defined in section \ref{secads3ansbcs}), we have the extra conserved quantity
\begin{align}\label{eesemmmsB2}
\text{Core}:\qquad \mathcal{E}_B
&=-\kappa \frac{M_{(1)}^2}{g^2}(1-2e^{-12\alpha})\,,\qquad (\varphi\to 0) \,.
\end{align}

Next recall from the $AdS_5$ boundary analysis, with $h > 0$ we deduced in \eqref {klesszero} that $n > 0$.
Since $M_{(1)}>0$ we deduce from \eqref{emmbcpoles} 
that if $0<n<\frac{1}{2}$ then $t$ is unrestricted, while
if $n \ge \frac{1}{2}$ then $(-1)^t=\kappa$ and hence $s=-\frac{1}{2} \kappa$.
Thus, we have two branches of solutions:
\begin{align}\label{constraintkgen}
\text{Branch 1 (main branch)}&: \qquad s=-\frac{\kappa}{2}, \qquad  \text{arbitrary $n$}\,,\nn
\text{Branch 2}&: \qquad s=+\frac{\kappa}{2}, \qquad  0<n<\frac{1}{2}\,.
\end{align}
Notice that the first branch contains solutions continuously connected to having no conical singularity (i.e. $n=1$) and hence
continuously connected to the vacuum with in addition the monodromy sources switched off; 
thus, we refer to this as the ``main branch" of solutions.
Later we present some numerically constructed solutions 
with $n=1$ and non-vanishing monodromy sources, with $\varphi\ne 0$. 
We have also shown that solutions for other values of $n$ exist, including solutions on the other
branch with $s=+\kappa/2$. 
We continue our analysis, for the most part, with both branches, sometimes restricting to the main branch $s=-\kappa/2$ for simplicity.

In particular, when
$s=-\kappa/2$,
at the core we have $M_{(1)}=2+\frac{1}{n}$ and $M_{(2)}=2\kappa(1+\frac{1}{n})$.
When there is no conical singularity on the boundary, $n=1$, we have 
$M_{(1)}=3$, $M_{(2)}=4\kappa$ and
\begin{align}\label{eesemmms2}
\text{$n=1$},\qquad \text{Core:}\qquad\mathcal{E}_R &= -\frac{9\kappa}{g^2} \left[1 - 4e^{-12\alpha}  \right]
\, ,\nn
\mathcal{E}_F&= \pm\frac{9}{g^2} \left[ 1 - 4e^{-12\alpha} \right]^{1/2}
\,.
\end{align}
Furthermore, from \eqref{n4scebpstext} and \eqref{LSscebpstext}, we also see that when 
$s=-\kappa/2$,
the R-symmetry source can be written, both for the $\mathcal{N}=$ SYM case and the LS case, as
\begin{align}
s=-\kappa/2:\qquad g\mu_R=\kappa(1-n)\,; \qquad g\mu^{LS}_R=\kappa(1-n)\,,
\end{align} 
and for the case of no conical singularity on the boundary
we have $g\mu_R=0$ and $g\mu^{LS}_R=0$ for the two cases.

For future use, note that using \eqref{betbc2} and \eqref{emmbcpoles}, at the core we have
\begin{align}\label{sumIscore}
(\mathcal{I}^1+\mathcal{I}^2)|_\text{core}=(\mathcal{I}^3)|_\text{core}=\frac{1}{g}(-n\kappa  +s)\,,
\end{align}
where the $\mathcal{I}^i$ are defined in \eqref{eq:IntegratedFluxesExpr1text}.


\subsection{Evaluating the conserved charges at the $AdS_5$ boundary}
We now consider the expansion at the $AdS_5$ boundary and evaluate the conserved charges. This allows us to relate the boundary currents to core quantities.
For both $\mathcal{N}=4$ and LS asymptotics,
from \eqref{veeachexp}, \eqref{xiexp} and the relation $ h e^{-V} = -n \sin\xi $, we deduce
\begin{equation}
h_0 e^{-V_0} = n \, ,
\end{equation}
and recall that as mentioned earlier $n$ must be positive.

\subsubsection{$\mathcal{N}=4$ SYM case, $\varphi\ne 0$}

We first consider the $\mathcal{N}=4$ case. The two conserved quantities (when $\varphi\ne 0$) 
are determined by the asymptotic data as (see 
\eqref{FGEREF})
\begin{align}\label{FGEREFtext}
\mathcal{E}_R &=\frac{16\pi^2e^{2V_0}L^2}{ nN^2}\langle{{J}_R^{\varphi}}\rangle, \qquad 
  \mathcal{E}_F= \frac{8\pi^2e^{2V_0}L^2}{ nN^2}\vev{J_F}\,.
\end{align}
For the restricted STU solutions with $\varphi=0$  we have the additional conserved quantity (see \eqref{FGEREFB})
\begin{align}\label{FGEREFBtext}
\mathcal{E}_B &=\frac{16\pi^2e^{2V_0}L^2}{ nN^2}\vev{J_B}\,,\qquad (\varphi\to 0) \,.
\end{align}
Comparing now with \eqref{eesemmms} 
we obtain the following relations between the two conserved currents and the core quantities: 
\begin{align}\label{eq:ConservedCurrentsFromCore}
 \langle{{J}_R^{\varphi}}\rangle &=\frac{nN^2}{64 \pi^2e^{2V_0}}{M_{(1)}^2 } \left( -\kappa + M_{(2)} e^{-12\alpha} \right)|_\text{core}\,,\nn
\vev{J_F} &=\pm\frac{nN^2}{32 \pi^2e^{2V_0}}  {M_{(1)}^2 } \sqrt{1-4e^{-12\alpha}}|_\text{core} \, ,
\end{align}
where once again $\pm$ corresponds to the sign of $\kappa \beta$. Note that, since $n>0$, we have 
$\operatorname{sign}(\vev{J_F}) = \kappa \operatorname{sign}(\beta)$.
Similarly, using \eqref{eesemmmsB2},
for the restricted STU solutions we have
\begin{align}\label{eq:ConservedCurrentsFromCoreB}
& \vev{J_B}=
-\frac{n \kappa N^2}{64 \pi^2e^{2V_0}}{ M_{(1)}^2}(1-2e^{-12\alpha})|_\text{core}\,,\qquad (\varphi\to 0) \,.
 \end{align}

In the case $n=1$, when the conformal boundary has no conical deficit angle in flat slicing, recalling that in this case $(-1)^t=\kappa$, these expressions simplify to:
\begin{align}
\langle{{J}_R^{\varphi}}\rangle&=- \frac{9\kappa N^2}{64\pi^2e^{2V_0}}  (1-4e^{-12\alpha})|_\text{core} \,, \nn
 \vev{{J}_F} &= \pm \frac{9\kappa N^2 }{32\pi^2e^{2V_0}}\sqrt{1-4e^{-12\alpha}}|_\text{core} \, ,\nn
\vev{{J}_B} &= -\frac{9\kappa N^2}{64\pi^2e^{2V_0}}  (1-2e^{-12\alpha})|_\text{core} \,,\qquad (\varphi\to 0) \,.
\end{align}

It is interesting to note that in the case of solutions of the \emph{LS truncation}, with $\beta=0$ and $A^{1}=A^{2}$
 (see section \ref{section:moretruncs}), we have
the core value of $\alpha$ is the same as that of the LS vacuum, $e^{6\alpha}|_\text{core}=2$,
as noted below \eqref{betbc2}. In this case, we have $g\mu_F= \vev{{J}_F}=0$. 
For the
$s=-\kappa/2$ branch
we have
$\langle{{J}_R^{\varphi}}\rangle=-\frac{n\kappa  N^2}{128\pi^2e^{2V_0}}(2+\frac{1}{n})^2(1-\frac{1}{n})$
and, when $\varphi=0$, $\vev{{J}_B}=-\frac{n\kappa  N^2}{128\pi^2e^{2V_0}}(2+\frac{1}{n})^2$.
For $n=1$, these become $\langle{{J}_R^{\varphi}}\rangle =0$ and $\vev{{J}_B} =-\frac{9\kappa N^2}{128\pi^2e^{2V_0}} $.

\subsubsection{LS case}
For the LS case, we find that the two conserved quantities 
can be expressed in terms of the asymptotic data as
\begin{align}\label{FGEREFtextLS}
\mathcal{E}_R &=\frac{16\pi^2e^{2V_0}L^2}{ nN^2}\langle{{J}_R^{\varphi}}\rangle, \qquad 
  \mathcal{E}_F= \frac{8\pi^2e^{2V_0}L^2}{ nN^2}\vev{J_F}\,.
\end{align}
which are \emph{exactly the same expressions} that we saw above \eqref{FGEREFtext} for $\mathcal{N}=4$ SYM asymptotics
with $\varphi\ne 0$. 
Since the $\mathcal{N}=4$ SYM case with $\varphi\ne 0$ and the LS case have the same symmetries one might have anticipated that the conserved charges are proportional to the same conserved currents, but it is remarkable that they also have the same proportionality constant.
Using \eqref{eesemmms} these are again related to core quantities as in \eqref{eq:ConservedCurrentsFromCore}
\begin{align}\label{eq:ConservedCurrentsFromCoreLS}
 \langle{{J}_R^{\varphi}}\rangle &=\frac{nN^2}{64 \pi^2e^{2V_0}}{M_{(1)}^2 } \left( -\kappa + M_{(2)} e^{-12\alpha} \right)|_\text{core}\,,\nn
\vev{J_F} &=\pm\frac{nN^2}{32 \pi^2e^{2V_0}}  {M_{(1)}^2 } \sqrt{1-4e^{-12\alpha}}|_\text{core} \, .
\end{align}

\section{Relating the currents to the sources}\label{sec:Iscore}
Continuing our analysis of solutions with $\varphi\ne0$,
we now explain how we can analytically relate the sources for the currents, $g\mu_i$, to quantities at the
core of the defect and hence obtain expressions for the currents in terms of the sources (without an explicit solution). Specifically, we will integrate the key result \eqref{effaiprimetext1} that follows from the BPS equations:
\begin{equation}\label{effaiprimet}
F^{i}_{yz} = (a^{i})' = (\mathcal{I}^{i})' \, ,
\end{equation}
where $\mathcal{I}^{i}$ are given in \eqref{eq:IntegratedFluxesExpr1text}.
Since we are working in a regular gauge with $\bar\theta=0$ and $a^{i}=0$ at the core, at the $AdS$ boundary $a^i$ defines the sources of the monodromy defect, $ga^{i}\to g\mu_i$.
We therefore deduce
\begin{equation}
\label{eq:GeneralIntegExpressionForFluxSources}
g\mu_i = g\mathcal{I}^{i}|_\text{bdy} - g\mathcal{I}^{i}|_\text{core} \, .
\end{equation}
We can now evaluate each of these expressions $\mathcal{I}^{i}$ at the boundary using the boundary expansion, 
and at the core using the boundary conditions at the core given in section \ref{sec:bcscore}

We first notice that we can write
\begin{align}\label{eyealbetv}
 \mathcal{I}^{1} \mathcal{I}^{2} \mathcal{I}^{3}=-\frac{n^3}{8}e^{3V}\cos^3\xi\,,\quad
 \frac{ \mathcal{I}^{1} }{ \mathcal{I}^{3}} = e^{-6\alpha+2\beta}\,,\quad
   \frac{\mathcal{I}^{1} }{\mathcal{I}^{2} }&=e^{4\beta}\,.
\end{align}
Then, at the boundary, both for $\mathcal{N}=4$ SYM and LS, 
we have $\cos\xi e^V|_\text{bdy}=\kappa R$ where $R$ is the radius of the
fixed point.
Hence
\begin{align}\label{eyealbetv2}
 e^{3V}|_\text{core}&=
 -2s \kappa 
 R^3(1-\frac{\mu_1}{I^1|_\text{bdy}})(1-\frac{\mu_2}{I^2|_\text{bdy}})(1-\frac{\mu_3}{I^3|_\text{bdy}})\,,\nn
 e^{-6\alpha+2\beta}|_\text{core}&=  e^{-6\alpha+2\beta}|_\text{bdy}(1-\frac{\mu_1}{I^1|_\text{bdy}})(1-\frac{\mu_3}{I^3|_\text{bdy}})^{-1}\,,\nn
e^{4\beta}|_\text{core}&=e^{4\beta}|_\text{bdy}(1-\frac{\mu_1}{I^1|_\text{bdy}})(1-\frac{\mu_2}{I^2|_\text{bdy}})^{-1}\,.
\end{align}
Notice that $(1-\frac{\mu_1}{I^1|_\text{bdy}})$, $(1-\frac{\mu_2}{I^2|_\text{bdy}})$,  $(1-\frac{\mu_3}{I^3|_\text{bdy}})$ and $-s \kappa$ all have the same sign.
Recall from \eqref{constraintkgen} that there are two branches of solutions, the main branch $s=-\kappa/2$, valid for all $n>0$ and
so all these quantities are positive, and branch 2 with
$s=+\kappa/2$, valid only for $0<n<1/2$, so all these quantities are negative.

As we will see, and remarkably, we will get the same results for $e^{3V}|_\text{core},  e^{-12\alpha}|_\text{core}$ and
$e^{4\beta}|_\text{core}$ for both $\mathcal{N}=4$ SYM and LS asymptotics, as a function of $n,s,\kappa$ and $g\mu_F$ (see \eqref{eyealbetv23again} and \eqref{eyealbetv23againLS}). To appreciate this we should recall
from section \ref{secads3ansbcs} that $g\mu_F$ labels the same source in both
the $\mathcal{N}=4$ SYM and LS cases.
This reveals an interesting universality at the core of the corresponding solutions.

\subsection{$\mathcal{N}=4$ SYM case, $\varphi\ne 0$}\label{subsecn4}
For the $\mathcal{N}=4$ SYM case, using the boundary expansion in appendix \ref{FGN=4}, we calculate from the
definitions \eqref{eq:IntegratedFluxesExpr1text} that $ g\mathcal{I}^{i}|_\text{bdy}=-n{\kappa }$
for all $i$. Recalling that the $AdS_5$ radius is $R=L$, we can thus write
\begin{align}\label{eyealbetv23}
 e^{3V}|_\text{core}&=-2s \kappa L^3(1+\frac{g\mu_1}{\kappa n})(1+\frac{g\mu_2}{\kappa n})(1+\frac{g\mu_3}{\kappa n})\,,\nn
 e^{-12\alpha}|_\text{core}&=  (1+\frac{g\mu_1}{\kappa n})(1+\frac{g\mu_2}{\kappa n})(1+\frac{g\mu_3}{\kappa n})^{-2}\,,\nn
 e^{4\beta}|_\text{core}&=(1+\frac{g\mu_1}{\kappa n})(1+\frac{g\mu_2}{\kappa n})^{-1}\,,
\end{align}
and from above $(1+\frac{g\mu_i}{\kappa n})$, $-s  \kappa$ have the same sign.\footnote{Notice that if we sum 
$(1+\frac{g\mu_i}{\kappa n})$ and use the condition
 on $g\mu_R$ in \eqref{eq:FluxSourcesFromCore}, we can also deduce there are two branches of solutions as
 in \eqref{constraintkgen}.}
Also, using the results that we established in \eqref{sumIscore} for $ \mathcal{I}^{i}|_\text{core}$ in \eqref{eq:GeneralIntegExpressionForFluxSources}
we deduce that sources for the broken symmetry and the R-symmetry currents are fixed given $n, \kappa$ and $s$ via
\begin{align}\label{eq:FluxSourcesFromCore}
&g\mu_B \equiv g\mu_1 + g\mu_2 - g\mu_3= -\kappa n \, ,\nn
&g\mu_R \equiv g\mu_1 + g\mu_2 + g\mu_3 
= -\kappa n -2 s \,,
\end{align}
consistent with the result \eqref{n4scebpstext} which is obtained from the expansion of the BPS equations near the boundary for $\mathcal{N}=4$.
The only independent sources are the flavour symmetry monodromy $g\mu_F \equiv g\mu_1 - g\mu_2$, as well as $\varphi_s$. 

We can then write the sources $g\mu_i$ as follows
\begin{align}\label{muinfour}
&g\mu_1 = \frac{1}{2}(-\kappa n-s+{g\mu_F})\,, \quad
g \mu_2 = \frac{1}{2}(-\kappa n-s-{g\mu_F})\ \,, \quad
 g \mu_3 = -s\, ,
\end{align} 
and hence we can obtain the following expressions for $V,\alpha$ and $\beta $ at the core using \eqref{eyealbetv23}:
\begin{align}\label{eyealbetv23again}
 e^{3V}|_\text{core}&=\frac{4s}{g^3 n^3} (-\kappa n+s -g\mu_F)(-\kappa n+s +g\mu_F)(-\kappa n+s)\,,\nn
 e^{-12\alpha}|_\text{core}&=  \frac{1}{4}(-\kappa n+s -g\mu_F)(-\kappa n+s +g\mu_F)(-\kappa n+s)^{-2}\,,\nn
 e^{4\beta}|_\text{core}&=(-\kappa n+s -g\mu_F)(-\kappa n+s +g\mu_F)^{-1}\,.
\end{align}
From the expression for $e^{4\beta}|_\text{core}$ we immediately deduce that $\mu_F$ is bounded:
\begin{equation}\label{eq:FlavourFluxSourceBound}
|g\mu_F| < 
 | -\kappa n + s|\, .
\end{equation}
This bound is saturated when $\alpha\to\infty$ at the core. 
For the $s=-\kappa/2$ main branch of solutions
this bound becomes 
$|g\mu_F| < 
 n+\frac{1}{2}$.
The case $\mu_F=0$ corresponds to the $SU(2)$ invariant (LS) case, in which $\alpha=\alpha_{LS}$ at the core
(and here it may be helpful to recall footnote \ref{alphafpcore}.)

For the case of $n=1$, associated with no conical singularity, these expressions all simplify. For example, we have
\begin{align}\label{bingo}
g\mu_R=0,\qquad g\mu_B=-{\kappa}\,,
\end{align}
and also 
\begin{align}\label{eyealbetv23kminone}
 e^{3V}|_\text{core}&= \frac{3}{g^3}(\frac{9}{4}-( g\mu_F)^2)\,,\nn
  e^{-12\alpha}|_\text{core}&= \frac{1}{9}(\frac{9}{4}-(g\mu_F)^2)\,,\nn
 e^{4\beta}|_\text{core}&=(\frac{3\kappa}{2}+ g\mu_F)(\frac{3\kappa}{2}- g\mu_F)^{-1}\,,
\end{align}
with the bound \eqref{eq:FlavourFluxSourceBound} becoming $|g\mu_F|<\frac{3}{2}$.
The value of $ \varphi|_\text{core}$ is determined by the value of $\varphi_s$, and can
be determined by constructing solutions numerically.

Next, combining \eqref{eq:ConservedCurrentsFromCore}, with \eqref{eyealbetv23again}
we can obtain expressions for 
the conserved currents $\langle{J_R^{\varphi}}\rangle$ and $\vev{J_F}$ in terms of $g\mu_F$,
 \begin{align}\label{allcurrentsintrmssces}
\langle{J_R^{\varphi}}\rangle&=\frac{N^2 }{32\pi^2 n^2e^{2V_0}}\Big[
       (g\mu_F)^2(2s-n\kappa)-(2s+n\kappa)(s-n\kappa)^2\Big]\,,\nn
\vev{J_F}&=
-\frac{N^2}{8\pi^2 ne^{2V_0}}(\kappa s-n)(g\mu_F)\,,
\end{align}
which is one of our main results. 
We also recall from the definitions \eqref{neq4currents}
that the finite counterterm contributions that appear in the currents in
\eqref{textbcurrents} are not present in $\langle{J_R^{\varphi}}\rangle$ and $\vev{J_F}$, which is associated with the fact that these are conserved currents
when $\varphi\ne0$. The constant counterterm parameter $\delta_\beta$ in \eqref{textbcurrents} only appears in $\vev{J_B}$. 

In general, in order to obtain an analogous expression for $\vev{J_B}$ as a function of $g\mu_F$,
we need to solve the BPS equations. However, in the limit that $\varphi\to 0$,
associated with a restricted STU solution, we can use 
\eqref{eq:ConservedCurrentsFromCoreB}, arising from the extra conserved quantity in this limit, and we find
\begin{align}
\vev{J_B}&=-\frac{N^2}{32\pi^2\kappa ne^{2V_0}}[(g\mu_F)^2+( s- n \kappa)^2],\qquad (\varphi\to 0)\,.
\end{align}

When 
$n=1$, with no conical singularities on the flat boundary, these expressions simplify to
\begin{align}\label{jrfbn4kminone}
\langle{J_R^{\varphi}}\rangle&=-\frac{\kappa N^2}{16\pi^2 e^{2V_0}}(g\mu_F)^2\,,\nn
\vev{J_F}&=\frac{3N^2}{16\pi^2 e^{2V_0}}(g\mu_F) \, ,\nn
\vev{J_B}&=-\frac{\kappa N^2}{32\pi^2 e^{2V_0}}[(g\mu_F)^2+\frac{9}{4}],\qquad (\varphi\to 0)\,.
\end{align}

We can now feed all of these results into the expressions for the stress tensor in \eqref{stresstextn4},
\eqref{stresstextn42} and the scalar one-point functions
in \eqref{vevstextn4}. Interestingly, since $\vev{\cO_\b}$  is proportional to $ \vev{J_F}$ we can immediately obtain an expression for $\vev{\cO_\b}$ in terms of $g\mu_F$. However, since $\vev{\cO_\a}, \vev{\cO_\varphi}$ and $h_D$ depend on $\vev{J_B}$, when $\varphi\ne 0$ we need to solve the BPS equations in order to get similar expressions for these quantities. However, for the restricted STU solutions we can obtain such expressions and, for example,
with $n=1$ and $\varphi\to 0$ we have 
\begin{align}\label{hdexpkmone}
h_D&= \frac{N^2}{\pi 192 e^{2 V_0}}\left(
28(g\mu_F)^2
+{9} \right),\qquad (\varphi\to 0)\,.
\end{align}

Unlike the solutions of the STU model with $\varphi=0$ that we consider in section \ref{sectstu}, we do not have analytic solutions when $\varphi\ne 0$.
However, we have numerically constructed some solutions, showing that they indeed exist. 
{For simplicity we focus on solutions with no conical singularity, setting $n=1$, and briefly comment upon the  $n\ne 1$ case below.
Setting $n=1$, which is necessarily on the main branch, we fix $\kappa=+1$
and hence for supersymmetry
we set $g\mu_R=0$ and $g\mu_B=-1$ (see \eqref{bingo}). We fix $g\mu_F$ and then find a one-parameter family of
solutions that can be specified by the value of $\varphi$ at the core, $\varphi_{\text{core}}$, or equivalently by  
specifying the asymptotic behaviour, $\varphi_s$, fixing the source in the $\mathcal{N}=4$ SYM theory. 
We find solutions with $\varphi_{\text{core}}\in (0,\varphi^{\text{crit}}_{\text{core}})$, with
the critical core value $\varphi^{\text{crit}}_{\text{core}}$ determined by $g\mu_F$. Equivalently, the solutions are specified 
by $\varphi_s\in (0,\infty)$, with $\varphi_s$ a monotonic function of $\varphi_{\text{core}}$.
We now illustrate by setting $g\mu_F=1$, and numerically we find $\varphi^{\text{crit}}_{\text{core}}\sim 0.46431$.
In figure \ref{fig:N4phi_vortex_num} we present a solution with $g\mu_F=1$
and with $\varphi_\mathrm{core}\sim  0.464$, somewhat near the critical value,
associated with $\varphi_s=25.1$. Notice that $\varphi(y)$ increases with $y$ and approaches close to the LS vacuum value 
$\frac{1}{2}\log{3}\sim 0.55$ (see \eqref{lsfpsc})
before dropping back down to zero at the boundary. That is, as the core value approaches $\varphi^{\text{crit}}_{\text{core}}$   
the solution develops an intermediate LS vacuum region (at the right hand side of the figure), before asymptoting to the $\mathcal{N}=4$ SYM theory.
\begin{figure}[htbp]
\begin{center}
\includegraphics[scale=.45]{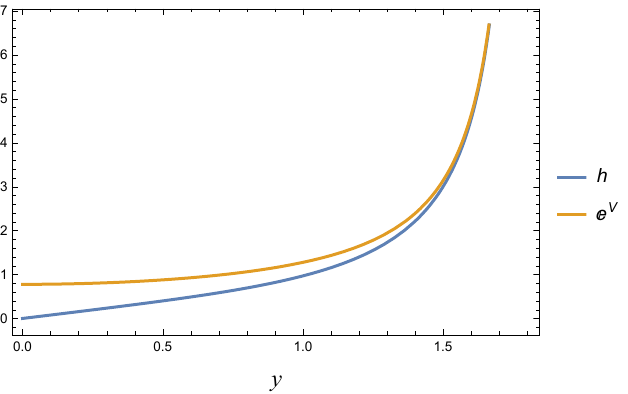}~
\includegraphics[scale=.45]{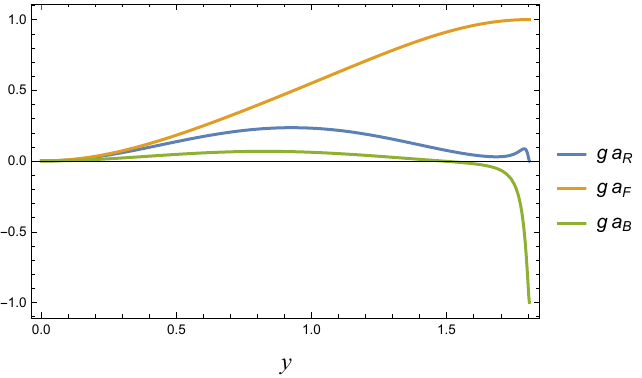}~
\includegraphics[scale=.45]{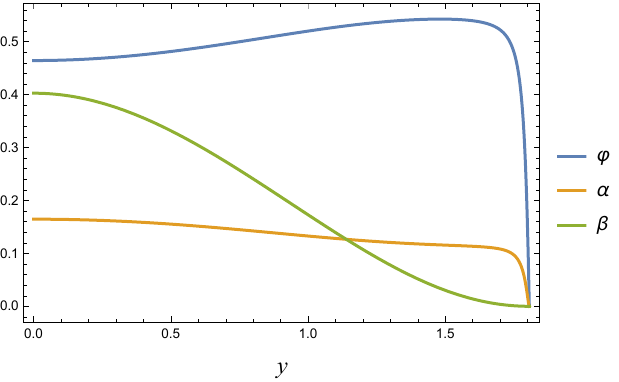}
\caption{
A solution with $\varphi\ne 0$ corresponding to a defect in $\mathcal{N}=4$ SYM theory with $n=1$ (i.e. no conical singularity), 
$\kappa=1$, $g\mu_R=0,~g\mu_B=-1$ (as required by supersymmetry) and $g\mu_F=1$, presented in conformal gauge \eqref{confgaugef}. 
In the left panel we have plotted the metric functions, in the middle panel the gauge field functions (associated with \eqref{lincsn4gf0}) and in the right panel the scalar functions. This 
solution exhibits an intermediate region where the solutions is nearly in the LS vacuum (recall from \eqref{lsfpsc} that the LS fixed point has $\varphi\sim 0.55$, $\alpha\sim 0.12$ and $\beta=0$).}
\label{fig:N4phi_vortex_num}
\end{center}
\end{figure}

We discuss solutions for defects in the LS theory in section \ref{subsecLS} and then STU solutions for defects in $\mathcal{N}=4$ SYM theory in
section \ref{sectstu}. Together we develop the following picture for $n=1$ solutions
(see figures \ref{fig:LS_vortex_num}, \ref{fig:rest_STU_vortex} and also
figure \ref{fig:sumsolutions}). 
Start with a restricted STU solution with $\varphi=0$ and
with $g\mu_R=0$, 
as required by supersymmetry, $g\mu_B=-\kappa $ imposed by hand and a fixed value of 
$g\mu_F$, consistent with the range \eqref{eq:FlavourFluxSourceBound}, which
here is $|g\mu_F|<\frac{3}{2}$. We can then switch on
$\varphi_s$ while preserving supersymmetry, associated with defects in $\mathcal{N}=4$ SYM theory 
with spatially dependent mass deformations. The values of $\alpha,\beta$ and $V$ approach a universal value at the core, given by
\eqref{eyealbetv23kminone}.

As we increase $\varphi_s$ the solution starts developing an intermediate scaling regime and
as $\varphi_s\to \infty$ the solution closely approximates the LS defect solution (which has $g\mu_B=0$), with the same value of $g\mu_F$, for values of $y$ from the core value out to a fixed value of $y$ before suddenly dropping back to the $\mathcal{N}=4$ SYM theory at the boundary value of $y$.
In particular, as $\varphi_s\to \infty$ the critical core value 
$\varphi^{\text{crit}}_{\text{core}}$, mentioned above,
is exactly the same as the core value in the LS defect solution, $\varphi^{\text{crit}}_{\text{core}}=\varphi^{LS}_{\text{core}}$.
Interestingly, we show in section \ref{susyrenyi} that
the on-shell action for this
one parameter family of solutions is actually \emph{independent} of $\varphi_s$ and has the same value as the restricted STU solution (with $g\mu_B=-n\kappa $)
and also as the LS defect solution (with $g\mu_B=0$).
We also recall that while $\langle J_R^\varphi\rangle $, $\langle J_F\rangle$ are independent
of $\varphi_s$, the conformal weight $h_D$ does depend on $\varphi_s$, since it depends on the non-conserved current $\langle J_B\rangle$.

Our numerical investigations support the above picture, when $n=1$ for all values of $|g\mu_F|<\frac{3}{2}$ in the bound
\eqref{eq:FlavourFluxSourceBound}. We now briefly comment on what happens when $n\ne 1$. 
Our numerical investigations indicate that the overall picture outlined above for the $n=1$ solution space and their properties,
continues to hold both for the main branch with $n\ne 1$ and also for branch 2 solutions 
(whenever these solutions actually exist for a given $g\mu_F$ and $n$). In particular,
the bound \eqref{eq:FlavourFluxSourceBound} is a necessary condition for the existence of defect solutions; it goes beyond the scope of this paper to determine sufficient conditions when $\varphi\ne 0$, but it appears that \eqref{eq:FlavourFluxSourceBound} might be sufficient
for the main branch of solutions, while for the branch 2 solutions it is not sufficient in general.}

\subsection{LS case}\label{subsecLS}

Consider now the case of a solution which is asymptotically LS, \emph{i.e.} a solution for which the scalar fields approach
\eqref{lsfpsc} and $R_{LS}=3/(2^{2/3}g)$. From the boundary expansion given in appendix \ref{FGLS}
we calculate $ \mathcal{I}^{1}|_\text{bdy}=\mathcal{I}^{2}|_\text{bdy}=-\frac{3\kappa n}{4g}$ and $\mathcal{I}^{3}|_\text{bdy}=-\frac{3\kappa n}{2g}$. Thus, \eqref{eyealbetv2} now gives
\begin{align}\label{eyealbetv23LS}
 e^{3V}|_\text{core}&=-2s \kappa R_{LS}^3(1+\frac{4g\mu_1}{3\kappa n})(1+\frac{4g\mu_2}{3\kappa n})(1+\frac{2g\mu_3}{3\kappa n})\,,\nn
 e^{-12\alpha}|_\text{core}&=  \frac{1}{4}(1+\frac{4g\mu_1}{3\kappa n})(1+\frac{4g\mu_2}{3\kappa n})(1+\frac{2g\mu_3}{3\kappa n})^{-2}\,,\nn
 e^{4\beta}|_\text{core}&=(1+\frac{4g\mu_1}{3\kappa n})(1+\frac{4g\mu_2}{3\kappa n})^{-1}\,.
\end{align}
Now
 $(1+\frac{4g\mu_1}{3\kappa n})$,  $(1+\frac{4g\mu_2}{3\kappa n})$,  $(1+\frac{2g\mu_3}{3\kappa n})$ and
 $-s  \kappa$ all have the same sign.\footnote{Notice that if we sum 
 $(1+\frac{4g\mu_1}{3\kappa n})$,  $(1+\frac{4g\mu_2}{3\kappa n})$,  $(1+\frac{2g\mu_3}{3\kappa n})$ and use the condition
 on $g\mu_R^{LS}$ in \eqref{eq:FluxSourcesFromCoreLS}, we can also deduce there are two branches of solutions as
 in \eqref{constraintkgen}.}
Using the results in \eqref{sumIscore} for $ \mathcal{I}^{i}|_\text{core}$,
we deduce that sources for the broken symmetry and the R-symmetry
 are fixed, given $n, \kappa$ and $s$, with:
 \begin{align}\label{eq:FluxSourcesFromCoreLS}
&g\mu_B \equiv g\mu_1 + g\mu_2 - g\mu_3= 0 \, ,\nn
&g\mu_R^{LS}\equiv \frac{4}{3}(g\mu_{1} +g\mu_{2}+\frac{1}{2}g\mu_{3})=2g\mu_3
= -\kappa n -2 s\,.
\end{align}
Since $g\mu_B=0$ we have $g\mu_R^{LS}=g\mu_1 + g\mu_2 + g\mu_3\equiv g\mu_R$ which is exactly the same value as for the $\mathcal{N}=4$ SYM case \eqref{eq:FluxSourcesFromCore}.

The only independent source is again the flavour symmetry flux $g\mu_F \equiv g\mu_1 - g\mu_2$.
We can now write the sources $g\mu_i$ as follows
\begin{align}\label{muinfourLS}
&g\mu_1 = \frac{1}{4}(-\kappa n-2s+{2g\mu_F})\,, \quad
g \mu_2 = \frac{1}{4}(-\kappa n-2s-{2g\mu_F})\ \,, \quad
 g \mu_3 = -\frac{1}{2}\kappa  n-s\, ,
\end{align} 
and obtain expressions for $V,\alpha$ and $\beta $ at the core using \eqref{eyealbetv23LS}.
In fact we find exactly the same expressions that we had for the $\mathcal{N}=4$ case given in \eqref{eyealbetv23again}:
\begin{align}\label{eyealbetv23againLS}
 e^{3V}|_\text{core}&=\frac{4s}{g^3 n^3} (-n\kappa +s -g\mu_F)(-n\kappa +s +g\mu_F)(-n\kappa +s)\,,\nn
  e^{-12\alpha}|_\text{core}&=  \frac{1}{4}(-\kappa n+s -g\mu_F)(-\kappa n+s +g\mu_F)(-\kappa n+s)^{-2}\,,\nn
 e^{4\beta}|_\text{core}&=(-\kappa n+s -g\mu_F)(-\kappa n+s +g\mu_F)^{-1}\,,
\end{align}
thus demonstrating a kind of attractor mechanism at play. 
Note that the LS solutions have $\varphi$ approaching the
LS vacuum value at the asymptotic boundary and correspondingly approach a
specific core value $\varphi^{LS}|_\text{core}$ for given values of $g\mu_F$ and $n$ (and fixed $\kappa,s$). 
Setting $g\mu_F=1$, $n=1$ (and $\kappa=1$) to illustrate, numerically
we find $\varphi^{LS}|_\text{core}\sim0.46431$, which as we mentioned above is also the same value as
the critical core value for the one-parameter family of solutions labelled by $\varphi_s$.
Thus, from \eqref{eyealbetv23againLS}
we deduce exactly the same bound for $g\mu_F$ that we saw before in \eqref{eq:FlavourFluxSourceBound}: 
\begin{equation}\label{eq:FlavourFluxSourceBoundLS}
|g\mu_F| < 
 | -\kappa n + s|\, .
\end{equation}
This bound is saturated when $\alpha\to\infty$ at the core. 
For the $s=-\kappa/2$ branch of solutions
this bound becomes 
$|g\mu_F| < n+\frac{1}{2}$.
The case $\mu_F=0$ corresponds to the $SU(2)$ invariant (LS) case, in which $\alpha=\alpha_{LS}$ at the core.

For the case of $n=1$ these expressions all simplify. For example, we have
\begin{align}\label{bingo2}
g\mu_B= g\mu_R=0\,,
\end{align}
with core values as in \eqref{eyealbetv23kminone}.
The bound \eqref{eq:FlavourFluxSourceBoundLS} is again $|g\mu_F|<3/2$.

By following the same steps as in the previous subsection we can obtain 
expressions for the conserved currents in terms of the monodromy sources.
Remarkably we find exactly the same result:
 \begin{align}\label{allcurrentsintrmsscesLS}
\langle{J_R^{\varphi}}\rangle&=\frac{N^2 }{32\pi^2 n^2e^{2V_0}}\Big[
       (g\mu_F)^2(2s-n\kappa)-(2s+n\kappa)(s-n\kappa)^2\Big]\,,\nn
     \vev{J_F}&=
-\frac{N^2}{8\pi^2 ne^{2V_0}}(\kappa s-n)(g\mu_F)\,.
\end{align}
Recalling \eqref{stresstextLS2},
\begin{align}\label{stresstextn423LS2}
h_D= -\frac{2\pi }{\kappa n}  \vev{J_R^{\varphi}}\,.
\end{align}
we can express $h_D$ in terms of the monodromy sources.
For example, when $n=1$ we get 
\begin{align}\label{hdnmone}
h_D=\frac{N^2}{8\pi e^{2V_0}}(g\mu_F)^2\,.
\end{align}
Given the constraint \eqref{eq:FlavourFluxSourceBoundLS}, one can also show for the main branch ($s=-\kappa/2$) that when 
$n\ge 1$ we have $h_D\ge 0$. For $n<1$ on this branch we can have $h_D<0$; for example in solutions
of minimal gauged supergravity (see appendix \ref{mingsugrasols}). For branch 2 solutions, we have $h_D>0$.

\begin{figure}[htbp]
\begin{center}
\includegraphics[scale=.45]{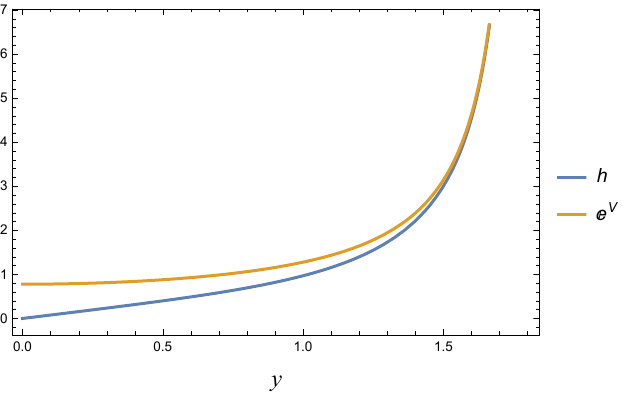}~
\includegraphics[scale=.45]{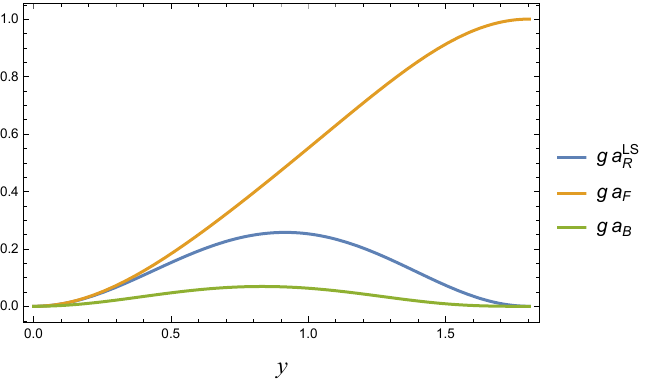}~
\includegraphics[scale=.45]{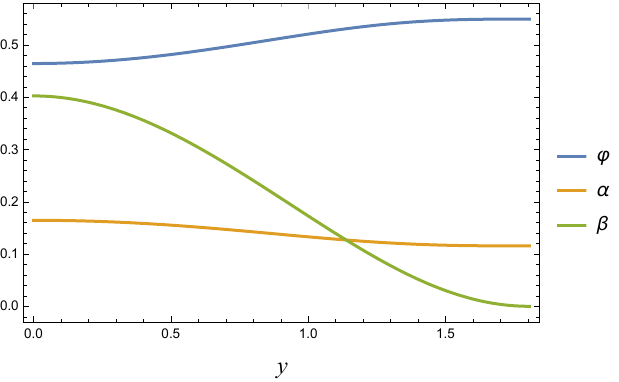}
\caption{A solution corresponding to a defect in the LS theory with $n=1$ (i.e. no conical singularity), $g\mu_R^{LS}=g\mu_B=0$ (as required by supersymmetry) and $g\mu_F=1$, presented in conformal gauge \eqref{confgaugef} with $\kappa=1$. 
In the left panel we have plotted the metric functions, in the middle panel the gauge field functions (associated with \eqref{lincsLSgf}, in contrast to
figure \ref{fig:N4phi_vortex_num}) and in the right panel the scalar functions. A comparison of the core behaviour of $V,\alpha$ and $\beta $
in this figure and in figure
\ref{fig:N4phi_vortex_num} reveals an attractor mechanism at play. The core value of $\varphi$ in figure \ref{fig:N4phi_vortex_num} was chosen to be close
to the core value of $\varphi$ in this figure.}
\label{fig:LS_vortex_num}
\end{center}
\end{figure}

As in the previous subsection we don't have analytic solutions, so we numerically verify that solutions exist. In figure \ref{fig:LS_vortex_num} we demonstrate a representative solution with $n=1$, $g\mu_F=1$ (and $\kappa=1$) and so, from \eqref{bingo2}, $g\mu_R=g\mu_B=0$.
Notice that the solution has exactly the same core values of $\alpha,\beta$ and $V$ as in figure \ref{fig:N4phi_vortex_num}, associated with the attractor mechanism - recall \eqref{allcurrentsintrmsscesLS}. We also note that for the solutions in figure \ref{fig:N4phi_vortex_num} we chose a core value for $\varphi$ close to the critical value,
$\varphi^{\text{crit}}_{\text{core}}=\varphi^{LS}_{\text{core}}$, and hence the whole solution is very close to that of the LS defect solution in
figure \ref{fig:LS_vortex_num} up to a fixed value of $y$, before it suddenly turns back to the $\mathcal{N}=4$ SYM boundary.
Our investigations are consistent with these solutions existing, when $n=1$, for all values of $g\mu_F$ consistent with the
bound \eqref{eq:FlavourFluxSourceBoundLS}, $|g\mu_F|<3/2$.
For $n\ne 1$, based on preliminary investigations
we anticipate that solutions on the main branch will exist for all values of ${g\mu_F}$ satisfying the
bound \eqref{eq:FlavourFluxSourceBoundLS}, while for branch 2 solutions we expect additional restrictions, but we leave a verification
of this to future work.

\section{Solutions of the STU model, $\varphi=0$}\label{sectstu}
The STU model, relevant for $\mathcal{N}=4$ SYM theory, 
is obtained when the complex scalar $\zeta=\varphi e^{i\theta}=0$. As we argue below, there
are again two possible branches of solutions, as we saw when $\varphi\ne 0$ in \eqref{constraintkgen}.
Analytic solutions\footnote{We also highlight that
the same local supergravity solutions of the STU model have been studied in \cite{Couzens:2021tnv,Suh:2021ifj}, but with 
a different global completion and a different physical interpretation.}
 for
the STU model are known (see e.g. \cite{Gauntlett:2006ns,Kunduri:2007qy,Huang:2014pda}) 
and we have checked that solutions for both branches in \eqref{constraintkgen} exist; 
we further comment on the solution space below.

Following our approach in previous sections, however,
we do not need to utilise the analytic solutions in order to compute various quantities of interest as we now discuss.
In fact our techniques greatly facilitate the computation of the current and stress tensor one-point functions
expressed in terms of the monodromy sources. Indeed, working with the explicit solutions, the most straightforward approach
requires dealing with roots of a cubic function, and this is presumably why these one point functions have not been presented previously.

We first highlight that from the expansion of the BPS equations at the  $AdS_5$ boundary, for the general STU model 
solutions we only have a single constraint
on the sources $g\mu_i$:
\begin{align}\label{n4scebpstextstu}
g\mu_R&\equiv g\mu_1+g\mu_2+g\mu_3=-{\kappa n}-2s\,,
\end{align}
In particular, in general we do not have  $g\mu_B=-\kappa n$ as in the restricted STU solutions
discussed in previous sections (recall \eqref{n4scebpstext}).

For the STU model and the $AdS_3$ ansatz, the equations of motion for the gauge fields now give rise
to three constants of motion $(\mathcal{E}_R, \mathcal{E}_F, \mathcal{E}_B)$ defined in
\eqref{gaugeintmot}, \eqref{extraone}.
It is more convenient to now use the linear combinations
\begin{align}
\mathcal{E}_1=\frac{1}{4}(\mathcal{E}_R+2\mathcal{E}_F+\mathcal{E}_B)\,,\quad
\mathcal{E}_2=\frac{1}{4}(\mathcal{E}_R-2\mathcal{E}_F+\mathcal{E}_B)\,,\quad
\mathcal{E}_1=\frac{1}{4}(\mathcal{E}_R-\mathcal{E}_B)\,.
\end{align}
Using the BPS equations, from \eqref{conschgestext}, \eqref{conschgestext2}
these
can be rewritten in the form 
\begin{align}
\mathcal{E}_1&=\frac{g}{2}e^{3V}\cos\xi-\kappa e^{2V+2\alpha-2\beta}\,,\nn
\mathcal{E}_2&=\frac{g}{2}e^{3V}\cos\xi-\kappa e^{2V+2\alpha+2\beta}\,,\nn
\mathcal{E}_3&=\frac{g}{2}e^{3V}\cos\xi-\kappa e^{2V-4\alpha}\,.
\end{align}
Notice that we can write
\begin{align}\label{stucaseei}
\mathcal{E}_i&=\frac{g\cos\xi e^{3V}}{2}(1+\frac{\kappa n}{g \mathcal{I}^i})\,,
\end{align}
where the $\mathcal{I}^i$ are defined in \eqref{eq:IntegratedFluxesExpr1text}.
We can also write the superpotential $W$, defined in \eqref{superpottext}, in the form
\begin{align}\label{superpotstu}
W=\frac{1}{ne^V\cos\xi}\sum_i \mathcal{I}^i\,.
\end{align}

We now evaluate the conserved quantities $\mathcal{E}_i$ at the $\mathcal{N}=4$, $AdS_5$ boundary. Using
the boundary expansion presented in appendix \ref{FGN=4} we find
\begin{align}\label{eeyeone}
\mathcal{E}_i
=\frac{4\pi^2 L^2 e^{2V_0}}{n N^2}\vev{J_i}\,,
\end{align}
and note that for the STU model all of the $\vev{J_i}$ are independent of the finite counterterms
in the renormalisation scheme
\eqref{bpsscheme} which we are using for $\mathcal{N}=4$ SYM theory.

We can also obtain an expression for $\mathcal{E}_i$ at the core from \eqref{stucaseei}, by using the results of
section \ref{subsecn4}.
Specifically, using 
\begin{equation}
\label{eq:GeneralIntegExpressionForFluxSourcesagain}
g\mu_i = g\mathcal{I}^{i}|_\text{bdy} - g\mathcal{I}^{i}|_\text{core} \, ,
\end{equation}
and the fact that for $\mathcal{N}=4$ SYM boundary we have $\mathcal{I}^{i}|_\text{bdy}=-\kappa n/g$, we obtain
\begin{align}\label{valuesorestukey}
g\mathcal{I}_i|_\text{core}&=-(\kappa n+g\mu_i)\,,\nn
 e^{3V}|_\text{core}&=-2s \kappa L^3(1+\frac{g\mu_1}{\kappa n})(1+\frac{g\mu_2}{\kappa n})(1+\frac{g\mu_3}{\kappa n})\,,
\end{align}
with $(1+\frac{g\mu_i}{\kappa n})$, $-s \kappa$ having the same sign, 
and this, along with the supersymmetry condition
\eqref{n4scebpstextstu}, 
determines the allowed range of the monodromies. 
By summing $(1+\frac{g\mu_i}{\kappa n})$ and using \eqref{n4scebpstextstu} we 
also deduce there are two branches of solutions as in \eqref{constraintkgen}. 
Thus, for the STU solutions we therefore have the necessary constraints on the space of solutions given by
\begin{align}\label{stubranches}
\text{Main branch}:\quad  &s=-\frac{\kappa}{2},\quad g\mu_R=\kappa(1-n),\quad (1+\frac{g\mu_i}{\kappa n})>0,\quad 0<n\,,\nn
\text{Branch 2}: \quad&s=+\frac{\kappa}{2}, \quad  g\mu_R=-\kappa(1+n),\quad (1+\frac{g\mu_i}{\kappa n})<0,\quad
0<n<\frac{1}{2}\,.
\end{align}
For branch 2 there are additional bounds on the value of $n$ as most simply illustrated in
appendix \ref{mingsugrasols} for solutions of minimal gauged supergravity where we show that 
branch 2 solutions exist for $0<n<1/3$. For the main branch it appears that all values of $n>0$ are allowed 
(in fact for minimal gauged supergravity this is explicitly proven in appendix \ref{mingsugrasols}).
We will refer to the bounds on $g\mu_i$ in
\eqref{stubranches} as ``naive bounds" as there are further restrictions in practise, as we illustrate for the main branch below.

For the conserved quantities, we can now write
\begin{align}\label{eeyetwo}
\mathcal{E}_1&=\frac{L^2}{n^3}(g\mu_1)(\kappa n+g\mu_2)(\kappa n+g\mu_3),\qquad \text{and cyclic}\,.
\end{align}
Equating \eqref{eeyeone} with \eqref{eeyetwo} we deduce an expression for $\vev{J_i}$ in terms of the $g\mu_i$ and hence we conclude
\begin{align}\label{litjistu}
\vev{J_1}&=\frac{e^{-2V_0}N^2}{4\pi^2}(g\mu_1)(1+\frac{g\mu_2}{\kappa n})(1+\frac{g\mu_3}{\kappa n})\,,\nn
\vev{J_2}&=\frac{e^{-2V_0}N^2}{4\pi^2}(g\mu_2)(1+\frac{g\mu_1}{\kappa n})(1+\frac{g\mu_3}{\kappa n})\,,\nn
\vev{J_3}&=\frac{e^{-2V_0}N^2}{4\pi^2}(g\mu_3)(1+\frac{g\mu_1}{\kappa n})(1+\frac{g\mu_2}{\kappa n})\,.
\end{align}
For the stress tensor from we find
\begin{align}\label{stresstextn4stu}
\vev{\cT_{ab}}dx^a dx^b&=\frac{h_D}{2\pi}\left[ds^2(AdS_3)+3n^2 dz^2\right]\,,
\end{align}
where, as in the first line of \eqref{stresstextn42},
\begin{align}\label{genstuhdexpress}
h_D=  -\frac{2\pi }{3\kappa n} \left( \vev{J_1} + \vev{J_2} + \vev{J_3}  \right)
=-\frac{2\pi }{\kappa n}\langle{J}_R^{\mathcal{N}=4}\rangle\,.
\end{align}
In particular, 
when $n=1$ we have 
 \begin{align}\label{hdeesnfour}
h_D=\frac{N^2}{6\pi e^{2V_0}}\Big[
\sum_i(g\mu_i)^2-3\kappa(g\mu_1)(g\mu_2)(g\mu_3)\Big]\,.
\end{align}
Recall that when $n=1$ we must have $(1+\kappa{g\mu_i})>0$ and so $h_D\ge 0$, with $h_D=0$ iff
there is no monodromy defect, $g\mu_i=0$. 
In fact, following the analysis of appendix E of \cite{toappear}, we can show for the main branch ($s=-\kappa/2$) that when 
$n\ge 1$ we have $h_D\ge 0$. For $n<1$ on this branch we can have $h_D<0$; for example this is the case for the solutions
of minimal gauged supergravity discussed in appendix \ref{mingsugrasols}. For branch 2 solutions, again following \cite{toappear},
we can show that $h_D>0$.

From \eqref{vevstextn4}, the expectation values of the scalar operators are given by
\begin{align}\label{vevstextn4again}
\vev{\cO_\a} &=- \frac{1}{\kappa n} \left( \vev{J_1}+  \vev{J_2}  - 2 \vev{J_3}  \right)\,,\nn
\vev{\cO_\b} &= \frac{1}{\kappa n} \left( \vev{J_1}  -\vev{J_2}   \right)\,,\nn
\vev{\cO_\varphi} &=0
\,.
\end{align}

\begin{figure}[htbp]
\begin{center}
\includegraphics[scale=.6]{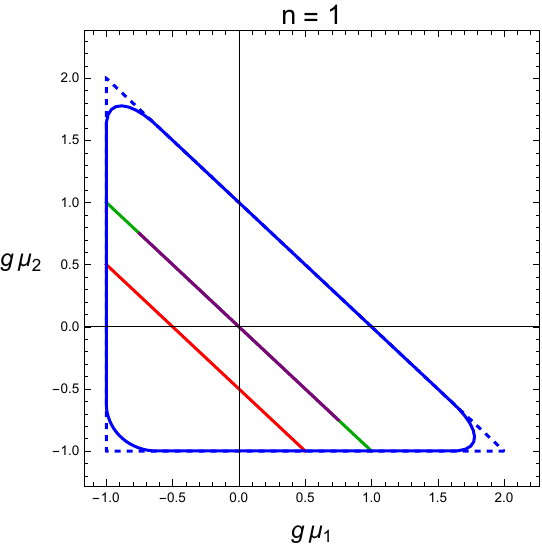}\qquad
\includegraphics[scale=.6]{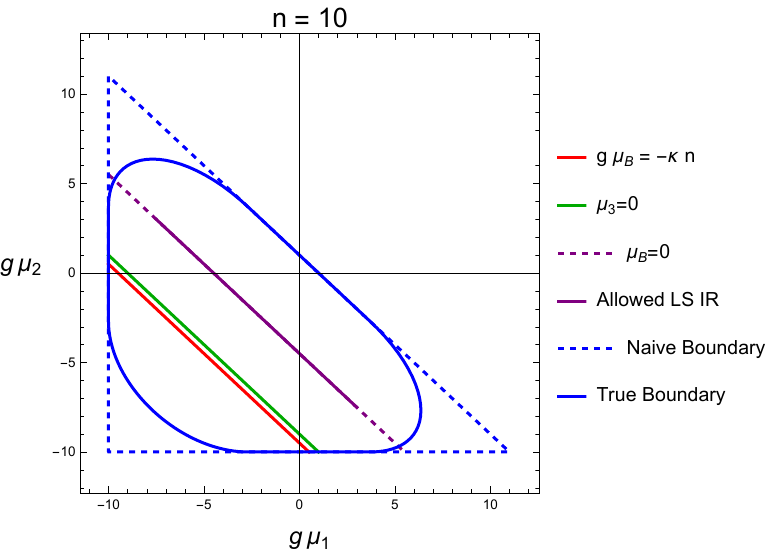}~
\caption{Solutions space of STU solutions as a function of two independent monodromy parameters $g\mu_1$, $g\mu_2$ for $n=1$ (left plot) and $n=10$ (right plot), for the main branch of solutions with $s=-\kappa/2$ (we have set $\kappa=+1$). }
\label{fig:stusolnsspace}
\end{center}
\end{figure}

We now make some additional comments on the space of solutions for the main branch. We seem to find that solutions
exist for all values of $n>0$ but, for a given $n$, for values of the $g\mu_i$ which are more restricted than the ``naive bounds" given in
\eqref{stubranches} as illustrated in figure \ref{fig:stusolnsspace} for $n=1$ and $n=10$ (the corresponding plot for $n=1/2$ is similar with the solid blue line closer to the dashed blue triangle).
In each figure the dashed blue triangle is the naive bound on $g\mu_i$ as given in \eqref{stubranches}, while the solid blue line is the boundary of the region
where the STU defect solutions actually exist.\footnote{There is an interesting interplay with the complementary existence of spindle type solutions as discussed in the context
of minimal gauged supergravity in
appendix \ref{mingsugrasols}.} We have also plotted the naive bounds on the $g\mu_i$ for various sub-classes of solutions:
(i) the restricted STU solutions with $g\mu_B=-\kappa n$ (red line) 
and (ii) the defect solutions with $g\mu_3=0$ and enhanced $\mathcal{N}=(2,2)$ supersymmetry (green line);
for both of these cases we find that the naive bound is actually the true bound on the space of solutions.
(iii) STU defects with $g\mu_B=0$ (dashed purple line); for this case when $n=1$, so $g\mu_R=0$, we have $g\mu_B=0$ implies that $g\mu_3=0$
and so the green line is coincident with the dashed purple line for $n=1$. Also for this case, we
see in the right plot that when $n=10$, the true bound for $g\mu_B=0$, which lies inside the blue solid curve, is 
more restricted than the naive bound. Finally we have also marked on top of the dashed purple line, the bound for the existence of LS solutions
lying on the main branch, $|g\mu_F| < n+\frac{1}{2}$, given in \eqref{eq:FlavourFluxSourceBound}.

\begin{figure}[htbp]
\begin{center}
\includegraphics[scale=.45]{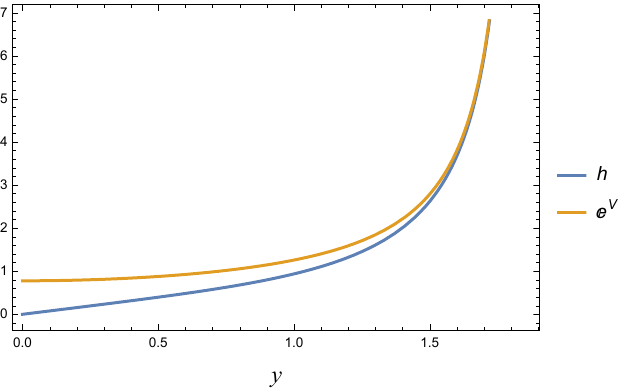}~
\includegraphics[scale=.45]{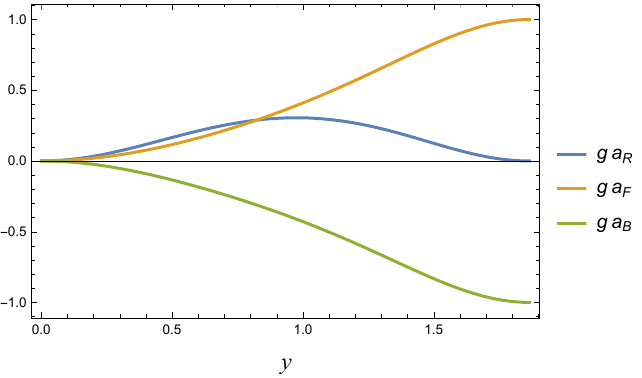}~
\includegraphics[scale=.45]{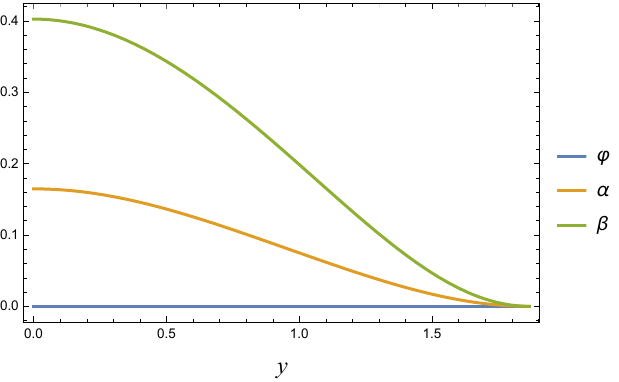}
\caption{A restricted STU solution with $\varphi=0$ and $n=1,~g\mu_R=0 ,~ g\mu_B=-1$ and $g\mu_F=1$, with $\kappa=1$. Note that
the core values of $\alpha,\beta$ and $V$ are the same as in figures \ref{fig:N4phi_vortex_num} and \ref{fig:LS_vortex_num}.}
\label{fig:rest_STU_vortex}
\end{center}
\end{figure}

A number of further comments are also now in order. Firstly, for the STU solutions, if in addition to
the constraint on the R-symmetry source given in \eqref{n4scebpstextstu} we also impose
$g\mu_B=-\kappa n$ then one recovers the results for the restricted STU solutions discussed in previous
sections, as expected. 
In particular, the core values of $\alpha,\beta$ and $V$ take the attractor values given in
\eqref{eyealbetv23again}.
We plot a restricted STU solution in figure \ref{fig:rest_STU_vortex} with $n=1$ for comparison 
with the previous solutions with $\varphi \neq 0$ in figures \ref{fig:N4phi_vortex_num} and \ref{fig:LS_vortex_num}. 
Also, as we have just noted, the restricted STU solutions appear to exist for values of the monodromy parameters consistent with the naive bound
(the red lines in figure \ref{fig:stusolnsspace}).

Second,
imagine one starts from STU monodromy defects in flat space, then set $g\mu_B=0$ by hand
and deform by a homogeneous relevant mass deformation \emph{away} from
the defect. This will induce an RG flow whose IR endpoint should be a defect in the LS theory;
we call this an ``$\mathcal{N}=4$/LS bulk defect RG flow" - see figure \ref{fig:sumsolutions}.
We next note that for the STU solutions with $g\mu_B=0$, the condition below \eqref{valuesorestukey} implies the $s=+\kappa/2$ branch 2 solutions
only exists for $0<n<1/3$ i.e. more restrictive than in \eqref{constraintkgen}. We now continue this discussion focussing
on the main branch $s=-\kappa/2$. For $n=1$ the STU solutions with $g\mu_B=0$ exist for $|g\mu_F|<2$, while
the LS defect solutions are allowed for a smaller range $|g\mu_F| <  \frac{3}{2}$. Thus, when $n=1$ the $\mathcal{N}=4$/LS bulk defect RG flows 
can only exist for the range $|g\mu_F|< \frac{3}{2}$ and we leave it as an interesting open question to determine what happens in the IR for
RG flows with $g\mu_F$ in the range $\frac{3}{2}\le|g\mu_F|< {2}$. Similar comments apply to the main branch solutions for other values of $n$
with the additional feature that the $g\mu_B=0$ solutions might only exist for a more restricted range than the naive space of solutions. This
is illustrated in figure \ref{fig:stusolnsspace} for the case of $n=10$: the dashed purple line is the naive set of STU solutions with $g\mu_B=0$,
while the true solutions lie inside the solid blue curve. Notice that this range is still larger than the range of LS solutions as marked by the solid purple line.
For the $\mathcal{N}=4$/LS bulk defect RG flow that do actually exist, we can make a comparison of $h_D$ in the UV and the IR.
For example, taking $n=1$ for simplicity, setting $g\mu_B=0$ in \eqref{hdeesnfour}
we obtain for the STU defects in the UV:
\begin{align}\label{hdbelowrgflowdosc}
h_D=\frac{N^2}{12\pi e^{2V_0}}(g\mu_F)^2\,.
\end{align}
In the IR, for the LS defect with $n=1$ we get the result \eqref{hdnmone} which is 3/2 the value in the UV.

Third,
if we set $A^1=A^2$ and $\beta=0$  in the STU model, we obtain a model that arises as a
sub-truncation of Romans $SU(2)\times U(1)$ supergravity theory. 
The associated defect solutions are obtained by setting $g\mu_1=g\mu_2$ and $\beta=0$, with $g\mu_1=g\mu_2$ associated with a
source for the $U(1)\subset SU(2)$ and $g\mu_3$ associated with a source for the $U(1)$. Generically, these defects preserve $\mathcal{N}=(0,2)$ 
supersymmetry, but if we further set $g\mu_3=0$, we would get $\mathcal{N}=(2,2)$ supersymmetry; however in this case the constraint
gives $g\mu_R=2g\mu_1=-\kappa n- 2s$ and since the solutions are parametrised by $g\mu_1\ne 0$, these solutions must have a conical deficit on the boundary, as pointed out in \cite{Gutperle:2019dqf}.

Fourth, by setting one of the $g\mu_i$ to zero, say $g\mu_3=0$, we obtain 
defect solutions that preserve $\mathcal{N}=(2,2)$ supersymmetry. This allows us to 
compare with the construction\footnote{The construction in \cite{Gutperle:2019dqf} used Romans theory coupled to
an extra vector multiplet.} of defect solutions in 
\cite{Gutperle:2019dqf}  who just considered $n=1$.
The condition \eqref{stubranches} now eliminates the possibility of branch 2 with $s=+\kappa/2$ and we must be on the main branch with $s=-\kappa/2$ as denoted by the green lines on figure \ref{fig:stusolnsspace}.
The supersymmetry constraint now gives $g\mu_R=g\mu_1+g\mu_2=\kappa(1-n)$.
If we further set $n=1$ we have $g\mu_1=-g\mu_2=g\mu_F/2$ and these are the $\mathcal{N}=(2,2)$ defect
solutions of \cite{Gutperle:2019dqf} without conical singularities, with
\begin{align}\label{jsvevsetcn22case}
\vev{J_1}&=  -\frac{N^2}{4\pi^2e^{2V_0}}\kappa g\mu_1(g\mu_1-\kappa)\,,\quad
\vev{J_2}=  -\frac{N^2}{4\pi^2e^{2V_0}}\kappa g\mu_1(g\mu_1+\kappa)\,,\quad
\vev{J_3}= 0\,,\nn
h_D&= \frac{N^2}{3\pi e^{2V_0}}(g\mu_1)^2\,, \nn
\vev{\mathcal{O}_\alpha}&= \frac{N^2}{2\pi^2 e^{2V_0}}(g\mu_1)^2\,,\quad
\vev{\mathcal{O}_\beta}=  \frac{N^2}{2\pi^2 e^{2V_0}}\kappa(g\mu_1)\,.
\end{align}
This appears to be in accord with (4.17) of \cite{Gutperle:2019dqf}, however the expression for $h_D$ differs,
because \cite{Gutperle:2019dqf} did not consider the finite counterterm $\delta_{\Delta R^2}$ that we used
in our renormalisation scheme. We also note that with $g\mu_3=0$ and $n=1$, we have $g\mu_B=0$ and these are the $\mathcal{N}=(2,2)$ defect solutions indicated in the UV of the RG flows illustrated in figure \ref{fig:sumsolutions}. We note, though, that the STU solutions
with $g\mu_B=0$ and $n\ne 1$ have $g\mu_3\ne0$ and preserve $\mathcal{N}=(2,0)$ supersymmetry.

Finally, we can also consider setting $A^1=A^2=\beta=0$, to obtain defect solutions with $g\mu_1=g\mu_2=0$ which preserve
$\mathcal{N}=(4,4)$ supersymmetry (c.f. section 6 of \cite{Suh:2021ifj}) and necessarily these defects have a conical singularity
and only exist for the main branch, $s=-\kappa/2$.
Note that these are solutions of Romans theory with the $SU(2)$ gauge fields set to zero and were not considered in \cite{Gutperle:2019dqf}.

\section{Defect Anomaly $b$}\label{sec:bcalc}
We now compute the defect anomaly coefficient $b$ as a function of the monodromy parameters for the STU model solutions and also for 
the LS case using \eqref{bcalpha}, \eqref{derivbwrtn}.
For simplicity, in this section we will consider the main branch of solutions with
\begin{align}\label{bsecbranch}
s=-\kappa/2\,,
\end{align}
which we recall is the branch that includes solutions with $n=1$ i.e. no conical singularity (in flat space).

\subsection{STU model}\label{secbstucasesec}
For the STU solutions
we first consider setting $n=1$,
returning to $n\ne 1$ below.
We write 
\eqref{bintone}, \eqref{bcalpha} in the form 
\begin{align}\label{dbexpress}
db={12\pi^2e^{2V_0}} \sum_i\vev{J_i} d[g\mu_i]\,,
\end{align}
where we have fixed the normalisation as described below (the appearance of 
$e^{2V_0}$ is due to the boundary metric as given in \eqref{ads3ansbdy}).
Now the $g\mu_i$ are not independent since
$g\mu_R=\sum_ig\mu_i=0$ (when $n=1$). Using
this to eliminate one of the $g\mu_i$ and also using the expressions for the currents in \eqref{litjistu} we can integrate to get
\begin{align}\label{bexpstumodelkfixed2}
{b=-{3N^2}[(1+\kappa g\mu_1)(1+\kappa g\mu_2)(1+\kappa g\mu_3)-1]}
\,,\qquad\text{when $n=1$}\,.
\end{align}
Here we have fixed an integration constant by demanding that $b=0$ when $g\mu_i=0$.

For the restricted STU solutions, with the extra constraint $g\mu_B=-\kappa$ 
on the sources, this reduces to
\begin{align}\label{bexpreststu}
b&=\frac{3N^2}{32}\left(5+12 (g\mu_F)^2\right)\,,\qquad\text{when $g\mu_B=-\kappa $, $n=1$}\,.
\end{align}

If instead we set one of the $g\mu_i=0$, say $g\mu_3=0$, we obtain defects in $\mathcal{N}=4$ SYM
preserving $\mathcal{N}=(2,2)$ supersymmetry and for $b$ we get
\begin{align}\label{n22bvalstu}
b&={3N^2}(g\mu_1)^2=\frac{3N^2}{4}(g\mu_F)^2\,,
\qquad\text{when $g\mu_3=0 $, $n=1$}\,.
\end{align}
Recalling the discussion at the end of section \ref{sectstu},
this case (with $n=1$) has 
$g\mu_B=0$ and is associated with the seed point for the RG flow to defects from $\mathcal{N}=4$ SYM theory to
defects in the LS theory.
From the results of \cite{Bianchi:2019umv} one anticipates that the value of $b$ should be independent of the 
$\mathcal{N}=4$ SYM coupling. Upon setting $e^{2V_0}=1$ (recall \eqref{ads3ansbdy}) we find that our result for $b$
for the $\mathcal{N}=(2,2)$ defects with $n=1$ in \eqref{n22bvalstu} precisely agrees with a free field computation for $b$ given in \cite{Bianchi:2021snj} and this fixes the normalisation in \eqref{dbexpress}.
This is further discussed in appendix \ref{app:freefield} where we also show that the value of $d_2=18\pi h_D$ (when $n=1$) also agrees.
That there is agreement for $h_D$ (and hence $d_2$) follows from the fact that $h_D$ can be obtained from the (constrained) derivative of 
$b$ with respect to $n$ as in \eqref{dbexpress2} below.
Moreover, making some natural assumptions concerning the boundary conditions on the free fields we 
also show in appendix \ref{app:freefield}
 that there is also exact agreement for $b,h_D,d_2$ for the defects preserving 
$\mathcal{N}=(0,2)$ supersymmetry, with $n=1$, as well.

We now discuss how \eqref{dbexpress}, \eqref{bexpstumodelkfixed2} is generalised for solutions of the STU model
when we also allow $n$ to vary, subject to the
constraint $g\mu_R=\sum_ig\mu_i=\kappa(1-n)$ imposed by supersymmetry (recall \eqref{bsecbranch}). 
Using \eqref{bcalpha}, \eqref{derivbwrtn} we now write
\begin{align}\label{dbexpress2}
db&=12\pi^2 e^{2V_0} \left(\frac{1}{n}\sum_i\vev{J_i} d[g\mu_i]-3 \frac{h_D}{2\pi}dn   \right)+12 a_{\mathcal{N}=4}d n\,,\nn
&=12\pi^2 e^{2V_0}\frac{1}{n}\left(\sum_i\vev{J_i} d[g\mu_i]+{\kappa }\sum_i\vev{J_i}dn   \right)+12 a_{\mathcal{N}=4}d n\,,
\end{align}
where $a_{\mathcal{N}=4}=N^2/4$ is the large $N$ limit of the central charge of $\mathcal{N}=4$ SYM theory.
It is a non-trivial consistency check that the right hand side is indeed a closed form,
given our expressions for $\vev{J_i}$ in \eqref{litjistu} and the supersymmetry constraint
$g\mu_R=\kappa(1-n)$.
After integration, fixing the integration constant to obtain \eqref{bexpstumodelkfixed2} when $n=1$,
we find the elegant expression
\begin{align}\label{genbstumod3}
b_{\mathcal{N}=4}= 
-12 a_{\mathcal{N}=4}n\left(\mathcal{F}^{\mathcal{N}=4}-1 \right)=
-{3N^2}n\left(\mathcal{F}^{\mathcal{N}=4}-1 \right)\,,
\end{align}
where we have defined
\begin{align}\label{calfdefn4}
\mathcal{F}^{\mathcal{N}=4}\equiv (1+\frac{g\mu_1}{\kappa n})(1+\frac{g\mu_2}{\kappa n})(1+\frac{g\mu_3}{\kappa n})\,.
\end{align}
One can show\footnote{For example, express $b$ in terms of $g\mu_F$ and $g\mu_3$ and then show that
for the allowed range, $(1+\frac{g\mu_i}{\kappa n})>0$, the minimum value of $b$, for fixed $n$, occurs when $g\mu_F=0$ and $g\mu_i=\kappa(1-n)/3$.
We can also argue as in appendix E of \cite{toappear}.}
that for $n\ge 1$ we have $b\ge 0$ and moreover, the minimal value occurs, for fixed $n$, when $g\mu_i=\kappa(1-n)/3$, associated with solutions of minimal gauged supergravity. 
For $n<1$ (still on the main branch) we can have $b<0$, as in the solutions of minimal gauged supergravity
discussed in appendix \ref{mingsugrasols}, for example.

For the special case of restricted STU solutions with ${g\mu_B=-\kappa n }$
we find
\begin{align}\label{bLScasewithkmonedisctext}
b_{\mathcal{N}=4}|_{g\mu_B=-\kappa n }&=\frac{3}{32n^2}\left({(-1-6n-12n^2+24n^3)}+ {4(1+2n)}(g\mu_F)^2\right)N^2\,.
\end{align}
On the other hand, for the special case when $g\mu_B=0$, relevant for RG flows for general $n$, 
we obtain
\begin{align}\label{bmubzerostu}
b_{\mathcal{N}=4}|_{g\mu_B=0}&=\frac{3N^2}{32 n^2}\left((n-1)(1+8n+23n^2)+   4(1+n)(g\mu_F)^2\right)\,.
\end{align}

\subsection{LS case}\label{bvaluesLS}
We can also compute $b$ for monodromy defects in the LS SCFT. In this case we
have the constraints $g\mu_B=0$ and $g\mu_R^{LS}= \kappa(1-n)=g\mu_R$, (recall \eqref{bsecbranch}).

We first consider $n=1$, when $g\mu_R=g\mu_B=0$ and consider 
\begin{align}
db
=12\pi^2 e^{2V_0} \langle J_F\rangle d[g\mu_F]\,.
\end{align}
After integrating we get
\begin{align}\label{bLScasewithkmone}
b=\frac{9N^2}{8}(g\mu_F)^2\,,
\end{align}
where we fixed the integration constant by demanding that $b=0$ when $g\mu_F=0$.

More generally, we can also consider allowing $n$ to vary. Recalling \eqref{dbexpress2}, we
consider
\begin{align}\label{dbexpress22}
db&=12\pi^2 e^{2V_0}\left(\frac{1}{n}
\left(\vev{J^{\varphi}_R} d[g\mu_R]+\vev{J_F}d[g\mu_F]\right)
-3 \frac{h^{LS}_D}{2\pi}dn   \right)+12 a_{LS}d n\,,\nn
&=12\pi^2  e^{2V_0}\frac{1}{n}\left(\vev{J^{\varphi}_R} d[g\mu^{LS}_R]+\vev{J_F}d[g\mu_F]
+3\kappa\vev{J^{\varphi}_R}  dn\right)+12 a_{LS}d n\,,\nn
&=12\pi^2  e^{2V_0}\frac{1}{n}\left(\vev{J_F}d[g\mu_F]
+2\kappa\vev{J^{\varphi}_R}  dn\right)+12 a_{LS}d n\,,
\end{align}
where the second line follows using the expression for $h_D$ for the LS case in
\eqref{stresstextLS2}, while the final line comes from setting
$g\mu_R^{LS}=g\mu_R=\kappa(1-n)$.
Also $a_{LS}=\frac{27}{128}N^2$ 
is the large $N$ limit of the central charge of the LS SCFT.
 It is again a good consistency check that the right hand side is indeed a closed form. Integrating this expression
we arrive at 
\begin{align}\label{genbLSmod3}
b_{LS}= 
-12a_{LS}
n\left(
\mathcal{F}^{LS}-1
 \right)
 =
-\frac{81N^2}{32}
n\left(
\mathcal{F}^{LS}-1
 \right)\,,
\end{align}
where we have defined
\begin{align}\label{calfdefLS}
\mathcal{F}^{LS}\equiv (1+\frac{4g\mu_1}{3\kappa n})(1+\frac{4g\mu_2}{3\kappa n})(1+\frac{2g\mu_3}{3\kappa n})\,,
\end{align}
with $g\mu_i$ constrained as in \eqref{muinfourLS}. 
We can also write
\begin{align}\label{blscasemuf}
b_{LS}=\frac{3N^2}{32n^2}\left( (n-1)(1+7n+19n^2)+4(1+2n)(g\mu_F)^2\right)\,.
\end{align}
For $n\ge 1$ we have $b_{LS}\ge 0$, but for $n<1$, $b_{LS}$ can be negative (as in solutions of minimal gauged supergravity, for example).

\section{The on-shell action and supersymmetric Renyi entropy}\label{susyrenyi}
We now compute the on-shell action for the defect solutions that we have been discussing.
After Wick rotating $ds^2(AdS_3)\to ds^2(\mathbb{H}_3)$ this also enables us
to compute supersymmetric Renyi entropies. 

A helpful observation is that the ansatz we are considering is invariant under a
combination of translations in the $z$ direction and gauge transformations. 
In the $\bar\theta=0$ gauge that we are using, $\partial_z$ is a Killing vector that
preserves the whole solution and, as discussed in appendix \ref{secosact},  this allows us to write
the bulk Lagrangian density, on-shell, in the form 
\begin{align}\label{secondosactiontext}
\sqrt{+g}\mathcal{L}=&\frac{1}{\rho^3}\partial_y\Sigma\,,
\end{align}
where
\begin{align}
\Sigma=-\frac{e^{3V}h'}{2 f}
-\frac{e^{3V}}{f h}\Big[
e^{4\alpha-4\beta}a_1 a_1' +e^{4\alpha+4\beta}a_2 a_2'
+ e^{-8\alpha}a_3 a_3'
\Big]\,.
\end{align}

As usual, the total action also gets contributions from the Gibbons-Hawking and counterterm action (including finite counterterms) as discussed in sections \ref{holrennn4} and \ref{holrenmLS}. Combining them
we can then obtain an expression
for the total on-shell action in the form
\begin{align}
S=S|_\infty+S_\text{core}\,.
\end{align}
For all cases, the contribution at the core takes the universal form 
\begin{align}
S_\text{core}&=
\frac{N^2}{2\pi^2}(2\pi)\vol(AdS_3)\frac{1}{2L^3}e^{3V}|_\text{core}\,.
\end{align}
We include a few details of this computation in appendix \ref{secosact}, as well as $S|_\infty$,
 and quote the final results for each class of solutions below.

In this section, for the most part we consider both branches of solutions, $s=\pm\kappa/2$, restricting to the main branch $s=-\kappa/2$ both when
we discuss expressions involving the $b$ central charge, and when we discuss supersymmetric Renyi entropy 
(which is naturally associated with $n\ge 1$).

\subsection{$\mathcal{N}=4$ SYM boundary, $\varphi\ne 0$}
By explicit computation
we find that $S|_\infty$ has contributions involving $\sum_i\vev {J_i}$ and $\sum_i g\mu_i \vev {J_i}$. Now the current one-point functions
$\vev {J_i}$ can be expressed in terms of $\vev{ {J}_R^{\varphi} }$, $\vev{{J}_F}$, $\vev{{J}_B }$ via 
\eqref{lincsn4gf0}. Remarkably, we find that when we impose the supersymmetry conditions
$g\mu_R=-\kappa n -2 s$, $g\mu_B=-\kappa n$, all dependence on $\vev{{J}_B }$ drops out and we can then use \eqref{allcurrentsintrmssces}
to express $\vev{ {J}_R^{\varphi} }$, $\vev{{J}_F}$ in terms of monodromy sources. It is also notable that the
finite counter term $\delta_\beta$ appearing in \eqref{textbcurrents} also drops out of $S|_\infty$, as well as any dependence on $\varphi_s$. We next notice that
$S_\text{core}$ can also be expressed in terms of monodromy sources via
\eqref{eyealbetv23}.

Combining these observations we find that the total on-shell action can be expressed as
\begin{align}\label{restrictedstuacttextg}
S&=-\vol(AdS_3)
\frac{1}{8\pi n^2}\kappa(s-\kappa n) [(s-\kappa n)^2-(g\mu_F)^2]       
N^2\,.
\end{align}
Strikingly, it is independent of $\varphi_s$. 
One might expect (for a related discussion see \cite{Herzog:2019rke}) that the on-shell action would depend on the scalars parametrising this defect ``conformal manifold" via $\delta S\sim \vev{\cO_\a}L^{-2}\delta\alpha_s+\vev{\cO_\b} L^{-2}\delta\beta_s+\vev{\cO_\varphi} L^{-1}\delta \varphi_s$. For the supersymmetric configurations, from 
\eqref{textconsusuyssces} we have $\betas=0$ and $\alphas = \frac{2}{3} \varphi_s^2$
and so we expect a dependence on $\varphi_s$ via $\delta S\sim \left(\frac{4}{3}\vev{\cO_\a}L^{-1}\varphi_s+\vev{\cO_\varphi}\right) L^{-1}\delta \varphi_s$. However, from \eqref{sconeptfns} the right hand hide side of this expression precisely vanishes; this is associated with variations of $\varphi_s$, along the line of bulk marginal deformations of the defect, giving the expectation value of a Q-exact operator.

Below we will see that \eqref{restrictedstuacttextg} is exactly the same expression for the on-shell action for the restricted STU
solutions, obtained from the STU solutions by setting $g\mu_B=-\kappa n$ by hand. In other words, \emph{starting with a restricted
STU solution and then switching on $\varphi_s$ does not change the on-shell action.}

Also, recall that as we increase $\varphi_s\to\infty$ 
the solution starts to approach the LS defect solution. Indeed one can see this behaviour in figures \ref{fig:N4phi_vortex_num} and \ref{fig:LS_vortex_num}. One might therefore
expect that the on-shell action \eqref{restrictedstuacttextg} is also the same as the on-shell action for the LS defect solution (which recall has
$g\mu_B=0$ rather than $g\mu_B=-\kappa n$). We shall verify this expectation in the next subsection.
Notice that the on-shell action vanishes at the boundary of the allowed region of defect solutions given in \eqref{eq:FlavourFluxSourceBound}
and moreover for the main branch $s=-\kappa/2 $ the action is positive inside the region.

\subsection{LS Boundary}
For solutions associated with defects in LS, we find that the total on-shell action can be written in the compact form
\begin{align}\label{oshactstuLS}
S
&=\frac{27 N^2}{64\pi}\vol(AdS_3)\Big[n(1+\frac{4g\mu_1}{3\kappa n})(1+\frac{4g\mu_2}{3\kappa n})(1+\frac{2g\mu_3}{3\kappa n})\Big]\nn
&=\frac{27N^2}{64\pi}\vol(AdS_3)\big[n\mathcal{F}^{LS}\big]\,,
\end{align}
where $\mathcal{F}^{LS}$ was defined in \eqref{calfdefLS} and we should impose the
supersymmetry constraints
$g\mu_B=0$ and $g\mu_R^{LS}= -\kappa n -2 s$. It can also be written in terms of $g\mu_F$ and we 
find
\begin{align}\label{restrictedstuacttextg2}
S&=-\vol(AdS_3)
\frac{1}{8\pi n^2}\kappa(s-\kappa n) [(s-\kappa n)^2-(g\mu_F)^2]       
N^2\,,
\end{align}
which is precisely the same as \eqref{restrictedstuacttextg}. 
Notice that the on-shell action vanishes at the boundary of the allowed region of defect solutions given in \eqref{eq:FlavourFluxSourceBoundLS} and
when $s=-1/2 \kappa$ (recall \eqref{constraintkgen}) the action is positive inside the region.

It is interesting to note that we can express the on-shell action in terms of $b$ for the LS defect given in
\eqref{genbLSmod3}, for $s=-\kappa/2$,
as follows
\begin{align}\label{abexpresforls}
S=\vol(AdS_3)\Big(\frac{1}{6\pi }\Big)\Big(12a_{LS}n-b_{LS}\Big)\,.
\end{align}

\subsection{STU model}
For the solutions of the STU model we can write the total on-shell action in terms of the boundary sources and we obtain 
\begin{align}\label{oshactstu}
S
&=\frac{N^2}{2\pi}\vol(AdS_3)\Big[n(1+\frac{g\mu_1}{\kappa n})(1+\frac{g\mu_2}{\kappa n})(1+\frac{g\mu_3}{\kappa n})\Big]\nn
&=\frac{N^2}{2\pi}\vol(AdS_3)\big[n\mathcal{F}^{\mathcal{N}=4}\big]\,,
\end{align}
where $\mathcal{F}^{\mathcal{N}=4}$ was defined in \eqref{calfdefn4}.
Recall that for the STU solutions we have the supersymmetry constraint $g\mu_R=\sum_ig\mu_i=-{\kappa n}-2s$. 
Notice, again, that the on-shell action vanishes at the boundary of the allowed region of monodromy parameters (recall the
comment below \eqref{valuesorestukey}) and 
when $s=-1/2 \kappa$ the action is positive inside the region (for the branch $s=+\kappa/2$ the on-shell action is negative).
Also, for the restricted STU model solutions we should further set $g\mu_B =-\kappa n$ and then we recover \eqref{restrictedstuacttextg}, as already noted.

It is interesting to note that we can again express the on-shell action in terms of $b_{\mathcal{N}=4}$ for the defects in $\mathcal{N}=4$ SYM theory given in
\eqref{genbstumod3}, for $s=-\kappa/2$,
as follows
\begin{align}\label{abexpresforn4}
S={\vol(AdS_3)}\Big(\frac{1 }{6\pi }\Big)\Big( 12a_{\mathcal{N}=4}n-b_{\mathcal{N}=4}\Big)\,,
\end{align}
where $a_{\mathcal{N}=4}=\frac{1}{4}N^2$
is the large $N$ limit of the central charge of the $\mathcal{N}=4$ SYM theory. Note that
this has exactly the same form as \eqref{abexpresforls}.

\subsection{Supersymmetric Renyi entropy}
The notion of supersymmetric Renyi entropy (SRE) for SCFTs with an R-symmetry was introduced in \cite{Nishioka:2013haa} and can be viewed as a particular case of charged Renyi entropies \cite{Belin:2013uta}. It has been studied
using holography in various spacetime dimensions, including
\cite{Huang:2014gca,Nishioka:2014mwa,Huang:2014pda,Crossley:2014oea,Alday:2014fsa,Hama:2014iea,Hosseini:2019and}.

For a four-dimensional SCFT in flat spacetime, consider the SRE associated with a spherical ball with an $S^2$ boundary, on a constant time slice. The Renyi entropy can be obtained by computing the partition function of the SCFT on a suitable branched cover of the sphere $S^4_n$, $n\in \mathbb{N}$ with the addition of a suitable background R-symmetry gauge field in order to preserve supersymmetry.
It is most convenient to perform this computation by conformally mapping $S^4_{n}$ to $H^3\times S^1$ 
with the period of the $S^1$ given by $2\pi n$. Notice that this is exactly the same as 
the boundary metric for the supersymmetric defects in \eqref{ads3poinc} after Wick rotating $AdS_3\to H^3$
(ignoring the overall $e^{2V_0}$ factor). In this subsection we take $n\ge 1$ and hence are on the main branch of solutions
with $s=-\kappa/2$.

After Wick rotating our defect solutions, we obtain
the Euclidean on-shell action from the Lorenztian action $S$ via $I_n(g\mu_i)=-\log Z(n;g\mu_i)=-S$.
The supersymmetric Renyi entropy (SRE) can be defined as 
\begin{align}\label{sregenform}
S^{SRE}=-\frac{I_{n}(g\mu_i)-n I_{n=1}(g\mu_i)}{1-n}\,,
\end{align}
where $I_n(g\mu_i)$ is understood to be a function of $n$ and any independent monodromy sources
for the flavour symmetries, \emph{after} imposing the supersymmetry constraints, and $I_{n=1}(g\mu_i)$ is the same quantity
after setting $n=1$ and holding the same independent monodromy sources fixed. In particular, we note that this definition
is supersymmetric and moreover has a well defined limit as $n\to 1$. 
We emphasise that in previous works different quantities have been held fixed in taking the limit $n\to 1$ in
$I_{n=1}(g\mu_i)$, and hence obtaining different
notions of SRE, as we will illustrate below.

Note that the on-shell action contains a $\vol(H^3)$ factor, associated with the usual power law divergences
for entanglement entropies. In the present setting there is a sub-leading log divergence and it is the coefficient of
this log that is universal\footnote{In particular it is independent of the choice of scheme that
we used in our holographic renormalisation.} and defines the SRE. As discussed in \cite{Hung:2011nu}, to extract this universal piece 
we should take $\vol(H^3)\to -2\pi\log(2R/\delta)$, where $R$ is
the radius of the entangling surface and $\delta\ll R$ is an IR cutoff in $H^3$. Below we take this point to be understood and
maintain the $\vol(H^3)$ factor in our expressions for the SRE.

We also highlight a general point that from \eqref{abexpresforn4} and \eqref{abexpresforls} we have
\begin{align}
I_{n}(g\mu_i)&=-\vol(H_3)\frac{1}{6\pi}\Big[12an-b(n;g\mu^i)\Big]\,,
\end{align}
and in particular we see that the Renyi entropy is independent of the $a$ central charge and is just determined by the $b$ central charge of the defect.

\subsubsection{STU model}\label{sturenyi}
We first consider the SRE for $\mathcal{N}=4$ SYM theory associated with the STU solutions. The supersymmetry constraint
is given by $g\mu_R=\sum_ig\mu_i=\kappa(1-n)$. It is convenient to write
\begin{align}
I_{n}(g\mu_i)&=n\mathcal{F}^{\mathcal{N}=4} F^{\mathcal{N}=4}\,,
\end{align}
where 
\begin{align}
\mathcal{F}^{\mathcal{N}=4}&=
(1+\frac{g\mu_1}{\kappa n})(1+\frac{g\mu_2}{\kappa n})(1+\frac{g\mu_3}{\kappa n})\,,\nn
F^{\mathcal{N}=4}&=-\frac{N^2}{2\pi}\vol(H_3)\,.
\end{align}

Let us first write an expression for $S^{SRE}$ where we consider the action $I_n(g\mu_i)$ to be a function of $n$ and
$g\mu_{F}$, $g\mu_{F'}$ (recall \eqref{lincsn4gfnatbas}). That is, we hold $g\mu_{F}$, $g\mu_{F'}$ fixed in defining $I_{n=1}(g\mu_i)$ appearing
in \eqref{sregenform}. Doing so, we obtain 
\begin{align}\label{sreffp}
S^{SRE}=\Big(-\frac{1+7n+19n^2}{27n^2}        &+\frac{1+3n +3n^2}{48n^2}(4(g\mu_F)^2+3(g\mu_{F'})^2)   \nn
&+\frac{1+n+n^2}{32 n^2}\kappa g\mu_{F'}((g\mu_{F'})^2-4(g\mu_F)^2  \Big)F^{\mathcal{N}=4}\,.
\end{align}

Another case we can consider is solutions of the STU model that arise in minimal gauged supergravity.
In this case we impose that $g\mu_1=g\mu_2=g\mu_3=\frac{\kappa}{3}(1-n)$ and then the action $I_n(g\mu_i)$
is just a function of $n$.
Constructing $S^{SRE}$ we obtain 
\begin{align}\label{stumin}
S^{SRE}=-\frac{1+7n+19n^2}{27n^2}F^{\mathcal{N}=4}\,,
\end{align}
in agreement with \cite{Huang:2014pda}.
Notice that this expression can also be obtained from \eqref{sreffp} by setting $g\mu_F=g\mu_{F'}=0$, which
we see from \eqref{lincsn4gfnatbas} does correspond to the minimal gauged supergravity case.

We next consider the case of solutions of the STU model with $g\mu_3=0$, which are solutions preserving
$\mathcal{N}=(2,2)$ supersymmetry. The supersymmetry condition implies that we can write
$g\mu_1=\frac{1}{2}(\kappa(1-n)+g\mu_F)$ and $g\mu_2=\frac{1}{2}(\kappa(1-n)-g\mu_F)$.
If we consider the action $I_n(g\mu_i)$ to be a function of $n$ and $g\mu_F$ we find 
\begin{align}\label{ensussre1}
S^{SRE}=-\frac{1+3n-(1+n)(g\mu_F)^2}{4n}F^{\mathcal{N}=4}\,.
\end{align}
We might be tempted to obtain this expression from \eqref{sreffp} by setting
$g\mu_{F'}=\frac{2}{3}(1-n)\kappa$, which along with $g\mu_R=\kappa(1-n)$ gives
$g\mu_3=0$. However, this would not be correct since
\eqref{sreffp} was derived assuming that $g\mu_F$, $g\mu_{F'}$ were held fixed and independent of $n$ in
the expression for $I_{n=1}(g\mu_i)$ in $S^{SRE}$.

In a similar vein, we can consider the case $g\mu_3=0$ and $g\mu_1=g\mu_2=\frac{\kappa}{2}(1-n)$ which arise as solutions of Romans' supergravity theory with $\beta=0$. Substituting this directly in the action and taking $I_n(g\mu_i)$ to be
then just a function of $n$, we can
obtain $S^{SRE}$ and we find
\begin{align}\label{romanssre}
S^{SRE}=-\frac{1+3n}{4n}F^{\mathcal{N}=4}\,,
\end{align}
in agreement with \cite{Crossley:2014oea}. This result can be obtained from \eqref{ensussre1}
by setting $g\mu_F=0$.

It is also interesting to consider the limit that $n\to 1$ of the SRE, bearing in mind 
the issues we have just mentioned regarding the subtlety in what we are holding fixed. 
From \eqref{sregenform}, we can Taylor expand $I_n$ about $n=1$ and then take the limit $n\to 1$ to get an expression for the supersymmetric charged entanglement entropy
\begin{align}\label{dersSRE}
\lim_{n\to 1}S^{SRE}=\frac{dI_{n}(g\mu_i)}{dn}\Big|_{n=1}-I_{n=1}(g\mu_i)\,,
\end{align}
where we are holding specific $g\mu_i$ fixed. 
For example, for the case where we hold 
$g\mu_{F}$, $g\mu_{F'}$ fixed we have the expression for $S^{SRE}$ as in \eqref{sreffp}. Setting $n\to 1$ in 
 \eqref{sreffp}, we find that we can write the expression in the form 
\begin{align}\label{sreffp2}
\lim_{n\to 1}S^{SRE}=\frac{\vol(H_3)}{2\pi}\left(4a_{\mathcal{N}=4}-\frac{1}{3}b_{\mathcal{N}=4}-\frac{2}{9}d^{\mathcal{N}=4}_2\right)\,,
\end{align}
where we recall $d^{\mathcal{N}=4}_2=18\pi h^{\mathcal{N}=4}_D$ and $a_{\mathcal{N}=4}=N^2/4$.
It is interesting to compare this with \eqref{entent} (see also \eqref{osactstubexpdicsec}).

We can also consider a parametrisation of the monodromy sources that uniformly vanishes as $n\to 1$. Specifically,
similar\footnote{Note that \cite{Huang:2014pda} computed these slightly differently,  
using thermodynamic considerations, via their 4.5 and 4.10.} to \cite{Huang:2014pda} 
we can consider 
\begin{align}
g\mu_i=\kappa (1-n)\frac{\Delta_i}{2}\,,\qquad \sum_i\Delta_i=2\,,
\end{align}
where the constraints on the $\Delta_i$ ensure that the supersymmetry condition is satisfied.
In this case we can write 
\begin{align}
S^{SRE}=-\left(1+\frac{1-n}{4n}(\Delta_1\Delta_2+\Delta_1\Delta_3+\Delta_2\Delta_3) +\frac{(1-n)^2}{8n^2}\Delta_1\Delta_2\Delta_3\right)F^{\mathcal{N}=4}\,.
\end{align}
We recover the result for minimal supergravity \eqref{stumin} by setting $\Delta_i=2/3$ and the result for Romans theory
\eqref{romanssre} by setting $\Delta_1=\Delta_2=1$ and $\Delta_3=0$. We also notice that taking the limit $n\to 1$ 
in this case gives $\lim_{n\to 1}S^{SRE}=-F^{\mathcal{N}=4}$.

Finally, by analogy with \cite{Hosseini:2019and} in a different context, we can also consider
\begin{align}
g\mu_i=\frac{3\kappa}{2}(\Delta_i-\frac{2}{3})+\kappa(1-n)(1-\Delta_i),\qquad \sum_i\Delta_i=2\,,
\end{align}
which is clearly non-vanishing as $n\to 1$.
In this case we find the Euclidean action has the form $I_n=\frac{(1+2n)^3}{8n^2}\Delta_1\Delta_2\Delta_3$ 
leading to 
\begin{align}
S^{SRE}=-\frac{1+7n+19n^2}{8n^2}(\Delta_1\Delta_2\Delta_3)F^{\mathcal{N}=4}\,.
\end{align}
Notice that if we set $\Delta_1=\Delta_2=\Delta_3=2/3$ we recover the expression for  minimal gauged supergravity
given in \eqref{stumin}.

\subsubsection{LS model}\label{lssecsre}
We now consider the SRE for LS theory. The supersymmetry constraint is given by $g\mu_R=\sum_ig\mu_i=\kappa(1-n)$ and we also 
have $g\mu_B=0$. 
It is convenient to write
\begin{align}
I_{n}(g\mu_i)&=n\mathcal{F}^{LS} F^{LS}\,,
\end{align}
where 
\begin{align}
\mathcal{F}^{LS}&=
(1+\frac{4g\mu_1}{3\kappa n})(1+\frac{4g\mu_2}{3\kappa n})(1+\frac{2g\mu_3}{3\kappa n})\,,\nn
F^{LS}&=-\frac{N^2}{2\pi}\frac{27}{32}\vol(H_3)\,.
\end{align}

Holding $g\mu_F$ fixed, we now find
\begin{align}\label{sreffpLS}
S^{SRE}=\Big(-\frac{1+7n+19n^2}{27n^2}        +\frac{(1+3n +3n^2)}{27n^2}4(g\mu_F)^2  \Big)F^{{LS}}\,.
\end{align}
Notice that if we set $g\mu_F=0$, we have $g\mu_1=g\mu_2=\frac{\kappa}{4}(1-n)$ and $g\mu_3
=\frac{\kappa}{2}(1-n)$, and we get the correct result for the minimal gauged supergravity case associated with 
the LS vacua (compare with \eqref{stumin}). We also record the limit as $n\to 1$
\begin{align}\label{sreffpLS2}
\lim_{n\to 1}S^{SRE}&=\Big(-1+\frac{28}{27}(g\mu_F)^2  \Big)F^{{LS}}\nn
&=
\frac{\vol(H_3)}{2\pi}\left(4a_{LS}-\frac{1}{3}b_{LS}-\frac{2}{9}d^{LS}_2\right)\,,
\end{align}
where $d^{LS}_2=18\pi h^{LS}_D$ and $a_{LS}=\frac{27}{128}N^2$, which has the same form as \eqref{sreffp2}.

We can also consider a parametrisation of the monodromy sources that uniformly vanishes as $n\to 1$. Specifically,
we can consider 
\begin{align}
g\mu_i=\kappa (1-n)\frac{\Delta_i}{2}\,,\qquad \sum_i\Delta_i=2\,, \qquad \Delta_1+\Delta_2-\Delta_3=0\,,
\end{align}
where the constraint on the $\Delta_i$ ensures that the supersymmetry condition is satisfied as well 
as $g\mu_B=0$. If we solve the constraints via
\begin{align}
\Delta_1=\frac{1}{2}(1+\Delta_F)\,,\qquad
\Delta_2=\frac{1}{2}(1-\Delta_F)\,,\qquad
\Delta_3=1\,,\qquad
\end{align}
In this case we now find
\begin{align}\label{sreffpLS3}
S^{SRE}=\Big(-\frac{1+7n+19n^2}{27n^2}        +\frac{(1+n -2n^2)}{27n^2}\Delta_F^2 \Big)F^{{LS}}\,.
\end{align}

\section{$\mathcal{N}=4$/LS bulk defect RG flows}\label{RGflowsec}
In this section we make some observations regarding the $b$ central charge and also the on-shell action for
$\mathcal{N}=4$/LS bulk defect RG flows. 
For simplicity in this section we just discuss the main branch of solutions
with 
\begin{align}
s=-\kappa/2\,.
\end{align}
Recall (above \eqref{hdbelowrgflowdosc}) that these RG flows
start from $\mathcal{N}=4$ SYM theory in the UV with a monodromy defect with 
$g\mu_R=\kappa(1-n)$, $g\mu_B=0$ and, in addition,
a homogeneous relevant mass deformation (not localised on the defect). As discussed in section \ref{sectstu},
provided $|g\mu_F|< |n+\frac{1}{2}|$, it is expected that the end point of this RG flow
is a monodromy defect in the LS theory with $g\mu_R=\kappa(1-n)$, $g\mu_B=0$. 

\begin{figure}[htbp!]
\begin{center}
\includegraphics[scale=.41]{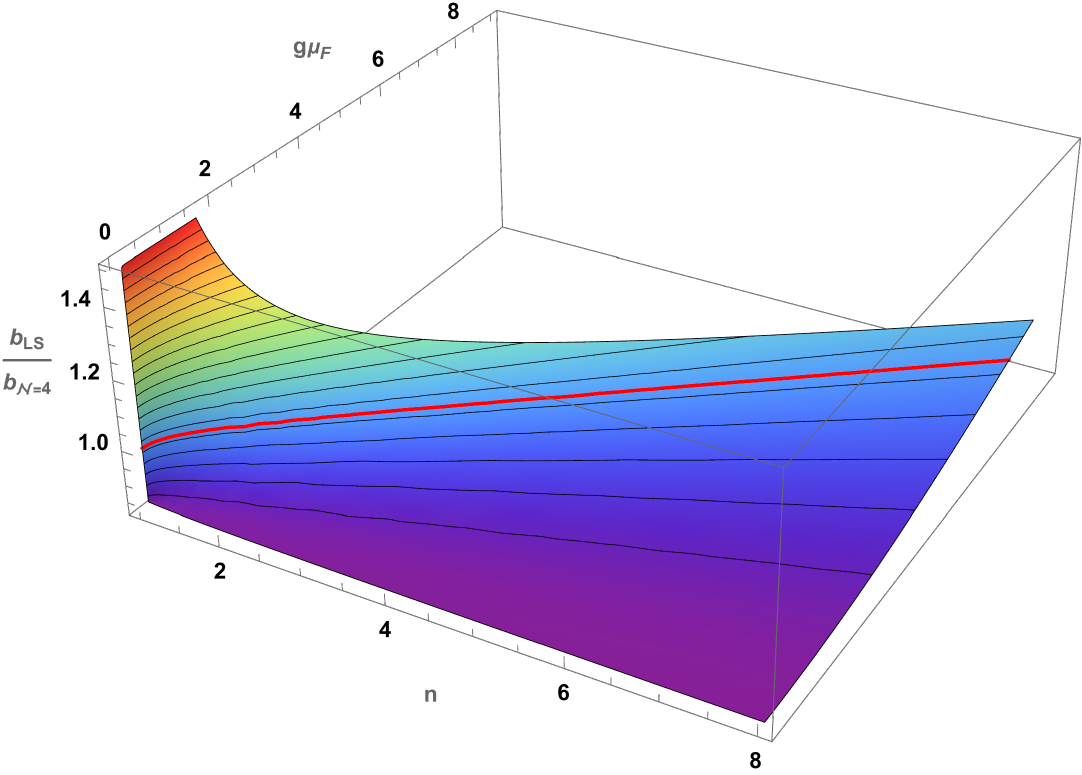}~
\includegraphics[scale=.41]{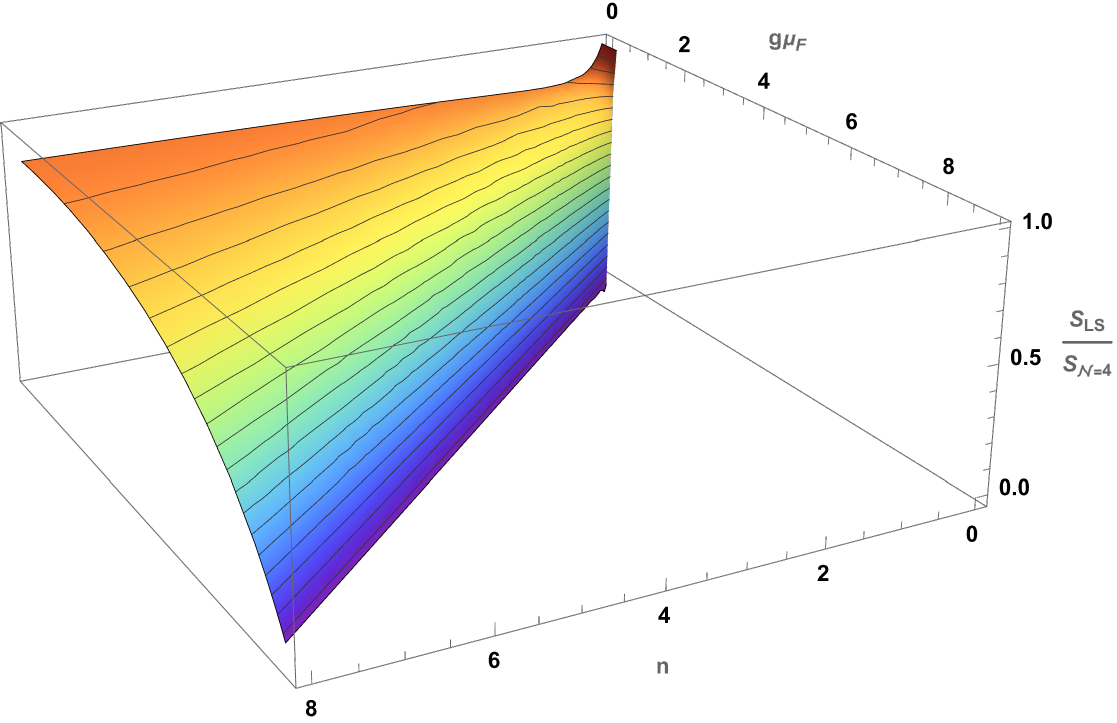}~
\caption{Behaviour of the $b$-central charge and the on-shell action for $\mathcal{N}=4$/LS bulk defect RG flows.
The left frame plots $b_{LS}/b_{\mathcal{N}=4}|_{g\mu_B=0}$ while the right frame plots $S_{LS}/S_{\mathcal{N}=4}|_{g\mu_B=0}$ as a function of $n$ and $g\mu_F$ in the relevant range $|g\mu_F|< | n+\frac{1}{2}|$ where the RG flows should exist (when $s=-\kappa/2$).
Note that $b_{LS}/b_{\mathcal{N}=4}|_{g\mu_B=0}$ can both be bigger and smaller than 1, or exactly equal to 1 as denoted by the red line, 
while $S_{LS}/S_{\mathcal{N}=4}|_{g\mu_B=0}<1$.}
\label{figbactrgflow}
\end{center}
\end{figure}

We first consider $b$ in the UV and in the IR for such an RG flow. 
From \eqref{bmubzerostu} and \eqref{blscasemuf} we have
\begin{align}\label{bmubzerostu2}
b_{\mathcal{N}=4}|_{g\mu_B=0}&=\frac{3N^2}{32 n^2}\left((n-1)(1+8n+23n^2)+   4(1+n)(g\mu_F)^2\right)\,,\nn
b_{LS}&=\frac{3N^2}{32 n^2}\left( (n-1)(1+7n+19n^2)+4(1+2n)(g\mu_F)^2\right)\,.
\end{align}
For $n=1$ (with the defects in the UV then having $\mathcal{N}=(2,2)$ supersymmetry),
we have $b_{LS}=\frac{3}{2} b_{\mathcal{N}=4}|_{g\mu_B=0}$ and so $b$ would increase under such an RG flow.
The monotonicity of $b$ along such bulk defect RG flows has also been considered
in different contexts in e.g. \cite{Green:2007wr,Sato:2020upl}
(and recall that the $b$-theorem of \cite{Jensen:2015swa,Casini:2016fgb} applies
to defect RG flows driven by deformations localised on the defect, in contrast to what we are considering here).
When $n\ne 1$, depending on the value of $g\mu_F$ in the relevant range $|g\mu_F|<| n+\frac{1}{2}|$, we find that $b$ can both increase or decrease along the RG flows 
as illustrated in figure \ref{figbactrgflow}.

We next consider the on-shell action. From \eqref{abexpresforn4}, \eqref{abexpresforls} we have
\begin{align}
S_{\mathcal{N}=4}&={\vol(AdS_3)}\Big(\frac{1 }{6\pi }\Big)\Big( 12a_{\mathcal{N}=4}n-b_{\mathcal{N}=4}\Big)\,,\nn
S_{LS}&=\vol(AdS_3)\Big(\frac{1}{6\pi }\Big)\Big(12a_{LS}n-b_{LS}\Big)\,.
\end{align}
with $a_{LS}=\frac{27}{128}N^2$ and $a_{\mathcal{N}=4}=\frac{1}{4}{N^2}$. Restricting the monodromy flux 
$g\mu_B=0$ for the defects in $\mathcal{N}=4$ SYM theory, as in \eqref{bmubzerostu2}, we find the
remarkable result that for such flows $S_{LS}/S_{\mathcal{N}=4}|_{g\mu_B=0}<1$, as also illustrated in figure \ref{figbactrgflow}.
Notice that as $n\to 0$ the ratio $S_{LS}/S_{\mathcal{N}=4}|_{g\mu_B=0}\to 1$,  
while taking $n\to\infty$ for fixed $g\mu_F$, we find $S_{LS}/S_{\mathcal{N}=4}|_{g\mu_B=0}\to 8/9$ (which
is slightly higher than the ratio of central charges $a_{LS}/a_{\mathcal{N}=4}=27/32$).
Our results also suggest, that $12an-b$ is a good monotonic quantity to consider in similar RG flows with defects in more general settings.

\section{Discussion}\label{Fincomments}
We have holographically computed various observables for $d=2$, $\mathcal{N}=(0,2)$ monodromy defects in two $d=4$ SCFTs,
$\mathcal{N}=4$ SYM theory and the $\mathcal{N}=1$ LS SCFT. The defects can be considered to be living in flat space or, after a
Weyl transformation, on $AdS_3\times S^1$.
From the flat space perspective
the defects are supported by monodromies for global symmetries of the SCFTs as one circles the defect on a constant time slice
and we also allow for the possibility of a conical singularity at the location of the defect, labelled by a parameter $n$. From the $AdS_3\times S^1$ perspective we have non-trivial monodromy for the global symmetry as we traverse the $S^1$ and the parameter $n$ gives the ratio of the radii of the $AdS_3$ and $S^1$ factors.
Supersymmetry imposes a constraint relating the $R$-symmetry monodromy, $g\mu_R$, to $n$ and in the special case when $n=1$ this
becomes the vanishing of the $R$-symmetry flux, $g\mu_R=0$.

For the case of $\mathcal{N}=4$ SYM theory with no mass deformations, associated with solutions of the STU model, 
we found expressions for the currents $\vev {J^i}$ and the defect conformal weight $h_D$ in terms of the monodromy parameters and the parameter $n$; see \eqref{litjistu} and
\eqref{genstuhdexpress}.
An expression for the $b$ central charge is
given in \eqref{bexpstumodelkfixed2}
along with an extension allowing for dependence on $n$ in \eqref{genbstumod3}.
Expressions for the on-shell action are given in \eqref{oshactstu}, \eqref{abexpresforn4}.
The finiteness of the on-shell action is
associated with the fact that the conformal anomaly vanishes on $AdS_3\times S^1$ and it is natural to assume that supersymmetry implies that it is a scheme independent and hence physically meaningful quantity.
The supersymmetric Renyi entropy is discussed in section \ref{sturenyi}, where we also highlighted
that there is some ambiguity in how it is defined.
In appendix \ref{app:freefield} we show that our holographic expressions for $b$ and $d_2=18\pi n h_D$ 
are in precise agreement with the large $N$ limit of a computation arising in $\mathcal{N}=4$ SYM theory in the free limit when there are
no conical singularities. 
For $b$ this result could have been anticipated from the results of 
\cite{Herzog:2019rke,Bianchi:2019umv}. The result for 
$d_2,h_D$ is more of a surprise, but should follow from the fact that these quantities can be obtained from
the expression for $b$ in the presence of conical singularities (see \eqref{derivbwrtn}). We expect this to be true for more general supersymmetric defects and it would be interesting to prove this.

For the LS case, we derived similar expressions to those we obtained for $\mathcal{N}=4$ SYM.
We obtained expressions for the conserved currents and 
the defect conformal weight $h_D$ in terms of the monodromy parameters and the parameter $n$;
see \eqref{allcurrentsintrmsscesLS} and \eqref{hdnmone}.
An expression for the $b$ central charge is
given in \eqref{bLScasewithkmone}
along with an extension allowing for dependence on $n$ in \eqref{genbLSmod3}.
Similar expressions for the on-shell action are given in \eqref{oshactstuLS}, \eqref{abexpresforls},
and the supersymmetric Renyi entropy
is discussed in section \ref{lssecsre}.

For the defects of both of these SCFTs it would be interesting to verify that the values of $b$ and $h_D$ that we have derived are
consistent with the expression for the entanglement entropy given in \eqref{entent}.
The entanglement entropy could be computed from the analytic STU solutions as in \cite{Jensen:2013lxa}.

For $n\ge 1$, necessarily on the main branch of solutions, we have seen that $h_D,b\ge 0$. We have also seen that it is possible
for $h_D,b< 0$ for $n<1$ on the main branch of solutions. 
When $n=1$ it has been argued that $h_D>0$ when the ANEC holds \cite{Jensen:2018rxu} and also 
when $\mathcal{N}=(0,2)$ supersymmetry is preserved \cite{Bianchi:2019sxz}. It would be of interest to understand
in more detail why these arguments breakdown when $n<1$.

It is remarkable that we have been able to obtain the above results without explicit solutions.
Instead we have utilised conserved quantities of the equations of motion along with additional
properties of the BPS equations, with the expressions
\eqref{hemvszeqtext} and \eqref{effaiprimetext1},\eqref{eq:IntegratedFluxesExpr1text}
playing a key role. The localisation techniques of \cite{BenettiGenolini:2024kyy} are expected to be useful for further illuminating this
fact. For the case of $\mathcal{N}=4$ SYM theory with no mass deformations
explicit solutions of the STU model are known. 
We saw that these STU solutions exist for values of the monodromy parameters that, in general, are 
more restrictive than the naive bounds given in \eqref{stubranches}.
For the defect solutions of the LS theory with $\varphi\ne 0$, we have constructed some explicit solutions numerically for different values of $g\mu_F$ and $n=1$. It appears that
when $n=1$ they exist for all values of $g\mu_F$ consistent with the bound that we found in \eqref{eq:FlavourFluxSourceBoundLS}; 
we anticipate a similar story for other values of $n$ (at least for the main branch of solutions) and it would be interesting to investigate this in more detail.

We discussed novel monodromy defects in $\mathcal{N}=4$ SYM theory with mass deformations, $\varphi_s\ne 0$,
that preserve the superconformal invariance of the monodromy defect.
On an $AdS_3\times S^1$ boundary these are constant deformations while on a flat boundary they are spatially modulated mass deformations.
For these solutions we derived an expression for
$h_D$ expressed in terms of currents \eqref{stresstextn42}. We also derived expressions for
the conserved currents $\langle{J_R^{\varphi}}\rangle$       
$\vev{J_F}$ in terms of monodromy parameters; see \eqref{allcurrentsintrmssces}. In order to find
an analogous expression for the current $\vev{J_B}$ associated with the symmetry broken by the mass deformation, we require an explicit solution. 
We have constructed some examples of
these solutions numerically when $n=1$ and for various values of $g\mu_F$ and it appears that they exist for
all values of $g\mu_F$ consistent with the bound \eqref{eq:FlavourFluxSourceBound};
again it would be interesting to investigate this further for other values of $n$.

We saw there are two branches of allowed solutions, the main branch with $s=-\frac{\kappa}{2}$, existing for $n>0$ and $s=+\frac{\kappa}{2}$ existing, at most, 
for $0<n<1/2$ (see \eqref{constraintkgen}).
For simplicity, in the remainder of this section we focus on $s=-\frac{\kappa}{2}$, 
illustrated (for $n=1$) in figure \ref{fig:sumsolutions}.
For given $g\mu_F,n$ there is a one parameter family of solutions
labelled by $\varphi_s$ that all have $g\mu_R=\kappa(1-n)$ and $g\mu_B=-\kappa n$, both of which are required 
to preserve supersymmetry when $\varphi_s\ne 0$. 
This family includes a restricted STU solution with $\varphi_s=0$, $g\mu_R=\kappa(1-n)$ and $g\mu_B=-\kappa n$, with the latter imposed by hand,
and can be viewed as a line of exactly marginal deformations in the sense of \cite{Herzog:2019bom}.
Remarkably as $\varphi_s\to\infty $ this family of solutions approaches the LS defect solution with the same value of $g\mu_F$ and
we recall that it has $g\mu_B=0$ in contrast to the one-parameter family of solutions associated with defects in 
$\mathcal{N}=4$ SYM theory. Furthermore, we have shown that the solutions reveal an attractor mechanism at work.
In particular, for the $\mathcal{N}=4$ SYM solutions with $\varphi_s\ne 0$ (including the restricted STU solutions) 
as well as the LS defect solutions, we saw that the
behaviour of the metric function $V$ and the scalars $\alpha,\beta$ (in the vector multiplets) at the core
have exactly the same dependence on the monodromy 
data $g\mu_F$ and $n$ ; see \eqref{eyealbetv23again} and \eqref{eyealbetv23againLS}. It would be
interesting to analyse this further using the approach of \cite{BenettiGenolini:2024kyy}. 
Strikingly, we also showed that the whole family of solutions\footnote{This can be contrasted with $N=2$, $d=4$ SCFTs on $S^4$, for example, where the finite part of the partition function depends on the K\"ahler potential of the 
(standard) conformal manifold \cite{Gomis:2015yaa}.} parametrised by $\varphi_s$
 as well the LS defect solution
have the same on-shell action as well as the same conserved currents. 
We also showed that the on-shell action can be expressed in terms of the $a$ and $b$ central charges in a uniform form:
\begin{align}\label{osactstubexpdicsec}
S&={\vol(AdS_3)}\Big(\frac{2}{\pi}\Big)\Big( a_{LS}n-\frac{1}{12}b_{LS}\Big)\,,\nn
&={\vol(AdS_3)}\Big(\frac{2}{\pi}\Big)\Big( a_{\mathcal{N}=4}n-\frac{1}{12}b_{\mathcal{N}=4}\Big)\,,
\end{align}
where $a_{LS}=\frac{27}{128}N^2$, $a_{\mathcal{N}=4}=\frac{1}{4}{N^2}$ and, when
$s=-\frac{\kappa}{2}$,
\begin{align}\label{bLScasewithkmonedisc}
b_{LS}&=\frac{3}{32n^2}\left((n-1)(1+7n+19n^2)+{4(1+2n)}(g\mu_F)^2\right)N^2\,,\nn
b_{\mathcal{N}=4}|_{g\mu_B=-\kappa n }&=\frac{3}{32n^2}\left({(-1-6n-12n^2+24n^3)}+ {4(1+2n)}(g\mu_F)^2\right)N^2\,.
\end{align}
Here we have included $n$ and for $b_{\mathcal{N}=4}$ we have imposed the extra constraint that $g\mu_B=-\kappa n$.
An interesting feature is that by computing $b_{\mathcal{N}=4}|_{g\mu_B=-\kappa n}$ and $a_{\mathcal{N}=4}$
just within ${\mathcal{N}=4}$ SYM theory with $\varphi_s=0$, 
exploiting both the $g\mu_F$ and the $n$ dependence,
we can deduce the value of 
$a_{LS}$ as well as $b_{LS}(g\mu_F,n)$.

We have also discussed novel ``$\mathcal{N}=4$/LS bulk defect RG flows" that flow from a monodromy defect of $\mathcal{N}=4$ SYM in
the UV to a monodromy defect in the LS theory in the IR (see figure \ref{fig:sumsolutions}). For simplicity, here we again focus the discussion in the particular setting when
$n=1$. In the UV the defect is described by an STU model solution with $g\mu_R=0$ (enforced by supersymmetry) and
$g\mu_B=0$ imposed by hand; with $n=1$ these defects preserve $\mathcal{N}=(2,2)$ supersymmetry.
The RG flow is driven by a bulk relevant deformation away from the defect, which is homogeneous 
from the flat space perspective\footnote{From the $AdS_3\times S^1$ Weyl frame this would be a spatially dependent deformation.} and is expected to flow to an LS defect with $g\mu_R=0$ (and necessarily $g\mu_B=0$) with
the same value of $g\mu_F$. Interestingly, the $\mathcal{N}=4$ STU defects with $g\mu_B=0$ are allowed for $|g\mu_F|<2$ while the LS defects can only exist
for the smaller range $|g\mu_F|<3/2$. Hence, only for $|g\mu_F|<3/2$ we expect that these RG flows will exist.
Our analysis revealed that for these specific RG flows the $b$ central charge of the defect would necessarily have to increase from the UV to the IR when $n=1$ (for $n\ne 1$ it can both increase and decrease depending on $g\mu_F,n$). 
It would be interesting to explicitly construct these RG flows holographically by
solving an associated set of PDEs. Such an analysis would also clarify what happens in the IR to the RG flows
starting in the UV with $\mathcal{N}=4$ defects with $g\mu_B=0$ and $3/2\le |g\mu_F|<2$.

We also obtained an interesting result for the ratio of the on-shell action in the UV and the IR for these RG flows. 
Specifically, the on-shell action is proportional to $12 an-b$, for general $n$. For ranges of $g\mu_F$ where
the RG flows should exist we found (for $s=-\kappa/2$) the remarkable result that 
$(12a_{LS}n-b_{LS})/(12a_{\mathcal{N}=4}n-b_{\mathcal{N}=4})|_{g\mu_B=0}<1$. 
It is natural to conjecture  
that the on-shell action will monotonically decrease along the whole of the RG flow. It is also natural to conjecture
that for arbitrary bulk RG flows in the presence of defects
$12an-b$ is a quantity that will always decrease in such RG flows. In addition, we also anticipate that 
$12an-b$ can be obtained by an extremisation principle in the supersymmetric setting.

In carrying out our holographic computations we employed a specific renormalisation scheme which involved
fixing certain finite counterterms as in \eqref{bpsscheme} and \eqref{lscterm}. Some of these were fixed
 using supersymmetry arguments coming from interfaces \cite{Arav:2020obl}.
Here we fixed the finite counterterm parametrised by $\delta_{\Delta R^2}$ by ensuring that $h_D$ is proportional to the R-symmetry current 
when $\varphi_s=0$, as in the supersymmetric analysis of \cite{Bianchi:2019sxz}. 
Interestingly, this leads to a cohomologically trivial $\Box R$ term in the conformal anomaly (see of (B.15) of \cite{Arav:2020obl}). 
We also fixed, by hand,
the condition \eqref{eliminphisqterms}, in order to eliminate $\varphi_s^2$ terms in $h_D$ (or by demanding that the on-shell action is independent of $\varphi_s$); since this choice is associated with non-Weyl invariant finite counterterms, it will also lead to cohomologically trivial terms in the conformal anomaly (see (B.15) of \cite{Arav:2020obl}). It would be interesting to further study the scheme we have used (in particular the additional condition \eqref{extraschemecond})
and the precise connection with supersymmetry for configurations with $\varphi_s\ne 0$.

We expect that the main features that we have observed in the specific context of monodromy defects
of $\mathcal{N}=4$ SYM theory and LS theory will arise more generally whenever there are
two $\mathcal{N}=1$, $d=4$ SCFTs that are related by a supersymmetric RG flow with the IR fixed point breaking some of
the flavour symmetry. In particular, we expect the analogue of \eqref{osactstubexpdicsec} to hold.
Moreover, we also expect similar solutions to be present in the setting of $\mathcal{N}=2$, $d=3$ SCFTs.
Indeed, a similar analysis to the one reported in this paper
can be carried out in the context of $D=4$ supergravity and now associated with ABJM theory as well as the $d=3$, $\mathcal{N}=2$ SCFT 
that is obtained from ABJM theory after a suitable mass deformation. We will report on this soon \cite{toappear}.

\section*{Acknowledgements}

\noindent 
We thank Lorenzo Bianchi, Jaume Gomis, Seyed Morteza Hosseini, Ioannis Papadimitriou, Brandon Robinson
and Yifan Wang for helpful discussions.
This work was supported in part by FWO project G003523N, STFC grants ST/T000791/1, ST/X000575/1
the National Research Foundation of Korea (NRF) grant funded by the Korea government (MSIT) (No. 2023R1A2C1006975), 
an appointment to the JRG Program at the APCTP through the Science and Technology Promotion Fund and Lottery Fund of the Korean Government,
the European  MSCA grant HORIZON-MSCA-2022-PF-01-01 and by the H.F.R.I call ``Basic research
Financing (Horizontal support of all Sciences)" under the National Recovery and Resilience Plan ``Greece 2.0" funded by the
European Union - NextGenerationEU (H.F.R.I. Project Number: 15384.).
JPG is supported as a Visiting Fellow at the Perimeter Institute.

\appendix
\section{BPS equations}\label{susyvarstext}
Here we recall the BPS equations for the $AdS_3$ ansatz of section \ref{sec:ads3}, along with some of the analysis of \cite{Arav:2022lzo}.
We use the orthonormal frame  $e^a=e^V \bar e^a$, $e^3=f dy$, $e^4=hdz$, where $\bar e^a$ is an orthonormal frame for $ds^2(AdS_3)$.
For the gamma matrices we take
\begin{align}\label{eq:5d_clifford_deftext}
\gamma^m = & \Gamma^m \otimes \sigma^3, \qquad  
\gamma^3 =  {1} \otimes i  \sigma^1, \qquad
\gamma^4 =   {1} \otimes  i  \sigma^2, 
\end{align}
with $\Gamma^m=(\sigma^2,  i\sigma^3, i\sigma^1)$ gamma matrices in $D=3$.
The Killing spinor is $\epsilon=\psi\otimes\chi$ with $\psi$ a two component spinor on $AdS_3$ which satisfies
$D_m\psi=\frac{i}{2}\kappa\Gamma_m \psi$
with $\kappa=\pm 1$.
The two component spinor $\chi$ can be
written in the form
\begin{align}\label{chiralityspin}
\chi  =e^{V/2}e^{is z}\begin{pmatrix}\sin\frac{\xi}{2} \\ \cos\frac{\xi}{2} \end{pmatrix}\,,
\end{align}
where the constant $s$ is the gauge-dependent charge of the spinor under the action of $\partial_z$. 

The solutions of interest to us have $\sin\xi$ not identically equal to zero and the associated 
BPS equations can be written in the form
\begin{align}\label{summbbpsapptext}
f^{-1}\xi'&=gW\cos\xi+2\kappa e^{-V}\,,\nn
f^{-1} V' &= \frac{g}{3} W \sin\xi\,,\nn
f^{-1}\alpha'&=-\frac{g}{12}\partial_\alpha W\sin\xi\,,\nn
f^{-1}\beta'&=-\frac{g}{4}\partial_\beta W\sin\xi\,,\nn
f^{-1}\varphi'&=-\frac{g}{2}\frac{\partial_{\varphi}W}{\sin\xi}\,,\nn
f^{-1}\frac{h'}{h}&=\frac{1}{\sin\xi}\Big(2\kappa e^{-V}\cos\xi+\frac{gW}{3}(1+2\cos^2\xi)\Big)\,,
\end{align}
along with the two constraint equations
\begin{align}\label{summconstext}
(s-Q_z)\sin\xi&=-\frac{1}{2}g  Wh\cos\xi-\kappa h e^{-V}\,,\nn
\frac{g}{2}\partial_{\varphi}W\cos\xi&=\partial_{\varphi} Q_z \sin\xi h^{-1}\,.
\end{align}

The frame components of the field strengths are given by
\begin{align}
F^{(i)}_{34}=f^{-1}h^{-1}(a^{i})'\,,
\end{align}
and the derivation of the BPS equations leads to
\begin{align}\label{fstrengthsbpstext}
e^{2\alpha-2\beta}F^{1}_{34}&=-\frac{g}{12}[4W-\partial_\alpha W+3\partial_\beta W]\cos\xi-\kappa e^{-V}\,,\nn
e^{2\alpha+2\beta}F^{2}_{34}
&=-\frac{g}{12}[4W-\partial_\alpha W-3\partial_\beta W]\cos\xi-\kappa e^{-V}\,,\nn
e^{-4\alpha}F^{3}_{34}
&=-\frac{g}{6}[2W+\partial_\alpha W]\cos\xi-\kappa e^{-V}\,.
\end{align}

\subsection{Integrals of motion}
By calculating the derivative of $he^{-V}$ we deduce that
\begin{align}\label{hemvszeq}
he^{-V}=-n \sin\xi\,,
\end{align}
where $n$ is a constant.
This shows that at points where $h$ vanishes, which 
will correspond to the core of the solutions of interest, $\sin\xi$ also vanishes. It is also helpful to notice that from \eqref{summbbpsapptext}, \eqref{summconstext}
we can then write the equation for $\xi'$ as
\begin{align}\label{xiderivnice}
\xi'&=2n^{-1}(s-Q_z)(e^{-V}f)\,,
\end{align}
while the two constraints \eqref{summconstext} can now be written in the form
\begin{align}\label{summconstext2}
(s-Q_z)&=n[\frac{1}{2}g  We^V\cos\xi+\kappa]\,,\nn
\frac{g}{2}\partial_{\varphi}W\cos\xi&=-n^{-1}e^{-V}\partial_{\varphi} Q_z\,.
\end{align}

With $\varphi\ne 0$, we can integrate two of the gauge equations of motion. Specifically, the 
gauge equations of motion are equivalent to
\begin{align}\label{gaugeintmot}
e^{3V}\left(e^{4\alpha-4\beta}F^{1}_{34}-e^{4\alpha+4\beta}F^{2}_{34}\right)&=\mathcal{E}_F\,,\nn
e^{3V}\left(e^{4\alpha-4\beta}F^{1}_{34}+e^{4\alpha+4\beta}F^{2}_{34}+2e^{-8\alpha}F^{3}_{34}\right)&=\mathcal{E}_R\,,\nn
(e^{3V-8\alpha}F^{3}_{34})'&=-e^{3V}fh^{-1}\frac{g}{4}\sinh^2 2\varphi D_z\theta \,,
\end{align}
where $\mathcal{E}_F$ and $\mathcal{E}_R$ are constants and $D_z\theta= (\bar\theta+ga^{(1)}+ga^{(2)}-ga^{(3)})$.
From \eqref{fstrengthsbpstext} we thus have
\begin{align}\label{conschges}
\mathcal{E}_R&=e^{2V}[2ge^V \cos\xi-2\kappa(e^{-4\alpha}+e^{2\alpha}\cosh 2\beta)]\,,\nn
\mathcal{E}_F&=2\kappa e^{2V}e^{2\alpha}\sinh 2\beta\,.
\end{align}
Note for the special case that $\varphi=0$, as in the STU solutions, we can also integrate the last gauge field equation in
\eqref{gaugeintmot} and there is an additional conserved quantity. It is convenient to define this extra conserved quantity as
\begin{align}\label{extraone}
\varphi=0:\qquad \mathcal{E}_B&=e^{3V}\left(e^{4\alpha-4\beta}F^{1}_{34}+e^{4\alpha+4\beta}F^{2}_{34}-2e^{-8\alpha}F^{3}_{34}\right)\,,\nn
&=2\kappa e^{2V}(e^{-4\alpha}-e^{2\alpha}\cosh 2\beta)\,,
\end{align}
where to get the second line we used the BPS equations \eqref{fstrengthsbpstext}.

The BPS equations \eqref{summbbpsapptext} can also be used\footnote{In \cite{Arav:2022lzo} this was noted in the context of the ``conformal gauge", but it is true in any gauge.} to re-express the field strengths \eqref{fstrengthsbpstext} 
in the form
\begin{equation}\label{effaiprime}
F^{(i)}_{yz} = (a^{i})' = (\mathcal{I}^{(i)})' \, ,
\end{equation}
where
\begin{align}\label{eq:IntegratedFluxesExpr1}
& \mathcal{I}^{(1)} \equiv -\frac{1}{2} n e^V \cos\xi \, e^{-2\alpha+2\beta} \, , \quad
\mathcal{I}^{(2)} \equiv -\frac{1}{2} n e^V \cos\xi \, e^{-2\alpha-2\beta} \, , \quad
 \mathcal{I}^{(3)} \equiv -\frac{1}{2} n e^V \cos\xi \, e^{4\alpha} \, .
\end{align}

At the core of the solution, for $\varphi_s\ne0$ (or restricted STU solutions) we have
\begin{align}\label{iidefs}
& \mathcal{I}^{(1)} |_{core}= \frac{1}{2} ( \mathcal{I}_0 \pm \mathcal{I}_\Delta )|_{core} \, , \nn
& \mathcal{I}^{(2)} |_{core}= \frac{1}{2} ( \mathcal{I}_0 \mp \mathcal{I}_\Delta ) |_{core}\, , \nn
& \mathcal{I}^{(3)} |_{core}= \mathcal{I}_0 |_{core}\, , 
\end{align}
where
\begin{align}\label{eq:IntegratedFluxesI0}
&\mathcal{I}_0|_{core} \equiv -\frac{1}{2g} n M_{(1)}|_{core} (-1)^{t}  \, ,\nn
&\mathcal{I}_\Delta|_{core} \equiv -\frac{1}{2g} n M_{(1)}|_{core} (-1)^{t} \sqrt{1-4e^{-12\alpha}}|_{core} \, ,
\end{align}
the $\pm$ sign in \eqref{iidefs} depends on the chosen sign of $\beta$ and
$M_{(1)}$, $M_{(2)}$ were defined in
\eqref{emmbc}.

Recall that at the core
\begin{align}
M_{(1)}&= 2 (-1)^{t} \kappa + \frac{1}{n}\,,
\end{align}
so
\begin{align}\label{eq:IntegratedFluxesI02}
&\mathcal{I}_0|_{core} \equiv -\frac{\kappa n}{g}-\frac{(-1)^t}{2g} \, ,\nn
&\mathcal{I}_\Delta|_{core} \equiv -\left(\frac{\kappa n}{g}+\frac{(-1)^t}{2g}\right) \sqrt{1-4e^{-12\alpha}}|_{core} \, .
\end{align}

\section{$\mathcal{N}=4$ SYM, $AdS_5$ boundary}\label{FGN=4}
In this appendix we construct a boundary expansion of the BPS equations
near the $\mathcal{N}=4$ SYM, $AdS_5$ boundary. We also carry out holographic renormalisation, including
a discussion of finite counterterms and the renormalisation scheme that we shall use. We focus on solutions with
$\varphi\ne 0$, but the procedure is very similar when $\varphi=0$.

To carry out the boundary expansion we work in the gauge
\begin{align}
f=\frac{L}{y}\,.
\end{align}
The gauge fields have the schematic expansion
\begin{align}\label{schematicexpapp}
a^{i} = & \mu_i +\dots   +j_i\frac{L^2}{y^2} + \ldots\,.
\end{align}
Recalling \eqref{lincsn4gf0}, it is convenient to define the following monodromy sources for the currents
\begin{align}\label{mun4appbn}
g\mu_R\equiv g\mu_1+g\mu_2+g\mu_3, \qquad g\mu_F\equiv g\mu_1-g\mu_2\,,\qquad g\mu_B\equiv g\mu_1+g\mu_2-g\mu_3\,. 
\end{align}
The $j_i$ give rise to the one-point currents $\langle J_i\rangle$ obtained by varying the on-shell action with respect 
to the monodromy sources $g\mu_i$; see \eqref{varyactappbmu}, \eqref{jsinapbmed}. We will also be interested
in the linear combinations ${J}_R^{\varphi}$, ${J}_F$, ${J}_B$ defined in \eqref{lincsn4gf0}.

\subsection{Expansion for equations of motion}
We solve the equations of motion in an expansion near the boundary at $y=+\infty$. 
The expansion for the scalars involves integration constants $(\phis,\alphas,\betas)$ and $(\phiv,\alphav,\betav)$, 
associated with, roughly, the sources and the vevs of the dual operators, respectively (we make this more precise below). Specifically,  
\begin{align}
\varphi &=  \phis\frac{ L}{y}  +  \left(\frac{L^2\phis}{2e^{2V_0}} -\frac{8\phis^3}{3} 
- \frac{   (g\mu_1+g\mu_2-g\mu_3)^2 L^2\phis}{2 h_0^2} \right) \frac{\lny L^3}{y^3} +\phiv \frac{L^3}{ y^3} + \ldots \,,\nn
\a &= \alphas \frac{ \lny L^2}{y^2} +\alphav  \frac{  L^2}{y^2}+\ldots \,, \nn
\b &=  \betas  \frac{ \lny L^2}{y^2} + \betav  \frac{ L^2}{y^2}+\ldots \,.
\end{align}
The expansion for the gauge fields involves integration constants $\mu_i$ and $j_i$, associated with the sources and, roughly, the expectation values for the currents, respectively, with 
\begin{align}
a_1 = & \mu_1 -2 (\mu_1+\mu_2 - \mu_3)\phis^2 \frac{\lny L^2}{y^2}+  j_1\frac{L^2}{y^2} + \ldots \,,\nonumber \\
a_2 = & \mu_2 -  2 (\mu_1+\mu_2 - \mu_3)\phis^2 \frac{\lny L^2}{y^2}+  j_2\frac{L^2}{y^2} + \ldots \,, \nonumber \\
a_3= & \mu_3 + 2 (\mu_1+\mu_2 - \mu_3)\phis^2 \frac{\lny L^2}{y^2}+  j_3\frac{L^2}{y^2} + \ldots  \,. 
\end{align}
Finally, the metric functions have the expansion
\begin{align}\label{veeachexp}
&e^{V} =~e^{V_0} \frac{ y}{L}  +\left( \frac{L^2}{4 e^{V_0}} - \frac{e^{V_0}\phis^2}{6}\right) \frac{L}{y} - e^{V_0}\left(\alphas^2 + \frac{\betas^2}{3}\right) \frac{\lny^2 L^3}{y^3}\nonumber \\
&- \left[2e^{V_0} \left(\alphas\alphav + \frac{\betas \betav}{3} - \frac{\phis^4}{3} \right) + \frac{L^2\phis^2}{6 e^{V_0}} \right] \frac{\lny L^3}{y^3}  + P\frac{L^3}{y^3}\ldots,\nn
 &\frac{h}{h_0} =  ~ \frac{y}{L} - \left(\frac{L^2}{4 e^{2V_0}} + \frac{\phis^2}{6} \right) \frac{L}{y} - \left(\alphas^2+ \frac{\betas^2}{3} \right) \frac{\lny^2L^3}{y^3}  \nonumber \\
&+ \left[ \frac{ (g\mu_1+g\mu_2-g\mu_3)^2L^2\phis^2}{2h_0^2}  - \left( {2\alphas\alphav}+  \frac{2 \betas \betav}{3}  - \frac{2\phis^4}{3}\right)  \right] \frac{\lny L^3}{y^3} \nonumber \\
&- \left[\frac{3P}{e^{V_0}} -  \frac{ (g\mu_1+g\mu_2-g\mu_3)^2L^2\phis^2}{8h_0^2}  + \frac{L^2 \phis^2}{8 e^{2V_0}}  
 + \frac{3\alphas^2 + 24\alphav^2 + \betas^2 + 8 \betav^2 + 6 \phis \phiv - 3 \phis^4}{6} \right] \frac{L^3}{y^3} + \ldots
\end{align}
Here $P$ is a constant of integration related to the stress tensor, while $V_0$ and $h_0$ give the sizes of the $AdS_3$ and $S^1$.
In particular the boundary metric has the form 
\begin{align}\label{bdmetfsig}
h_{ab}dx^a dx^b= e^{2V_0} \left(ds(AdS_3) - e^{-2V_0}h_0^2 dz^2\right)\,.
\end{align}

\subsection{Expansion for BPS equations}

We can also develop a boundary expansion for the BPS equations. 
With $\bar\theta=0$, we find
\begin{align}
\label{eq:genl_k_s_theta}
n&=e^{-V_0}h_0 \,,
\end{align}
while the sources for the gauge fields are constrained via 
\begin{align}\label{n4scebps}
g\mu_R&\equiv g\mu_1+g\mu_2+g\mu_3=-{\kappa n}-2s\,, \nn
g\mu_B&\equiv g\mu_1+g\mu_2-g\mu_3 =-{\kappa n} \qquad (\varphi \neq 0)\,,
\end{align}
noting that for BPS configurations  when $\varphi_s= 0$ 
we automatically have $\varphi=0$ for the whole solution (see \eqref{constabfveves} below).
We also find that the sources for the scalar fields are constrained via 
\begin{align}\label{gensusysscsconsappb}
\betas&=0\,,\qquad\qquad \alphas = \frac{2}{3} \varphi_s^2\,.
\end{align}
The six integration constants ($\phiv,\alphav,\betav,j_i$) satisfy three relations 
which can be written as 
\begin{align}\label{constabfveves}
\phiv &= \frac{L^2 \phis}{4e^{2V_0}} - \phis^3
 -\frac{2} {3\kappa n L} (j_1+j_2-2j_3) \phis\,,\nn
\alphav &= \frac{1} {6\kappa n L} (j_1+j_2-2j_3), \qquad \betav = -\frac{1 }{2\kappa n L} (j_1-j_2)\,,
\end{align}
while the constant $P$ is fixed via
 \begin{align}
P = &
\frac{23 e^{V_0} \phis^4}{72} - \frac{L^2 \phis^2}{24 e^{V_0}} 
- \frac{\kappa L}{6 h_0} \left({j_1}+{j_2}+j_3 \right) + \frac{\kappa e^{2V_0}\phis^2}{6h_0 L}  \left({j_1}+{j_2}-2j_3 \right)  \nonumber \\
& - \frac{e^{3V_0}}{9h_0^2 L^2} \left({j_1}^{2}+{j_2}^{2}+j_3^2-j_1 j_2-j_1 j_3-j_2 j_3 \right) \,.
\end{align}
 The phase appearing in the Killing spinor takes the form
\begin{align}\label{xiexp}
\xi = -\frac{\pi}{2} + \frac{\kappa L}{e^{V_0}}\frac{L}{y} -\frac{2L\phis^2}{3\kappa e^{V_0} }\frac{\ln[\frac{y}{\Lambda}]L^3}{y^3}
+\left(-\frac{2}{3h_0}(j_1+j_2+j_3)-\frac{L^3\kappa}{12e^{3V_0}}+\frac{L \kappa \phis^2}{6e^{V_0}}\right)\frac{L^3}{y^3}
+...
\end{align}

These expansions of the BPS equations can be used in \eqref{conschges} to obtain the following expressions for the
constants of motion $\mathcal{E}_R$, $ \mathcal{E}_F$:
\begin{align}\label{FGEREF}
\mathcal{E}_R &
= - \frac{2 e^{3V_0}}{h_0 L} (j_1+j_2+2j_3)\,,\qquad
 \mathcal{E}_F
  = - \frac{2e^{3V_0}}{h_0 L} (j_1-j_2)\,.
\end{align}
For the restricted STU solutions with $\varphi=0$, from \eqref{extraone} we find 
\begin{align}\label{FGEREFB}
\mathcal{E}_B &=-\frac{2e^{2V_0}}{nL}(j_1+j_2-2j_3)\,,\qquad (\varphi=0) \,.
\end{align}

 \subsection{Holographic renormalisation}\label{holrennn4}
 In order to carry out holographic renormalisation we consider the following action:
\begin{align}
S=S_{Bulk} +S_{GH}+S_{ct}+S_{finite}\,.
\end{align}
with 
\begin{align}\label{bulkacn4}
S_{Bulk}+S_{GH} =\frac{1}{4\pi G }\int d^5x \sqrt{g} \mathcal{L}
+\frac{1}{8\pi G }\int  \sqrt{|\g|} d^4x K\,,
 \end{align}
with $\mathcal{L}$ given in \eqref{model1text}. Note that 
the bulk metric approaches the boundary at $y\to \infty$ as
 \begin{align}
ds^2 = \g_{ab} dx^a dx^b - \frac{L^2}{y^2}dy^2 + \ldots, \qquad \gamma_{ab} = \frac{y^2}{L^2}h_{ab}\,.
\end{align}
Here $\gamma$ is the induced metric on a constant $y$ slice and $h_{ab}$ is the field theory boundary metric on $AdS_3\times S^1$ given by
$h_{ab}dx^a dx^b=e^{2V_0}ds^2(AdS_3)-h_0^2dz^2$.
The Newton constant is normalised as follows:
\begin{align}\label{n4normnnewton}
\frac{L^3}{16\pi G} =\frac{N^2}{8\pi^2}\,,
\end{align}
with the $AdS_5$ vacuum of radius $R_{\mathcal{N}=4}=L=2/g$, dual to $SU(N)$, $\mathcal{N}=4$ SYM theory.
The $a$ central charge of $\mathcal{N}=4$ SYM then reads
\begin{align}\label{acentchgen4}
a_{\mathcal{N}=4}=\frac{\pi L^3}{8G}=\frac{N^2}{4}\,.
\end{align}

The counterterm boundary action required to remove divergences in the on-shell action\footnote{We explain in appendix \ref{secosact}
how the on-shell action can be computed.}
is given by
\begin{align}\label{countertactterms}
S_{ct} = \frac{1}{16\pi G} \int d^4 x \sqrt{|\g|}&\left[ 
 \left(-\frac{6}{L} + \frac{L R}{2} \right) - \frac{2\varphi^2}{L} - \frac{8}{L}\left(1-\frac{1}{2\lny} \right) \left(3\a^2 + \b^2 \right) 
\right. \nonumber  \\
&\left. -\lny \frac{LR\varphi^2}{3}+ \lny \frac{16 \varphi^4}{3L} - 2 L\lny  (D\varphi)^2+\dots 
 \right]\,,
\end{align}
 where $\Lambda$ is an arbitrary reference scale and the dots refer to terms that are not needed in our analysis.
We will consider a number of finite counterterms given by the boundary action
\begin{align}\label{finitetermsgen}
S_{finite} = \frac{1}{16 \pi G} \int d^4 x \sqrt{|\g|} & \left[
-\delta_{\Delta R^2} \frac{L^3}{4} \left(R^{ab}R_{ab} + R^2/3 \right) - \delta_\b \frac{8}{\lny^2 L} \left(3 \a^2 + \b^2 \right)  \right. \nonumber \\
& \left. - \delta_{D\varphi^2}2L (D\varphi)^2- \delta_{R\varphi^2} \frac{L}{3}R \varphi^2 + \delta_4 \frac{16}{3L} \varphi^4+\dots 
\right]\,,
\end{align}
where the five constants $\delta_{\Delta R^2}, \delta_{R\varphi^2},  \delta_4,  \delta_\b$ and $\delta_{D\varphi^2}$ define a renormalisation scheme.
In these expressions all quantities utilise the metric $\gamma_{ab}$ evaluated in the limit $y\to\infty$.
They are a straightforward generalisation
 of those considered in appendix B of \cite{Arav:2020obl}, which did not consider gauge fields, and we have replaced\footnote{\label{footscheme}
 Note that in comparing with appendix B of \cite{Arav:2020obl} we should make the identification
 \begin{align}
\varphi^{here} = 2 \atanh[\tan[\phi_1^{there}/2]] = \phi_1^{there} + (\phi_1^{there})^3/6+\cO(\phi_1^5)
\end{align}
In particular the finite counterterm $\delta_{4(1)}$ in \cite{Arav:2020obl} can be identified via 
$\delta_{4(1)} = \delta_4-\frac{1}{8}$, a point we return to below.} 
 \begin{align}
 (\nabla \varphi)^2 \rightarrow (D\varphi)^2 = (\nabla \varphi)^2+ \frac{1}{4} \sinh^2(2\varphi) (D\theta)^2.
\end{align}
We have omitted $R_{abcd}R^{abcd}$ terms since for conformally flat backgrounds they can be expressed in terms
of $R_{ab}R^{ab}$ and $R^2$.
We have also omitted terms of the form $(R_{ab}R^{ab} - R^2/3)$ in \cite{Arav:2020obl}, since they play no role for our ansatz with 
boundary metric on $AdS_3 \times S^1$. Similarly, we have also omitted the usual $\int d^4 x \lny F^2$ counterterms, and associated finite counterterms 
as they also do not affect the correlators of interest since for our ansatz $F|_{y=\mathrm{const}}=0$.

With the above set-up we find that an on-shell variation of the total action depends on the metric parameters, 
$h_0,V_0$, the scalar source parameters $\varphi_s, \alpha_s,\beta_s$ as well as the sources for the gauge field $g\mu_i$.
From these we can obtain the expectation values of the stress tensor, the dual scalar operators and the currents, respectively. 
For the scalar operators we have in general 
\begin{align}
\delta_\Phi S &= \int d^4 x \sqrt{|h|} \vev{\cO_\Phi} \delta \Phi_s\,.
\end{align}
Consider $\alpha$, for example, which is dual to an operator with scaling dimension $\Delta=2$.
Correspondingly, we consider variations with respect to 
$L^{-2}\alpha_s$, which are given by
\begin{align}
\delta_{\a_s} S= \frac{1}{4\pi G} \int d^4 x \frac{h_0e^{3V_0}}{\rho^3}6L \left[{\alphav} - {2 \delta_\b\alphas}\right]\delta ( L^{-2}\alphas)\,.
\end{align}
By equating these expressions with 
$ \delta \Phi_s= \delta ( L^{-2}\alphas)$
and noting that $\sqrt{|h|}=h_0e^{3V_0}/\rho^3$ we can immediately deduce $\vev{\cO_\alpha}$.
With $\beta,\varphi$ dual to operators with $\Delta=2,3$, respectively, we consider corresponding
variations with respect to $L^{-2}\beta_s$ and $L^{-1}\varphi_s$
and deduce 
\begin{align}
\vev{\cO_\alpha}&= \frac{L^3}{4\pi G} 6\left[(L^{-2}{\alphav})- {2 \delta_\b (L^{-2}\alphas)}\right], \nn
\vev{\cO_\beta}&= \frac{L^3}{4\pi G} 2\left[(L^{-2}{\betav}) - {2 \delta_\b(L^{-2}\betas)}\right],\nn
\vev{\cO_\varphi} &= \frac{L^3}{4\pi G} \Big[
{2(L^{-3}\phiv)} + \left( \frac{8}{3} + \frac{16 \delta_4}{3} \right){(L^{-1}\phis)^3 }
- \left( \frac{1}{2} + \delta_{R\varphi^2} \right) \frac{L^{-1} \phis}{e^{2V_0}}\nn
&\qquad\qquad
+ \left( \frac{1}{2} + \delta_{D\varphi^2} \right) \frac{{\hat\theta}^2L^{-1}\phis}{h_0^2}
\Big]\,,
\end{align}
where, temporarily reinstating $\bar\theta$, we have defined
\begin{align}\label{thetahatdef}
\hat\theta = \bar\theta + g (\mu_1 + \mu_2 -\mu_3)\,.
\end{align}

The canonical normalisation of the stress tensor is given by
\begin{align}
\delta_{h_{ab}} S = -\frac{1}{2} \int d^4 x \sqrt{|h|} \langle \cT^{ab} \rangle \delta h_{ab}\,.
\end{align}
With a diagonal stress tensor
of the form
\begin{align}\label{diagsttens}
\vev{\cT^a{}_b} = \diag{\vev{\cT^t{}_t} ,\vev{\cT^t{}_t} ,\vev{\cT^t{}_t} ,\vev{\cT^z{}_z} }\,,
\end{align}
we find that $\vev{\cT^t{}_t} ,\vev{\cT^z{}_z}$ can be obtained by varying with respect to $V_0$ and $h_0$ and identifying 
\begin{align}
\delta_{V_0} S =\int d^4 x \sqrt{|h|}\left[-3 \vev{\cT^t{}_t} \right] \delta V_0, \qquad
\delta_{h_0} S =\int d^4 x \sqrt{|h|}\left[-\vev{\cT^z{}_z}   \right] \frac{\delta h_0}{h_0}\,.
\end{align}
Specifically, we find 
\begin{align}
\vev{\cT^t{}_t} = &
\frac{L^3}{4\pi G}
\Bigg[
- \frac{2L^{-4}P}{e^{V_0} }-\left(\frac{1}{8} + \delta_{\Delta R^2} \right) \frac{1}{2 e^{4V_0} }  - \delta_{D\varphi^2} \frac{ {\hat\theta}^2 (L^{-1}\phis)^2}{2 h_0^2} \nn
&- \left( \frac{1}{2} - \delta_{R\varphi^2} \right) \frac{(L^{-1} \phis)^2}{6 e^{2V_0} }
+(6 \delta_\b-1)L^{-4} \frac{3\alphas^2 + \betas^2}{3 } - L^{-4}({\phis \phiv +3\alphas\alphav + \betas \betav}) \nn&
- L^{-4}\frac{6\alphav^2 + 2 \betav^2}{3 } -\left(\frac{1}{4 } + 4 \delta_4 \right)\frac{(L^{-1}\phis)^4}{3}
\Bigg]\,,
\end{align}
and
\begin{align}
\vev{\cT^z{}_z}  =&  \frac{L^3}{4\pi G} \Bigg[
 \frac{6L^{-4}P}{e^{V_0} }+
\left(\frac{1}{8} + \delta_{\Delta R^2} \right) \frac{3}{2 e^{4V_0} }  + \delta_{D\varphi^2} \frac{ {\hat\theta}^2 (L^{-1}\phis)^2}{2 h_0^2} \nn&+ \left( \frac{1}{2} + \delta_{R\varphi^2} \right) \frac{ (L^{-1}\phis)^2}{2 e^{2V_0} }
+2 \delta_\b L^{-4}({3\alphas^2 + \betas^2})+ L^{-4}({\phis \phiv - 3\alphas\alphav - \betas \betav}) \nn&
+ L^{-4}({6\alphav^2 + 2 \betav^2})- \left(\frac{13}{12 } + \frac{4\delta_4}{3} \right){(L^{-1}\phis)^4}
\Bigg]\,.
\end{align}
Finally, for the currents we have, (including a factor of $g=2L^{-1}$),
\begin{align}\label{varyactappbmu}
\delta_A S &= \int d^4 x \sqrt{|h|} \langle J_i^\mu \rangle \delta (g A^{i}_\mu)\,.
\end{align}
Lowering the index with the boundary metric $h_{ab}$ we have 
\begin{align}\label{badjdef}
\vev{J_i}\equiv \vev{(J_i)_z}=-h_0^2\vev{J_i^z}\,.
\end{align}
where the minus sign arises because of the signature we are using in \eqref{bdmetfsig}.
Specifically, we find
\begin{align}\label{jsinapbmed}
\vev{J_1}&=-\frac{L^3}{4\pi G}\Big[{ L^{-3}j_1}+(\frac{1}{2}+\delta_{D\varphi^2}){\hat\theta (L^{-1}\varphi_s)^2}\Big]\,,\nn
\vev{J_2}&=-\frac{L^3}{4\pi G}\Big[{ L^{-3}j_2}+(\frac{1}{2}+\delta_{D\varphi^2}){\hat\theta (L^{-1}\varphi_s)^2}\Big]\,,\nn
\vev{J_3}&=-\frac{L^3}{4\pi G}\Big[{ L^{-3}j_3}-(\frac{1}{2}+\delta_{D\varphi^2}){\hat\theta (L^{-1}\varphi_s)^2}\Big]\,,
\end{align}
and notice the sign difference in the last expression.
It is interesting to note that using \eqref{lincsn4gf0} we see that
the conserved currents $\vev{ {J}_R^{\varphi}}, \vev{{J}_F}$ can be expressed in terms of just the $j_i$, while 
the non-conserved current $\vev{{J}_B}$ depends on $\varphi_s$, $\delta_{D\varphi^2}$ as well as the $j_i$.

With these results in hand we can then deduce the conformal anomaly $\cA$. Specifically,
the trace of the stress tensor (with respect to $h_{ab}$) satisfies
\begin{align}\label{trwid}
\vev{\cT^a{}_a}  + \vev{\cO_\varphi} (L^{-1}\phis)  + 2\vev{\cO_\a}(L^{-2}\alphas) + 2\vev{\cO_\b} (L^{-2}\betas) \equiv  \cA\,,
\end{align}
with the anomaly polynomial on the backgrounds we are considering given by 
\begin{align}\label{trwidanom}
 \cA =\frac{L^3}{4\pi G}[ -3(L^{-2}\alphas)^2 - (L^{-2}\betas)^2 + \frac{4}{3} (L^{-1}\phis)^4 
+ \frac{1}{2}\left(- \frac{1}{6} R^{(h)}  - 
(D\theta)_{(h)}^2\right)    (L^{-1}\phis)^2]\,.
\end{align}
Here $R^{(h)}=6e^{-2V_0}$ is the Ricci tensor with respect to the field theory metric $h_{ab}$ on $AdS_3\times S^1$ in \eqref{bdmetfsig}
and $(D\theta)_{(h)}^2\equiv D_a\theta D_b\theta h^{ab}=-\hat\theta^2/h_0^2$. 

\subsection{BPS sources}
Note that when considering BPS sources there are a number of simplifications in the one-point functions. We have
$\hat\theta=-\kappa h_0e^{-V_0}$ and this can be substituted into \eqref{jsinapbmed} to get the BPS currents.
We also find the scalar vevs are related to the current vevs as 
\begin{align}\label{bpsvevsappbfinal}
\vev{\cO_\a} &= -{\kappa h_0^{-1} e^{V_0}} \left( \vev{J_1} + \vev{J_2}  - 2 \vev{J_3}  \right)
+ \frac{L^3}{4\pi G}\left( 1+2 \delta_{D\varphi^2}-4\delta_\b \right){2(L^{-1}\phis)^2}\,,\nn
\vev{\cO_\b} &=  {\kappa h_0^{-1} e^{V_0}} \left( \vev{J_1}  -\vev{J_2}   \right)\,,\nn
\vev{\cO_\varphi} &= \frac{4\kappa h_0^{-1} e^{V_0}}{3} \left( \vev{J_1}  +\vev{J_2}  - 2 \vev{J_3}  \right)(L^{-1}\phis)
 + \frac{L^3}{4\pi G}\left( \frac{1}{2}+ \delta_{D\varphi^2}-\delta_{R\varphi^2} \right)\frac{(L^{-1}\phis)}{e^{2V_0}} \nn
&
-  \frac{L^3}{4\pi G}\left( 1- \frac{8}{3}\delta_4 + \frac{8}{3}\delta_{D\varphi^2} \right){2(L^{-1}\phis)^3}\,.
\end{align}
Furthermore the stress tensor components can be written
{\footnotesize
\begin{align}
\label{eq:bps_stress_current_relation}
&\vev{\cT^t{}_t}=
 - \frac{\kappa}{  3h_0e^{V_0}} \left( \vev{J_1} + \vev{J_2} + \vev{J_3} \right)
 \nonumber \\
  &+\frac{L^3}{4\pi G}\Bigg[ - \left( \frac{1}{8} + \delta_{\Delta R^2} \right) \frac{1}{2e^{4V_0}}
  - \left( \frac{1}{2} + \delta_{D\varphi^2} - \delta_{R\varphi^2} \right) \frac{(L^{-1}\phis)^2}{6e^{2V_0}}
  - (1+8 \delta_4 - 16 \delta_\b) \frac{(L^{-1}\phis)^4}{6}\Bigg]\,,\nn
&  \vev{\cT^z{}_z}=
  \frac{\kappa }{h_0e^{V_0}} \left( \vev{J_1} + \vev{J_2} + \vev{J_3} \right)
  \nonumber \\
  &
 +\frac{L^3}{4\pi G}\Bigg[ + \left( \frac{1}{8} + \delta_{\Delta R^2} \right) \frac{3}{2e^{4V_0}}
  - \left( \frac{1}{2} + \delta_{D\varphi^2} - \delta_{R\varphi^2} \right) \frac{(L^{-1}\phis)^2}{2e^{2V_0}}
  - (1+8 \delta_4 - 16 \delta_\b) \frac{(L^{-1}\phis)^4}{6}\Bigg]\,.
\end{align}
}

We notice the remarkable fact that the anomaly polynomial vanishes for BPS configurations
\begin{align}
\cA=0.
\end{align}

\subsection{Renormalisation scheme}\label{appbrenschemenfour}
It is desirable to choose the finite counterterm constants, appearing in \eqref{finitetermsgen},
that are associated with
a renormalisation scheme that preserves supersymmetry. In \cite{Arav:2020obl}, by considering
the local energy density for BPS Janus configurations, it was argued that
a supersymmetric regularisation scheme requires\footnote{To see this, recall footnote \ref{footscheme} and note it was shown in 
\cite{Arav:2020obl} that $\delta_{4(1)} = -\frac{1}{4}+2\delta_\b$.}
\begin{align}\label{delfourcond}
 \delta_4 =2 \delta_\b - \frac{1}{8}\,,
\end{align}
as well as fixing $\delta_{D\varphi^2}$ to a counterterm which involves scalars which we do not have in the truncation we are considering. 
This condition eliminates the contact terms proportional to $\phis^4$ appearing the expression for the stress tensor 
given in \eqref{eq:bps_stress_current_relation}. 

We also know from the work of \cite{Bianchi:2019sxz} that there is a scheme for which supersymmetric defects, preserving $\mathcal{N}=(0,2)$ supersymmetry,
and \emph{not} in the presence of mass deformations (i.e. setting $\phis=\alphas=\betas=0$), the stress tensor should be proportional to the currents. Hence
from \eqref{eq:bps_stress_current_relation} we conclude
that we should take  
\begin{align}\label{deltarsqct}
\delta_{\Delta R^2} = -\frac{1}{8}.
\end{align}
We are not aware of a similar argument in the presence of mass sources. However, 
the fact that \eqref{delfourcond} eliminates the contact terms proportional to $\varphi_s^4$ in \eqref{eq:bps_stress_current_relation} 
is suggestive that supersymmetry also implies the remaining contact term proportional to $\varphi_s^2$ should vanish.
We will work with this assumption and use a scheme with \footnote{Notice, by contrast, that the condition $\delta_{D\varphi^2} - \delta_{R\varphi^2}=0$ would give a Weyl invariant combination of the associated finite counterterms.}
\begin{align}\label{eliminphisqterms}
\frac{1}{2} + \delta_{D\varphi^2} - \delta_{R\varphi^2}=0\,.
\end{align}
We also highlight that the scheme conditions \eqref{delfourcond}-\eqref{eliminphisqterms} are sufficient to argue
the on-shell action is independent of $\varphi_s$, as explained below
\eqref{n4inftyact}.

We next notice that \eqref{eliminphisqterms} also eliminates the contact term proportional to $\varphi_s$ in $\vev{\mathcal{O}_\varphi}$ in \eqref{bpsvevsappbfinal}.
We then notice that the current independent term proportional to $\varphi_s^3$ in $\vev{\mathcal{O}_\varphi}$ and
the current independent term proportional to $\varphi_s^2$ in $\vev{\mathcal{O}_\alpha}$ both vanish if we also impose
\begin{align}\label{extraschemecond}
1+2 \delta_{D\varphi^2}-4\delta_\b=0\,.
\end{align}

Combining these observations, in summary, we use the scheme
\begin{align}\label{bpsscheme}
\delta_4 = 2 \delta_\b - \frac{1}{8}, \qquad \delta_{D\varphi^2} = 2 \delta_\b - \frac{1}{2}, \qquad \delta_{R\varphi^2} = 2 \delta_\b, \qquad \delta_{\Delta R^2} = -\frac{1}{8}\,,
\end{align}
which depends on just one free constant $\delta_\beta$.
In this scheme, for the BPS configurations of interest we have
\begin{align}\label{jaysn4our}
\vev{J_1}&=-\frac{L^3}{4\pi G}\left[{ L^{-3}j_1}-2\delta_\beta\kappa{h_0 e^{-V_0}} { (L^{-1}\varphi_s)^2}\right]\,,\nn
\vev{J_2}&=-\frac{L^3}{4\pi G}\left[{ L^{-3}j_2}-2\delta_\beta\kappa {h_0 e^{-V_0}}{ (L^{-1}\varphi_s)^2}\right]\,,\nn
\vev{J_3}&=-\frac{L^3}{4\pi G}\left[{ L^{-3}j_3}+2\delta_\beta\kappa {h_0 e^{-V_0}}{ (L^{-1}\varphi_s)^2}\right]\,,
\end{align}
with
\begin{align}\label{sconeptfns}
\vev{\cO_\a} &= -{\kappa h_0^{-1} e^{V_0}} \left( \vev{J_1}+  \vev{J_2}  - 2 \vev{J_3}  \right)\,,\nn
\vev{\cO_\b} &=  {\kappa h_0^{-1} e^{V_0}} \left( \vev{J_1}  -\vev{J_2}   \right)\,,\nn
\vev{\cO_\varphi} &= \frac{4}{3}\kappa h_0^{-1} e^{V_0} \left( \vev{J_1}  +\vev{J_2}  - 2 \vev{J_3}  \right)(L^{-1}\phis)
\,,
\end{align}
and
\begin{align}
\label{eq:bps_stress_current_relation2}
&\vev{\cT^t{}_t}=
 -\frac{\kappa }{3h_0e^{V_0}} \left( \vev{J_1} + \vev{J_2} + \vev{J_3} \right)\,,\nn
&  \vev{\cT^z{}_z}=
  \frac{\kappa }{h_0e^{V_0}} \left( \vev{J_1} + \vev{J_2} + \vev{J_3} \right)\,.
\end{align}

While we only have partial arguments that this scheme is supersymmetric, we conjecture that it is.

\section{Boundary expansion for LS fixed point}\label{FGLS}

In this appendix we construct a boundary expansion of the BPS equations near the LS $AdS_5$ boundary
and derive various one point functions.

On the LS background, the complex scalar and gauge field $A_B$ give rise to a massive vector field which is dual to a
vector operator with scaling dimension $\Delta=2+\sqrt{7}$; as this is an irrelevant operator we do not consider any associated source terms.
The scalars $\varphi,\a$ mix into scalar operators with scaling dimensions $\Delta=1+\sqrt{7}$ and 
$\Delta=3+\sqrt{7}$. The latter operator is irrelevant and so we again do not
consider turning on a source. To simplify the analysis we will also turn off the
source for the dimension $\Delta =1+\sqrt{7} \sim 3.65$ operator as well as source terms for the scalar $\beta$ which is dual to 
an operator with dimension $\Delta=2$. A consequence of this significant simplification is that
we will not obtain expressions for the expectation values of the $\Delta=1+\sqrt{7}$ and $\Delta=2$
operators.

\subsection{Expansion for equations of motion}
We work in the gauge
\begin{align}
f=\frac{\Lt}{y},\qquad \Lt=\frac{3}{2^{2/3}g}=\frac{3L}{2^{5/3}}\,,
\end{align}
where $\Lt=R_{LS}$ is the radius of the LS $AdS_5$ fixed point.
For the gauge fields, as in \eqref{lincsLSgf}, we use the following linear combinations
\begin{align}\label{lincsLSgf2}
a^{LS}_R&\equiv \tfrac{4}{3}(a^1+a^2+\tfrac{1}{2}a^3)\,,\qquad
a_F\equiv a^1-a^2\,,\qquad
a_B\equiv a^1+a^2-a^3\,.
\end{align}
We now develop a boundary expansion of the equations of motion (with the above sources set to zero).
For the gauge fields we have as $y\to\infty$
\begin{align}\label{expoffieldsappc}
a^{LS}_R= \mu^{LS}_R + \tilde j_R \left(\frac{\tilde L}{y}\right)^2 + \ldots\,, \nonumber \\
a_F = \mu_F +\tilde j_F  \left(\frac{\tilde L}{y}\right)^2 + \ldots\,, \nonumber \\
a_B =  \tilde j_B \left(\frac{ \Lt}{y}\right)^{1+\sqrt{7}} +\ldots\,,
\end{align}
with, in particular, $\mu_B=0$. The one point functions for the conserved currents ${J}^{\varphi}_R$, ${J}_F$ defined in
\eqref{lincsLSgf} are obtained by varying the on-shell action with respect to the monodromy sources 
$g\mu^{LS}_R$, $g\mu_F$ and are proportional to $\tilde j_R$, $\tilde j_F$, respectively; see 
\eqref{appcvaractmus}, \eqref{appcvaractmusb}.

As in the text we work in a gauge for which the phase of the complex scalar vanishes $\bar\theta=0$.
For the metric we have the expansion 
\begin{align}\label{LSmetexpseom}
e^{2V} = e^{2V_0}\left(\frac{y}{\Lt}\right)^2 + \frac{\Lt^2}{2} + P_v  \left(\frac{\Lt}{y}\right)^2+\ldots, \nonumber \\
\frac{h^2}{h_0^2}=  \left(\frac{y}{ \Lt}\right)^2 - \frac{ \Lt^2}{2 e^{2V_0}} + P_h  \left(\frac{ \Lt}{y}\right)^2+\ldots, 
\end{align}
while for the scalars we have
\begin{align}
\varphi &= \frac{\ln 3}{2} + \varphi_v \left(\frac{ \Lt}{y}\right)^{1+\sqrt{7}}+\ldots\,, \nonumber \\
\alpha  &= \frac{\ln 2}{6} -\frac{1+\sqrt{7}}{6}\varphi_v  \left(\frac{ \Lt}{y}\right)^{1+\sqrt{7}}+\ldots\,, \nonumber \\
\beta &= \beta_v  \left(\frac{ \Lt}{y}\right)^{2}+\ldots\,.
\end{align}
 There is one constraint on the expansion parameters given by
\begin{align}
e^{2V_0} P_h + 3 P_v = \frac{ \Lt^4}{4e^{2V_0}} - \frac{8}{3} e^{2V_0} \beta_v^2\,.
\end{align}

\subsection{Expansion for BPS equations}
The above expansion can be adapted to provide an expansion of the BPS equations. 
 With $\bar\theta=0$, we again have
\begin{align}
\label{eq:genl_k_s_thetaLS}
n&=e^{-V_0}h_0 \,,
\end{align}
We should also impose the following condition on the monodromy sources
\begin{align}
g \mu^{LS}_R =g \mu_R = -n\kappa  - 2s,
\end{align}
and we recall we also have set $g\mu_B=0$. In addition we impose the following condition on
the subleading coefficients in the expansions:
\begin{align}\label{pbetalscaseapc}
P_v &= \frac{\Lt^4}{16e^{2V_0}} - \frac{  e^{4V_0} (\tilde j_F)^2}{2^{1/3} 3h_0^2\Lt^2}- \kappa \frac{\tilde j_R e^{V_0}\Lt}{2^{5/3} h_0 },\nn
 \beta_v &= - \kappa \frac{e^{V_0} \tilde j_F}{2^{2/3} h_0 \Lt}\,.
 \end{align}

The phase appearing in the Killing spinor can be expanded as
\begin{align}
\xi=-\frac{\pi}{2}+\frac{\kappa\tilde L}{ e^{V_0}}\left(\frac{\tilde L}{y}\right)-\left(\frac{\kappa \tilde L^3}{12e^{3V_0}}+\frac{\tilde j_R}{2^{2/3}h_0} \right)   \left(\frac{\tilde L}{y}\right)^3+\dots\,.
\end{align}

\subsection{Holographic renormalisation}\label{holrenmLS}
As in section \ref{holrennn4} we consider the following action:
\begin{align}
S=S_{Bulk} +S_{GH}+S_{ct}+S_{finite}\,,
\end{align}
with
$S_{Bulk} +S_{GH}$ as in \eqref{bulkacn4}. With \eqref{LSmetexpseom}
 the bulk metric approaches the boundary at $y\to \infty$ as
 \begin{align}
ds^2 = \g_{ab} dx^a dx^b - \frac{\Lt^2}{y^2}dy^2 + \ldots, \qquad \gamma_{ab} = \frac{y^2}{\Lt^2}h_{ab}\,,
\end{align}
with $h_{ab}dx^a dx^b=e^{2V_0}ds^2(AdS_3)-h_0^2dz^2$ the field theory boundary metric on $AdS_3\times S^1$.
With the normalisation of the Newton constant as in \eqref{n4normnnewton} we have
\begin{align}\label{LSnormnnewton}
\frac{\tilde L^3}{16\pi G} =\frac{27}{32}\frac{N^2}{8\pi^2}\,,
\end{align}
and the 
$a$ central charge of the LS fixed point then reads
\begin{align}\label{acentchgen4LS}
a_{LS}=\frac{\pi \tilde L^3}{8G}=\frac{27}{32}a_{\mathcal{N}=4}\,.
\end{align}

With our simplifying assumption of switching off several source terms, it is sufficient to consider the following counterterm actions:
\begin{align}
S_{ct} &=\frac{1}{16\pi G} \oint \sqrt{|\g|} \left( - \frac{6}{ \Lt} + \frac{\Lt R}{2} - \frac{8}{\Lt} \beta^2 \right)\,,
\nn
S_{finite} &= \frac{1}{16 \pi G} \oint \sqrt{|\g|}  \left[
-\delta_{\Delta R^2} \frac{\Lt^3}{4} \left(R^{ab}R_{ab} + \frac{1}{3}R^2 \right) \right].
\end{align}
For the boundary conditions we are considering, the $\beta^2$ term
is actually a finite counterterm, with a coefficient fixed to ensure that we have Dirichlet boundary conditions for the field $\beta$, i.e. 
to make sure that $\delta \beta_v$ does not appear in $\delta S_{full}$. But allowing for more general source terms on
the boundary, it would be required to cancel the log branch divergence.

For solutions of the equations of motion we find a diagonal stress tensor
of the form \eqref{diagsttens} with
\begin{align}
\vev{\cT^t{}_t}&= \frac{\Lt^3}{4\pi G}\Big(- \frac{\Lt^{-4}P_v }{ e^{2V_0}} - \frac{2  \Lt^{-4}\b_v^2}{3} - \delta_{\Delta R^2} \frac{1}{2e^{4V_0} }\Big)\,, \nn
 \vev{\cT^z{}_z}  &=\frac{\Lt^3}{4\pi G}\Big(  \frac{3\Lt^{-4}P_v }{e^{2V_0}} + {2 \Lt^{-4} \b_v^2} + \delta_{\Delta R^2} \frac{3}{2e^{4V_0} }\Big)\,.
\end{align}
For the currents associated to the sources $g\mu^R, g\mu^F$ we have
\begin{align}\label{appcvaractmus}
\delta_A S = \oint \sqrt{h} \Big(\vev{J_R^{\varphi z}} \delta (g \mu^{LS}_R)+\vev{J_F^z} \delta (g \mu_F)\Big)\,,
\end{align}
and we find, after lowering the index with $h_{ij}$, 
\begin{align}\label{appcvaractmusb}
\vev{J_R^\varphi} \equiv\vev{(J^\varphi_R)_z} =-\frac{1}{4\pi G} \frac{\tilde j_R}{2^{5/3}}, \qquad 
\vev{(J_F)}\equiv\vev{(J_F)_z} =-\frac{1}{4\pi G} \frac{2^{4/3} \tilde j_F}{3}\,.
\end{align}

For solutions of the BPS equations we find that the stress tensor and the currents are related via
\begin{align}
\vev{\cT^t{}_t} &=
-\frac{\kappa }{ h_0 v_0} \vev{J^\varphi_R} - \frac{1+8 \delta_{\Delta R^2}}{4\pi G} \frac{\Lt^3}{16 e^{4V_0}}\,,\nn
\vev{\cT^z{}_z} &= \frac{3\kappa}{ h_0 v_0} \vev{J^\varphi_R}+\frac{1+8 \delta_{\Delta R^2}}{4\pi G} \frac{3\Lt^3}{16 e^{4V_0}}\,.
\end{align}
Using the scheme
\begin{align}\label{lscterm}
\delta_{\Delta R^2}=-\frac{1}{8}\,,
\end{align}
we see that the stress tensor is directly proportional to the conserved current $\vev{J_R^\varphi}$, consistent with the supersymmetric results
of \cite{Bianchi:2019sxz}, as in appendix \ref{appbrenschemenfour}.

Finally, we note that the expansion of the equations of motion implies that the conserved
quantities in \eqref{gaugeintmot} are given by
\begin{align}
\cE_R&= -\frac{3}{2^{4/3}}\frac{2e^{2V_0}}{\Lt n} \tilde j_R  
=\frac{16\pi^2e^{2V_0}L^2}{ nN^2}\langle{{J}_R^{\varphi}}\rangle\,,\nn
\cE_F&= -{2^{2/3}}\frac{2e^{2V_0}}{\Lt n} \tilde j_F
=\frac{8\pi^2e^{2V_0}L^2}{ nN^2}\vev{J_F}\,,
\end{align}
where we used \eqref{lsfpsc} and \eqref{n4normnnewton}.
The latter expressions are equally valid for $\mathcal{N}=4$ SYM theory asymptotics with $\varphi\ne 0$.

\section{Calculating the on-shell action}\label{secosact}
In this appendix we discuss how the on-shell action may be computed. 

We first discuss how the bulk on-shell Lagrangian can be written as a total derivative in two different ways. Writing the 
$AdS_3$ metric in Poincar\'e coordinates, as in \eqref{ads3poinc}, we see that 
$\partial_t$ is a Killing vector that preserves the whole $AdS_3$ ansatz. As a consequence\footnote{Consider the theory
in arbitrary dimensions with \emph{mostly minus} signature, 
\begin{align}\label{emax}
\mathcal{L}=-\gamma R-V(\phi)-\frac{Z(\phi)}{4}F^2+\frac{1}{2}\nabla_\mu\phi\nabla^\mu\phi+X(\phi)(D_\mu \theta D^\mu\theta)\,,
\end{align}
with $D\theta=d\theta+A$ and $\gamma$ a constant. Define the two form
$q_{\mu\nu}=-2\gamma\nabla_{\mu}k_{\nu}+Z(\phi)(A_\rho k^\rho)F_{\mu\nu}$
where $k^\mu$ is a Killing vector. Then, assuming $\mathcal{L}_kA_\mu=\mathcal{L}_k\phi=0$,
on-shell 
$\nabla_\nu q^{\mu\nu}=k^\mu \mathcal{L}-2XD^\mu\theta (k^\rho \partial_\rho\theta)$.
}  
we can rewrite the on-shell bulk
Lagrangian in the form:
\begin{align}\label{firstosaction}
\sqrt{+g}\mathcal{L}=\partial_y\left(-\frac{e^{3V}h V'}{2\rho^3 f}\right)+\partial_\rho\left(\frac{e^Vf h}{2\rho^2}\right)\,.
\end{align}
When $\varphi=0$ we also have another Killing vector $\partial_z$ that leaves the solution invariant
and in this case we know that we can write the on-shell Lagrangian as a total derivative in a second way. In fact for $\varphi\ne 0$, and temporarily reinstating $\bar\theta\ne 0$, the solution is invariant under a combination
of gauge transformations and translations in the $z$ direction and we find we can write
the on-shell bulk Lagrangian as
\begin{align}\label{secondosaction}
\sqrt{+g}\mathcal{L}=&\frac{1}{\rho^3}\partial_y\Sigma
-\frac{e^{3V} f}{4 \rho^3 h}\sinh^22\varphi\bar\theta D_z\theta\,,\nn
=&\frac{1}{\rho^3}\partial_y\Sigma
+\frac{1}{\rho^3}\partial_y\left(\frac{\bar\theta}{g}e^{3V-8\alpha}F^3_{34}\right)\,,
\end{align}
where we used \eqref{gaugeintmot} to get the second line, and
\begin{align}\label{expressionSigma}
\Sigma=-\frac{e^{3V}h'}{2 f}
-\frac{e^{3V}}{f h}\Big[
e^{4\alpha-4\beta}a_1 a_1' +e^{4\alpha+4\beta}a_2 a_2'
+ e^{-8\alpha}a_3 a_3'
\Big]\,.
\end{align}
Note that in the gauge $\bar\theta=0$, which we use in the paper, the last term in \eqref{secondosaction} vanishes;
this arises because in this gauge $\partial_z$ then preserves the ansatz. 

We continue to work in the $\bar\theta=0$ gauge and integrate\footnote{At the end of the section we will comment
on what happens if we instead used \eqref{firstosaction}.} \eqref{secondosaction} over all coordinates to deduce that
the total on-shell action is given by
\begin{align}\label{partfngen}
S&=S|_\infty+S_\text{core}\,,
\end{align}
where 
\begin{align}\label{partfngen2}
S|_\infty &=\frac{1}{4\pi G}(2\pi)\vol(AdS_3)
(\Sigma)|_\infty+S_{GH}+S_{ct}+S_{finite}\,,\nn
S_\text{core}&=\frac{1}{4\pi G}(2\pi)\vol(AdS_3)(-\Sigma)|_\text{core}\,.
\end{align}

We can make a further general observation regarding $S_\text{core}$. In the gauge $f=1$, regularity
of the metric requires that $h\sim (y-y_\text{core})$. For a regular gauge field we also require that $a_i \sim (y-y_\text{core})^2$ and hence the second
term in \eqref{expressionSigma} vanishes. Noting that $a_i'/f$ is gauge invariant, we see that the second term vanishes in any gauge and hence we can write
\begin{align}\label{expressionSigma2}
\Sigma|_\text{core}=-\frac{e^{3V}h'}{2 f}|_\text{core}.
\end{align}

\subsection{Solutions with $\varphi\ne 0$}
To evaluate \eqref{expressionSigma2},
we can use the BPS equations \eqref{summbbpsapptext} as well as \eqref{hemvszeq}
 to write
\begin{align}\label{hprimeatcalc}
\frac{e^{3V}h'}{2 f} =-\frac{e^{3V} n}{2}\Big(2\kappa \cos\xi+\frac{ge^{V}W}{3}(1+2\cos^2\xi)\Big)\,.
\end{align}
Thus, for configurations with $\varphi\ne 0$ 
we can use 
\eqref{betbc2} and hence deduce
\begin{align}
-\Sigma_\text{core}=\frac{e^{3V}h'}{2 f}|_{\text{core}} =-\frac{e^{3V} n}{2}\Big(2\kappa (-1)^t-{ge^{V+4\alpha}}\Big)|_{\text{core}}\,.
\end{align}
Recall that we defined $M_{(1)}\equiv g e^{4\alpha} e^V$ and so using \eqref{emmbcpoles} we simply get
\begin{align}\label{sigmaetvrel}
-\Sigma_\text{core}=\frac{1}{2}e^{3V}|_{\text{core}}\,.
\end{align}

\subsubsection{$\mathcal{N}=4$ SYM case, $\varphi\ne 0$}\label{appd1osact}
For this case we have the constraints $g\mu_R=-{\kappa n}-2s$ and $g\mu_B=-\kappa n$.
 We can evaluate \eqref{sigmaetvrel} for the $\mathcal{N}=4$ case with $\varphi\ne 0$ using
\eqref{eyealbetv23}  to get
\begin{align}
S_\text{core} &=\frac{L^3}{4\pi G}(2\pi)\vol(AdS_3)(-s\kappa) (1+\frac{g\mu_1}{\kappa n})(1+\frac{g\mu_2}{\kappa n})(1+\frac{g\mu_3}{\kappa n})\,,\nn
&=\frac{L^3}{4\pi G}(2\pi)\vol(AdS_3)\Big[\frac{1}{4n^3}[(-\kappa n+s)^2-(g\mu_F)^2](-\kappa n +s)s\Big]\,.
\end{align}
We can also compute the contribution $S|_\infty$ in \eqref{partfngen2} at the asymptotic boundary for the $\mathcal{N}=4$ case, using the results for the boundary BPS expansion in appendix \ref{FGN=4}, and we
find 
\begin{align}\label{n4inftyact}
S|_\infty&=\frac{1}{4\pi G}(2\pi)\vol(AdS_3)
\Big[\kappa e^{2V_0} \sum_i j_i +\frac{e^{2V_0}}{n }\sum_i (g\mu_i) j_i\nn
&-\frac{3n L^3}{2}(\frac{1}{8}+\delta_{\Delta R^2}  )
-\frac{\varphi_s^2 e^{2 V_0}L}{2}[    (g\mu_B)\kappa +n(\frac{1}{2}+\delta_{R\varphi^2}-\delta_{D\varphi^2}   )]\nn
&\frac{\varphi_s^4 n e^{4V_0}}{6L}(1-8(2\delta_\beta-\delta_4))\,.
\end{align}

We next substitute the expressions for $j_i$ given in \eqref{jsinapbmed}, with no assumptions on the finite counter terms in the renormalisation scheme yet, and set $\hat\theta=g\mu_B$
(i.e. $\bar\theta=0$ in \eqref{thetahatdef}). Imposing the supersymmetry condition $(g\mu_B)=-\kappa n$ (when $\varphi\ne 0$), we find the remarkable fact that
\begin{align}
\kappa e^{2V_0} \sum_i j_i +\frac{e^{2V_0}}{n }\sum_i (g\mu_i) j_i
=-4\pi G e^{2V_0}\left(\kappa\sum_i\vev{J_i} +\frac{1}{n}\sum_i\vev{J_i} g\mu_i \right)\,,
\end{align}
and, in particular, the dependence on $\delta_{D\varphi^2}$ and $\varphi_s$ has dropped out.
This can then be expressed in terms of $\langle{J}_R^{\varphi}\rangle,\langle{J}_F\rangle,\langle{J}_B\rangle$ using
\eqref{lincsn4gf0} and recall that when $\varphi\ne 0$ we require an explicit solution to obtain $\vev{J_B}$. 
However, also imposing the supersymmetry constraint $g\mu_R=-\kappa n-2 s$,
we see that $\vev{J_B}$ drops out (recall also the comment below \eqref{jsinapbmed}).

If we now impose the renormalisation scheme conditions \eqref{delfourcond}-\eqref{eliminphisqterms} (but not \eqref{extraschemecond}), as well as $(g\mu_B)=-\kappa n$, the last 
two lines in \eqref{n4inftyact} vanish. Thus, imposing
\eqref{delfourcond}-\eqref{eliminphisqterms} and the supersymmetry conditions
\eqref{n4scebps}, we find that the total on-shell
action can be written
\begin{align}\label{jpiecesaction}
S&=-(2\pi)\vol(AdS_3)
e^{2V_0}\left(\kappa\sum_i\vev{J_i} +\frac{1}{n}\sum_i\vev{J_i} g\mu_i \right)+S_\text{core}\nn
&=(2\pi)\vol(AdS_3)\frac{e^{2V_0}}{n}
\Big[2(-\kappa n+s)\vev{J^\varphi_R}- g\mu_F \vev{J_F}\Big]+S_\text{core}\,.
\end{align}
Using \eqref{allcurrentsintrmssces} we can then express the currents $\vev{J^\varphi_R}$ and $\vev{J_F}$ in terms of $g\mu_F$ and we obtain
\begin{align}\label{restrictedstuact}
S&=\frac{N^2}{2\pi^2}(2\pi)\vol(AdS_3)
\Big[-\frac{1}{8n^2}\kappa(s-n\kappa) ((s-n\kappa )^2-(g\mu_F)^2)       
\Big]\,.
\end{align}

\subsubsection{LS case}
We can do a similar computation for the LS defect solutions for which we now have the constraints
$g \mu^{LS}_R = g\mu_R= -n\kappa  - 2s$, $g\mu_B=0$.
From \eqref{sigmaetvrel} and using the results of section \ref{subsecLS}, we get the core contribution
\begin{align}
S_\text{core} &=\frac{1}{4\pi G}(2\pi)\vol(AdS_3)(-s  \kappa) R_{LS}^3(1+\frac{4g\mu_1}{3\kappa n})(1+\frac{4g\mu_2}{3\kappa n})(1+\frac{2g\mu_3}{3\kappa n})\,,\nn
&=\frac{L^3}{16\pi G}(2\pi)\vol(AdS_3)
\frac{s}{n^3} [(s-n\kappa )^2 -(g\mu_F)^2](s-n\kappa )\,,
\end{align}
where we used $R_{LS}=3/(2^{2/3}g)$.
We next compute the contribution at the asymptotic boundary for the LS case, $S|_\infty$, 
using the boundary expansion
of the BPS equations discussed in appendix \ref{FGLS}. We find
\begin{align}
S|_\infty 
= &-\vol(AdS_3)(2\pi) e^{2V_0} \Big[ 
\frac{1}{n} \left( \vev{J_F} g\mu_F+\vev{J^\varphi_R} g\mu_R^{LS}\right)+3 \kappa \vev{J_R^\varphi}\nn
&\qquad\qquad\qquad\qquad\qquad\qquad+\frac{3^4 n}{2^3 g^3 e^{2V_0}} \left( \frac{1}{8} + \delta_{\Delta R^2} \right)
\Big]\,.
\end{align}
We can then express this in terms of the monodromy sources using \eqref{allcurrentsintrmsscesLS}.
After using the scheme $\delta_{\Delta R^2}=-\frac{1}{8}$, as in \eqref{lscterm}, we find
\begin{align}
S|_\infty 
= &\vol(AdS_3)\frac{(2s+n\kappa)(s-n \kappa)(g\mu_F^2-(s-n \kappa)^2)}{8\pi n^3}N^2 \,,
\end{align}
and notice, in particular, that $S|_\infty =0$ when $n=1$, which necessarily has $s=-\kappa/2$.
Summing the contributions form the core and at infinity, using 
\eqref{n4normnnewton} we obtain
\begin{align}\label{oshactstuLSappd}
S
&=\frac{27 N^2}{64\pi}\vol(AdS_3)\Big[n(1+\frac{4g\mu_1}{3\kappa n})(1+\frac{4g\mu_2}{3\kappa n})(1+\frac{2g\mu_3}{3\kappa n})\Big]\,.
\end{align}
Furthermore, when written in terms of $g\mu_F,s$ and $\kappa$ it is the same as \eqref{restrictedstuact}.

\subsection{The STU model}
We now consider the general defect solutions of the STU model (with
 $g\mu_R=-{\kappa n}-2s$ but without the restriction that $g\mu_B=-\kappa n$).
At the core we can use \eqref{hprimeatcalc} as well as the expression for $W$ given in \eqref{superpotstu} to write
\begin{align}
-\Sigma_\text{core}=\frac{e^{3V}h'}{2 f}|_\text{core} =2se^{3V}
\Big(\kappa n +\frac{g}{2}\sum_i \mathcal{I}_i\Big)|_\text{core}\,,
\end{align}
and hence using \eqref{valuesorestukey}
we again obtain 
\begin{align}
-\Sigma_\text{core}=\frac{1}{2}e^{3V}|_{\text{core}}\,.
\end{align}
Hence, we can write
\begin{align}
S_\text{core} &=\frac{L^3}{4\pi G}(2\pi)\vol(AdS_3)(-s\kappa) (1+\frac{g\mu_1}{\kappa n})(1+\frac{g\mu_2}{\kappa n})(1+\frac{g\mu_3}{\kappa n})\,.\end{align}
The remaining piece $S|_\infty$ in \eqref{partfngen2} can be obtained exactly as above; see
\eqref{n4inftyact}. We can then substitute values for $j_i$ in terms of the sources (see \eqref{litjistu}) and
combining these ingredients we obtain
\begin{align}\label{stufinact}
S&=\frac{1}{4\pi G}(2\pi)\vol(AdS_3)
\Big[\kappa e^{2V_0}\sum_i j_i +\frac{e^{2V_0}}{n }\sum_i (g\mu_i) j_i\Big]+S_\text{core}\,,\nn
&=\frac{N^2}{2\pi^2}(2\pi)\vol(AdS_3)\Big[\frac{1}{2n^3}({\kappa n}+2s)(\kappa n+g\mu_1)(\kappa n+g\mu_2)(\kappa n+g\mu_3)\Big]
+S_\text{core}\,,\nn
&=\frac{N^2}{2\pi^2}(2\pi)\vol(AdS_3)\Big[\frac{\kappa}{2n^2}(\kappa n+g\mu_1)(\kappa n+g\mu_2)(\kappa n+g\mu_3)\Big]\,.
\end{align}
Here we have used that for the STU solutions we have the constraint $g\mu_R=\sum_ig\mu_i=-{\kappa n}-2s$.
We also highlight that when $n=1$ we are necessarily on the main branch with $s=-\kappa/2$ and so from the second line we see that $S|_\infty=0$ for this case,
with the on-shell action $S$ only getting a contribution from the core.

We can now make a number of observations in special sub-cases.
\begin{itemize}

\item $g\mu_B =-\kappa n$: this corresponds to the restricted STU model solutions and we exactly recover \eqref{restrictedstuact}, as expected.

\item $g\mu_3=0$, so $g\mu_2=- g\mu_1-2s-\kappa n$: this corresponds to monodromy defects preserving $\mathcal{N}=(2,2)$ supersymmetry and we get
\begin{align}\label{gutvicactcomp}
S&=\frac{N^2}{2\pi^2}(2\pi)\vol(AdS_3)[-\frac{1}{2n}(\kappa n +g\mu_1)(2s+g\mu_1)]\,,\nn
&=\frac{N^2}{2\pi^2}(2\pi)\vol(AdS_3)[-\frac{1}{2n}(\kappa n +g\mu_1)(-\kappa+g\mu_1)]\,,\qquad s=-\frac{\kappa}{2}\nn
&=\frac{N^2}{2\pi^2}(2\pi)\vol(AdS_3)\frac{1}{2}(1-(g\mu_1)^2),\qquad n=1\,.
\end{align}
The result for $n=1$ differs from (4.14) of \cite{Gutperle:2019dqf}; the difference arises\footnote{\label{footnotefct}If we had not included the finite counterterm $\delta_{\Delta R^2}$ then we would have got
$S\to S-\frac{N^2}{2\pi^2}(2\pi)\vol(AdS_3)(3n/16)$ in \eqref{stufinact}. The last line of \eqref{gutvicactcomp}
would then become $S\to \frac{N^2}{2\pi^2}(2\pi)\vol(AdS_3)\frac{1}{2}(\frac{5}{8}-(g\mu_1)^2)$.
} from the fact that with $\varphi_s=0$, we still get a contribution from the finite counterterm $\delta_{\Delta R^2}=-1/8$, we use in our renormalisation scheme (see \eqref{n4inftyact}), which 
\cite{Gutperle:2019dqf} didn't include. 

\item $g\mu_3=0$ and $g\mu_1=g\mu_2=-s-\frac{1}{2}\kappa n$: these are solutions of Romans supergravity theory with $\beta=0$ and\footnote{This appears to be in agreement with the result (3.26) of \cite{Crossley:2014oea}. However, this is puzzling because \cite{Crossley:2014oea} did not include the finite counterterms $\delta_{\Delta R^2}=-1/8$ and recalling footnote \ref{footnotefct} we would expect a difference of the form $-\frac{N^2}{2\pi^2}(2\pi)\vol(AdS_3)(3n/16)$.
An additional point is that \cite{Crossley:2014oea} did not include the counterterms proportional to $\alpha^2$ in \eqref{countertactterms}, which are required in order to have the correct variational principle (and, when $\alpha_s\ne0$, to cancel divergences).}
\begin{align}
S&=\frac{N^2}{2\pi^2}(2\pi)\vol(AdS_3)[\frac{1}{2n}(-\frac{\kappa n}{2}+s)^2]\nn
&=\frac{N^2}{2\pi^2}(2\pi)\vol(AdS_3)[\frac{(n+1)^2}{8n}],\qquad s=-\frac{\kappa}{2}\,.
\end{align}

\item $g\mu_1=g\mu_2=g\mu_3=-\frac{1}{3}(\kappa n+2s)$: these are solutions of minimal gauged supergravity
with $\alpha=\beta=0$ and we find
\begin{align}
S&=\frac{N^2}{2\pi^2}(2\pi)\vol(AdS_3)\Big[-\frac{4\kappa}{27n^2}(-\kappa n+s)^3\Big]\,,\nn
&=\frac{N^2}{2\pi^2}(2\pi)\vol(AdS_3)\Big[\frac{4}{27n^2}(n+\frac{1}{2})^3\Big]\,,\qquad s=-\frac{\kappa}{2}\,,\nn
&=\frac{N^2}{2\pi^2}(2\pi)\vol(AdS_3)\Big[\frac{1}{2}\Big]\,,\qquad n=1\,.
\end{align}

\end{itemize}

\subsection{Comment}
We obtained the above results for the on-shell action using \eqref{secondosaction}. It is natural to
ask what happens if we instead
use \eqref{firstosaction}. Using the BPS equation $f^{-1} V' = \frac{g}{3} W \sin\xi$ in \eqref{summbbpsapptext}
and the fact that $h\to 0$ at the core of the solution, we see that the first term in \eqref{firstosaction} will only get a contribution
from the $AdS_5$ boundary. Therefore the total on-shell action can also be written in the form
\begin{align}
S=\frac{1}{4\pi G}(2\pi)\vol(AdS_3)\Big[
\left(-\frac{g}{6}W\sin\xi h e^{3V}\right)|_\infty-\int_\text{core}^\infty dy e^V f h
\Big]\nn
+S_{GH}+S_{ct}+S_{finite}\,.\nn
\end{align}
One can check that there are divergent pieces in the bulk integral in this expression, which of course will cancel with
divergences in other terms. 
Nevertheless, by equating this expression with what we just derived for the total on-shell action above,
we can obtain an explicit expression for the bulk integral in terms of boundary data. 
Note this expression indicates a convenient choice of the function $f$: for example one might choose
$e^V f h\propto y$ so that one carry out the integral explicitly; indeed this is exactly the situation for
the explicitly known solutions of the STU model (in the standard coordinates). Furthermore, this approach to computing the on-shell
action was used in \cite{Crossley:2014oea}.

\section{Comparison with free $\mathcal{N}=4$ SYM theory}\label{app:freefield}
It is interesting to compare\footnote{We thank Yifan Wang for suggesting this comparison.} our holographic results for the STU solutions with free field computations
in $\mathcal{N}=4$ SYM theory, utilising the results of \cite{Bianchi:2021snj}. For simplicity we consider the case of no conical singularities, $n=1$,
and, recalling \eqref{ads3ansbdy}, in this appendix we set 
\begin{align}
e^{2V_0}=1\,.
\end{align}

We first consider defects preserving $\mathcal{N}=(2,2)$ supersymmetry, obtained by setting one of the $g\mu_i$ to zero,
and for definiteness we set $g\mu_3=0$ (and using the supersymmetry constraint, hence $g\mu_B=0$).
In this case, from \eqref{bexpstumodelkfixed2} with $n=1$, we obtain
\begin{align}\label{enhancedsusyb}
b={3N^2}(g\mu_1)^2\,.
\end{align}
In addition, from \eqref{d2hd} and \eqref{jsvevsetcn22case}, for these defects we have 
$d_2=18\pi h_D={6N^2}(g\mu_1)^2$.

We now compare these results with a perturbative computation in $\mathcal{N}=4$ SYM theory.
We can view $\mathcal{N}=4$ SYM theory as an $\mathcal{N}=2$ vector multiplet
coupled to a hypermultiplet, both in the adjoint of an $SU(N)$ gauge group,
with manifest $SU(2)_F\times SU(2)_R\times U(1)_R\subset SU(4)$ 
global symmetry where $SU(2)_R\times U(1)_R$ is the $\mathcal{N}=2$ R-symmetry. 
We consider having a defect with
non-trivial monodromy for $U(1)_F\subset SU(2)_F$ flavour symmetry, which breaks $SU(2)_F$ to $U(1)_F$. 
The $\mathcal{N}=2$  vector multiplet is a singlet with respect to 
$SU(2)_F$ while the fermions and the bosons in the hypermultiplet both transform as doublets. 
In the limit that $g_{YM}\to 0$ we can obtain expressions for $b$ and $h_D$ from the free field theory computations
of \cite{Bianchi:2021snj}. Specifically, taking the periodicity of the monodromy as $g\mu_1\in [0,1]$, 
the result for a free hypermultiplet in Table 1 of \cite{Bianchi:2021snj} gives
$b=3(N^2-1)(g\mu_1)^2$ and $d_2=6(N^2-1)(g\mu_1)^2$.
This value of $b$ and the inferred value of $h_D=d_2/(18\pi)$, from \eqref{jhrelation},
precisely agree with our holographic computations \eqref{enhancedsusyb} and \eqref{jsvevsetcn22case}
in the large $N$ limit.

In fact, we can similarly compare the full space of STU defect solutions to free field theory computations. However, this requires some assumptions, as the results of \cite{Bianchi:2021snj} depend on the boundary conditions chosen for the fields at the defect. In order to make this comparison we therefore make the following assumptions. First, we assume that the parameters $\xi$ and $\tilde{\xi}$ for the various fields, which fix the boundary conditions
of the fields in \cite{Bianchi:2021snj}, depend on the monodromies $g \mu_i$, but the dependence is piecewise constant. Second, we assume that the boundary conditions respect the $\mathcal{N}=(0,2)$ supersymmetry, and that when one of the monodromies vanishes (say $g \mu_3\to 0$) we should recover the above $\mathcal{N}=(2,2)$ case. Finally, we assume the choice of boundary conditions should respect the $S_3$ symmetry permuting the three chiral multiplets, as well as the CP symmetry of $\mathcal{N}=4$ SYM (such that taking $g\mu_i \to -g\mu_i$ corresponds to switching the conserved defect supersymmetry from $\mathcal{N}=(0,2)$ to $\mathcal{N}=(2,0)$). 
We are then lead to make the following choice:
\begin{itemize}
\item For each positive monodromy ($g\mu_i > 0$), we choose $\xi=1, \tilde{\xi}=0$ for the scalar charged under it and $\xi=0$ for the associated fermion.
\item For each negative monodromy ($g\mu_i < 0$), we shift it by $+1$ to a positive value ($g\mu_i \to g\mu_i +1$), then choose $\xi=\tilde{\xi}=0$ for the scalar charged under it and $\xi=1$ for the associated fermion.  
\end{itemize}
Alternatively, this prescription can be expressed as taking the monodromy parameter in the formulas from \cite{Bianchi:2021snj} to be $\alpha = |g \mu_i| $, then choosing for the scalar fields charged under the associated $U(1)$ to be $\xi = \Theta(g\mu_i), \tilde{\xi}=0$ (where $\Theta(g\mu_i)$ is the Heaviside function) and for the associated fermion $\xi=0$. 
Upon making these choices, we have the free field contribution to
$b$ and $d_2$ as a function of $g\mu_i$, with $g\mu_1+g\mu_2+g\mu_3 =0$, 
  to be given by
\begin{align}
b &= (N^2-1)\sum_{i=1}^3\left[ b^{\text{scalar}} (|g\mu_i|,\Theta(g\mu_i),0) + b^{\text{fermion}} (|g\mu_i|,0) \right]\,,\nn
d_2 &= (N^2-1)\sum_{i=1}^3\left[ d_2^{\text{scalar}} (|g\mu_i|,\Theta(g\mu_i),0) + d_2^{\text{fermion}} (|g\mu_i|,0) \right]\,,
\end{align}
where
\begin{align}
b^{\text{scalar}}(\alpha,\xi,\tilde{\xi}) &= \frac{1}{2}(1-\alpha)^2 \alpha^2 + 2\xi\alpha^3 + 2\tilde{\xi}(1-\alpha)^3\,,\nn
d_2^{\text{scalar}}(\alpha,\xi,\tilde{\xi}) &= \frac{3}{2}(1-\alpha)^2\alpha^2 + 6\xi\alpha^3 + 6\tilde{\xi}(1-\alpha)^3\,,\nn
b^{\text{fermion}}(\alpha,\xi) &= \frac{1}{2}\alpha^2 \left( 2-\alpha^2 - 2\xi(3-2\alpha) \right) + \frac{1}{2}\xi \,,\nn
d_2^{\text{fermion}}(\alpha,\xi) &= \frac{3}{2} \alpha(1-\alpha) \left( \alpha(1+\alpha)+2\xi(1-2\alpha)\right)\,.
\end{align}
Note that the formulas used for the fermion here have an extra $\frac{1}{2}$ factor in comparison to the results of \cite{Bianchi:2021snj}, since that paper uses Dirac fermions whereas here we are using Weyl fermions.
A computation then gives
\begin{align}
b &= -3 (N^2-1)[(1-g\mu_1)(1-g\mu_2)(1-g\mu_3)-1] \,,\nn
d_2 &= 3(N^2-1) \left[\sum_{i=1}^3 (g\mu_i)^2 + 3 (g\mu_1) (g\mu_2)(g\mu_3)\right] \,,
\end{align}
which precisely agrees with our results for $b$ in \eqref{bexpstumodelkfixed2} 
and $d_2=18\pi h_D$ from \eqref{hdeesnfour}, in
the large $N$ limit,  after setting $\kappa=-1$.

It is striking that $b$ and $d_2$ agree with the free field results (with the above assumptions).
In general, $b$ can depend on bulk marginal couplings, like the SYM coupling constant. However, when
the defect preserves $\mathcal{N}=(0,2)$ supersymmetry, $b$ should be independent of the coupling constant \cite{Bianchi:2019umv},
in accord with above discussion. We are not aware of similar arguments for $d_2$ (or $h_D$). However,
the arguments made in \cite{Bianchi:2019umv} should extend to $b$ being independent of the bulk marginal
couplings in the presence of a conical singularity i.e. $n\ne 1$. Assuming this to be true, then 
the result \eqref{derivbwrtn} that $h_D$ and $d_2$ can be obtained as a constrained derivative of $b$ with respect to $n$ (see \eqref{dbexpress2}), would explain the fact that $h_D,d_2$ are independent of the coupling constant.
It is natural to conjecture that this is true for defects preserving $\mathcal{N}=(0,2)$ supersymmetry in general.

\section{Periodicity of the monodromy}\label{appperiod}
In this appendix we discuss our holographic results with regard
to the periodicity of the monodromy parameters. 

For simplicity, consider solutions with no conical singularity, i.e. $n=1$.
Naively, one would expect the STU solution space to be invariant under the transformations 
$ g\mu_i \to g\mu_i + m_i$ where $ m_i \in \mathbb{Z}$,
as they correspond\footnote{The correct periodicity here can be deduced from the coupling of the gauge fields to the boundary CFT fields.} to proper $U(1)$ gauge transformations on the $S^1$ surrounding the defect, as long as $ m_1 + m_2 + m_3 = 0$
(to maintain the supersymmetry constraint $g\mu_1+g\mu_2+g\mu_3=0$). Similarly, for the $\varphi\neq 0$ and LS solutions one would expect invariance under $g\mu_F \to g\mu_F + 2m_F$ with $m_F \in \mathbb{Z}$. However, our holographic solutions do not have these periodicities. 

\begin{figure}[htbp]
\begin{center}
\includegraphics[scale=.5]{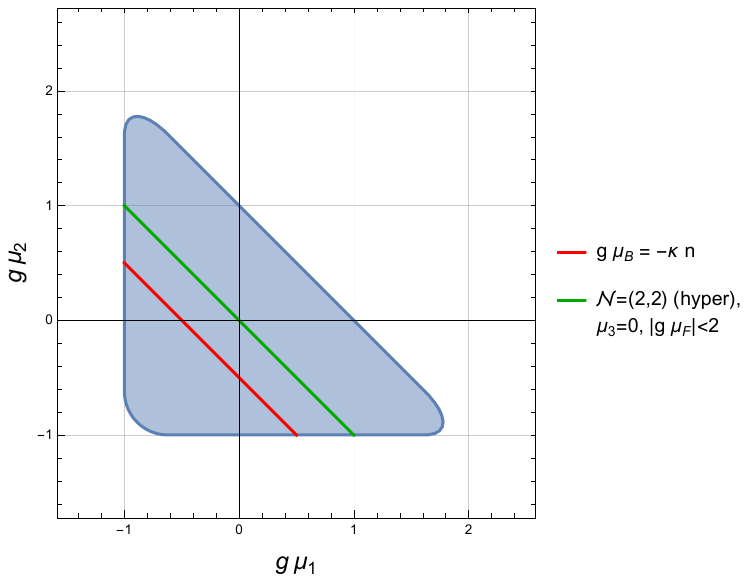}
\includegraphics[scale=.5]{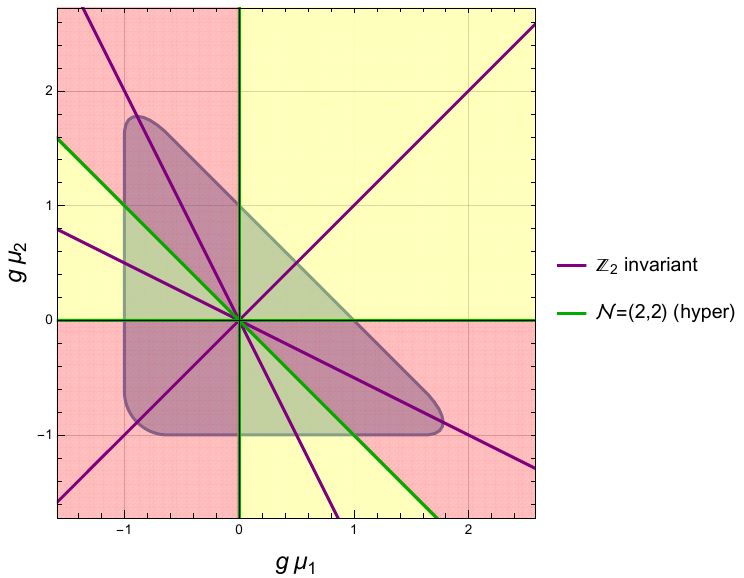}
\caption{The STU defect solution space for $n=1$ and $\kappa=1$. The blue region represents the values of $ g\mu_1, g\mu_2$ that correspond to a regular solution. 
On the left panel, the red line represents the restricted STU solutions and the green line represents the $\mathcal{N}=(2,2)$ solutions with $g\mu_3 = 0$ corresponding to an $\mathcal{N}=2$ hypermultiplet. 
On the right panel, the purple lines represent the solutions respecting a $\mathbb{Z}_2$ symmetry switching between two chiral multiplets, whereas the green lines represent the three branches of $\mathcal{N}=(2,2)$ solutions. The yellow regions are all related to each other via $S_3$ permutations, and so are the red regions. The yellow regions are related to the red regions via CP conjugation (which also flips the sign of $\kappa$).
}
\label{fig:STUSolutionSpace1}
\end{center}
\end{figure}

As an example, consider the allowed $n=1$ STU solution space, depicted as the blue region in Figure \ref{fig:STUSolutionSpace1}
(recall figure \ref{fig:stusolnsspace}). 
The blue region clearly contains solutions that are related by the above gauge transformation, yet our results for the corresponding solutions (for example the value of $b$ and $h_D$) are not periodic. 
However, as discussed in \cite{Bianchi:2021snj} in the context of free fields, one must specify boundary conditions for the various CFT fields near the defect and 
these can depend on the values of the monodromy parameters and break the above invariance and hence periodicity. 
Related discussion also appears in \cite{Alford:1989ie}.

Pursuing this idea, for the STU defects we might expect the supersymmetry and R-symmetries in the bulk and on the boundary to force certain boundary condition choices, that respect the residual $S_3$ symmetry which permutes the three $\mathcal{N}=4$ SYM chiral multiplets, as well as the CP symmetry (which takes $g\mu_i\to-g\mu_i$ and $\kappa\to-\kappa$). Thus, at most we might expect the periodicity to hold just within each of the yellow and red regions on the right panel in Figure \ref{fig:STUSolutionSpace1} individually.
Interestingly, the intersection of each of these six regions with the blue region does not contain solutions related by a gauge transformation, potentially resolving the issue with periodicity. 
Note, however, that restricting to the blue region with appropriate discontinuous boundary conditions means
that there are values of $(g\mu_1,g\mu_2)$ outside the blue region
which aren't equivalent to any point within it, while still restricting to a given yellow or red region. 
Why the holographic solutions seem to restrict\footnote{Recall that the blue region arose from demanding that our holographic solutions are regular in the bulk.} the possible values of monodromies remains unresolved here and we leave this for future work. 
It would also be interesting to further explore related issues when we allow for conical singularities, relaxing the $n=1$ condition.
For example, when $n=10$ from figure \ref{fig:stusolnsspace} we see that the blue region is considerably larger and hence a new feature is
that there can be, in a given analogous yellow or red region, several values of $g\mu_i$ related by integer shifts and yet label different solutions;
this can be interpreted as being associated with multiple saddle points.

\section{Solutions of minimal gauged supergravity}\label{mingsugrasols}
For minimal gauged supergravity, associated with the STU model, we take $\varphi = \alpha=\beta= 0$, as well as $a^1=a^2= a^3  \equiv \frac{1}{3}a_R$. The BPS equations then take the form
\begin{align}
\label{eq:ads4_minsugra_genl}
f^{-1} \xi' &=-\frac{3g}{2}\cos\xi+2\kappa e^{-V}\,, \nonumber \\
f^{-1} V' &=-\frac{g}{2} \sin\xi \,, \nonumber \\
h &= - n e^{V} \sin\xi\,, \nonumber \\
a_R &= \frac{2}{g}(-s+\kappa n)-\frac{3}{2}n e^{V} \cos\xi \,,
\end{align}
as follows from \eqref{summbbpsapptext}, \eqref{summconstext}, \eqref{xiderivnice}.
We want to examine the known solution (e.g. \cite{Ferrero:2020laf}) in our conventions, so we consider
the ansatz
\begin{align}
e^V = e^{V_0} y\,, \qquad f = \frac{h_0}{h}\,.
\end{align}
The regularity condition for the metric at the core reads
\begin{align}
\label{eq:ads4_core_reg}
(h^2)'|_{y=y_*} = 2 h_0\,.
\end{align}
The second and third equation of \eqref{eq:ads4_minsugra_genl} then fix $h_0$ to be
\begin{align}
g h_0 = 2 n e^{V_0}\,.
\end{align}
Substituting into the first equation of \eqref{eq:ads4_minsugra_genl} we obtain
\begin{align}
\xi'= \frac{3}{y\tan\xi}-\frac{4 \kappa }{g e^{V_0}y \sin\xi}\,,
\end{align}
with solution
\begin{align}
\xi=\arccos\Big[\frac{2\kappa}{e^{V_0 }g y}+\frac{Q}{y^3}\Big]
\,,
\end{align}
for some constant $Q$.

It is convenient to introduce a new coordinate, $\bar y$, and rescaled constant, $\bar Q$, as follows:
\begin{align}
y=\frac{4\kappa}{3g e^{V_0}}\bar y^{1/2}\,,\qquad
Q=-\frac{1}{2}\left(\frac{4\kappa}{3g e^{V_0}}\right)^3 {\bar Q}\,.
\end{align}
The full solution then reads
\begin{align}
ds^2&=\left(\frac{2}{g}\right)^2\left[\frac{4\bar y}{9}ds^2(AdS_3)
-\frac{\bar y}{q}d\bar y^2-\frac{q}{9\bar y^2} n^2 dz^2\right]\,,\nn
g a_R&=-2s-n\kappa+n\kappa\frac{{\bar Q}}{\bar y}\,,
\end{align}
where $q$ is the cubic
\begin{align}
q=4\bar y^3-(3\bar y-{\bar Q})^2\,.
\end{align}
When ${\bar Q}=0$ and ${\bar Q}=1$ we have a double root and for $0<{\bar Q}<1$ there are three distinct positive roots (where the spindle solutions of \cite{Ferrero:2020laf} appear).
We take $\bar y$ to range from the largest root, $\bar y_*$, to infinity. An examination of the metric at the root $\bar y_*$
reveals that with
$\Delta z=2\pi$, as we have taken in the text, regularity of the metric determines $n$ via 
\begin{align}
n=\frac{6\bar y^{3/2}}{q}\Big|_{\bar y=\bar y_*}\,.
\end{align}

For $-\infty<{\bar Q}<1$ we find regular solutions with $s=-\kappa/2$, i.e. lying on the main branch.
As ${\bar Q}$ increases from $-\infty$ to $1$ we find that regular solutions have $n$ monotonically increasing and lying
in the range $0<n<\infty$,
with $n=1$ corresponding to ${\bar Q}=0$. A closer examination of the ${\bar Q} =0$ solution reveals that it is
in fact just the vacuum $AdS_5$ solution as one sees after making the coordinate transformation
$\bar y=\frac{9}{4}\cosh^2 \rho$; this is as expected
since there is no defect for $n=1$ in
the minimal gauged supergravity theory. As $n\to \infty$ we approach the $a\to 1 $ solution discussed below.
For these main branch solutions we find from \eqref{genstuhdexpress}, \eqref{litjistu} and \eqref{genbstumod3}
\begin{align}
h_D=\frac{(n-1)(1+2n)^2}{n^3}\frac{N^2}{54e^{2V_0}\pi}\,,\qquad
b=\frac{(n-1)(1+7n+19n^2)}{9n^2}N^2
\end{align}
In particular we see that for $n>1$ we have $h_D,b>0$, while for $0<n<1$ we have $h_D,b<0$.

For ${\bar Q}>1$ we find that the regular solutions now have $s=+\kappa/2$ i.e. lying on branch 2. We now
find regular solutions with $n$ lying in the range $0<n<1/3$ with $n$ monotonically decreasing as
${\bar Q}$ increases from $1$ to $\infty$. In particular, in the minimal gauged supergravity theory the branch 2 solutions
exist for a range of $n$ that is smaller than the possible allowed range $0<n<\frac{1}{2}$ that we deduced in section \ref{sectstu}.
For these branch 2 solutions we find from \eqref{genstuhdexpress} and \eqref{litjistu}
\begin{align}
h_D=\frac{(n+1)(2n-1)^2}{n^3}\frac{N^2}{54e^{2V_0}\pi}\,,
\end{align}
and we see that $h_D>0$ for the allowed range.

We now return to the ${\bar Q}=1$ case.
The metric can be written (dropping the bar on $\bar y$ and setting $g=2$ for simplicity):
\begin{align}\label{aonecase}
ds^2=\frac{y^2}{4}ds^2(AdS_3)-\frac{y^2 dy^2}{(y-1)^2(y^2+2y-1)}-\frac{(y-1)^2(y^2+2y-1)}{4y^2} n^2 dz^2\,,
\end{align}
with $F=-(n\kappa/2 y^2) dy\wedge dz$.
As $y\to 1$ we have
\begin{align}
ds^2_4&\to \frac{1}{4}ds^2(AdS_3)-\frac{1}{2}[\frac{dy^2}{(y-1)^2}-(y-1)^2 n^2 dz^2]\,,
\end{align}
with $F=-(n\kappa/2 ) dy\wedge dz$. If we take $ nz$ to be a non compact coordinate (as is natural in thinking of this solution
as arising as the $n\to\infty$ limit of the main branch solutions above), this is the standard supersymmetric $AdS_3\times H^2$ solution (supported by a topological twist), with
the $H^2$ written in coordinates adapted to the upper half plane \cite{Klemm:2000nj,Naka:2002jz}. Also, as $y\to\infty$ we approach 
the $AdS_5$ vacuum with $AdS_3\times\mathbb{R}$ boundary. 
This solution can be contrasted with the standard magnetically charged black string with $H^2$ horizon 
(\cite{Klemm:2000nj} and e.g.  
(3.11) of \cite{Gauntlett:2007sm})
that interpolates between $AdS_5$, with 
$\mathbb{R}^{1,1}\times H^2$ boundary and the above $AdS_3\times H^2$ solution.


\providecommand{\href}[2]{#2}\begingroup\raggedright\endgroup

\end{document}